\documentclass[epj]{svjour}
\usepackage{graphics}
\usepackage{amsfonts}
\usepackage{amsmath}

\begin{document}
\title{
Stochastic quantum dynamics beyond mean-field
}
\author{Denis Lacroix\inst{1} \and Sakir Ayik\inst{2}
}                     
\offprints{}          
\institute{Institut de Physique Nucl\'eaire, IN2P3-CNRS, Universit\'e Paris-Sud, F-91406 Orsay Cedex, France \and Physics Department, Tennessee Technological University, Cookeville, TN 38505, USA}
\date{Received: date / Revised version: date}
%
\abstract{Mean-field approaches where a complex fermionic many-body problem is replaced by an ensemble of independent particles 
in a self-consistent mean-field can describe many static and dynamical aspects. It generally provides a rather good approximation for 
the average properties of one-body degrees of freedom. However, the mean-field approximation  generally fails to produce 
quantum fluctuations of collective motion. To overcome this difficulty, noise can be added to the mean-field theory leading to a stochastic description of the 
many-body problem. In the present work, we summarize recent progress in this field and discuss approaches where fluctuations have been 
added either to the initial time, like in the Stochastic Mean-Field theory or continuously in time as in the Stochastic Time-Dependent Hartree-Fock. In some cases, the initial problem can even be re-formulated exactly by introducing Quantum Monte-Carlo methods in real-time. The 
possibility to describe superfluid systems is also invoked.
Successes and shortcomings of the different beyond mean-field theories are discussed and illustrated.    
} 
\maketitle

\tableofcontents


\section{Introduction}
\label{intro}

The Time-Dependent Hartree-Fock (TDHF) or its Density Functional Theory (DFT) variant, 
by replacing the many-body dynamical problem of interacting particles on the dynamics of independent 
particle moving in an average self-consistent mean-field are certainly among the most useful tools ever to describe 
many facets of mesoscopic systems. Almost 50 years after its first application in nuclei \cite{Bon76}, we have observed in the last decade a renewal of interest 
in the development of dynamical mean-field theories for the nuclear many-body problem \cite{Kim97,Sim01,Nak05,Mar05,Uma05,Was08}. 
Most advanced applications of TDHF have  reached a certain level of maturity. Nowadays, calculations are performed in three dimensions without specific symmetries and including all components of the effective interaction. For a recent review, 
please see \cite{Sim10,Sim12a,Sim12b}. Recently, efforts have been made to include pairing correlations \cite{Has07,Ave08,Eba10,Ste11,Sca12,Eba12,Sca13,Bul13} 

Despite these important progress made on the applicability of time-dependent mean-field approaches including pairing 
or not, it is known already from the early time of this field \cite{Neg82} that mean-field alone cannot describe all 
aspects of nuclei.  Generally speaking, mean-field theory suffers from the underestimation of quantal effects in 
collective space and from the absence of dissipative effects induced by the coupling between single-particle degrees 
of  freedoms (DOF) with more complex internal DOF. The former limitation is well know from nuclear structure 
mean-field practitioners where configuration mixing techniques are generally employed to recover the effect of quantal collective fluctuations \cite{Rin80,Ben03}. Missing dissipative aspects are clearly pointed out, for instance in the context 
of nuclear giant resonances, where the small amplitude limit of time-dependent mean-field (RPA or QRPA) is widely used. 
In that case, it is well known that mean-field alone is able to describe qualitatively the mean collective energy but 
most often fails to reproduce the fragmentation and damping of collective excitations \cite{Lac04,Sca13a}.

Many theories, called hereafter beyond-mean field approaches and discussed briefly here, have been proposed 
 to increase the predictive power of microscopic methods using the mean-field as a building block. However, 
most often due to the important numerical effort required, most of them have never been really applied to 
realistic situations.  The goal of the present review is to answer to the following question:

{\it While some effects cannot be included using the standard approach with only one mean-field trajectory, might it 
be possible to include some of the missing effects by considering an ensemble of mean-field trajectories, each of them being independent from the others?}

A positive answer to this question is an important step, since in that case (i) the required technology is the existing one 
since only standard mean-field codes should be used (ii) performing several independent trajectories is possible on several independent computers that are used everyday in our scientific life. Therefore, contrary to other methods that are limited by the handling of large matrices, a stochastic approach turns out to be more practical and timely.  

The present review is devoted to the description of recent progress in the field of stochastic quantum mechanic applied 
to the fermionic many-body problems. Depending on the physical effect that one wants to introduce beyond the mean-field, several approaches have been proposed. Some of them, like the Stochastic mean-field (SMF) are able to treat initial collective quantum fluctuations and have already been applied to realistic physical systems. Some others, most often 
dedicated to correlations beyond the independent particle picture that built up in time are still at the stage of formal development, requiring future efforts for practical applications.

This review is organized as follows. Basic concepts and methods associated to mean-field are first introduced. 
The introduction we make is oscillating between the usual academic one and arguments based on information theory. The latter is particularly useful to grasp the physical interpretation behind mean-field approaches. Deterministic approaches beyond mean-field are then briefly discussed. The rest of the review is devoted to the introduction of stochastic approaches where a correlated dynamical evolution is replaced by a set of independent particle evolutions.


%

\section{Mean-field in many-body systems}

\label{sec:meanfield}

We consider here an ensemble of $N$ particles interacting through the following  
Hamiltonian  
\begin{eqnarray}
\label{eq:hamil} H &=&   \sum_{ij} T_{ij} \, a^\dagger_i \, a_j
  + \frac{1}{4} \sum_{ijkl} \tilde{v}^{(2)}_{ijkl} \, a^\dagger_i \, a^\dagger_j \, a_l \, a_k  \nonumber \\
&&  + \frac{1}{36} \sum_{ijklmn} \tilde{v}^{(3)}_{ijklmn} \, a^\dagger_i a^\dagger_j a^\dagger_k a_n a_m a_l + \cdots,
\label{eq:hamitgen}
\end{eqnarray}
where $\{a^\dagger_i,a_i \}$ are creation/annihilation operators associated to a complete single-particle basis. 
$T$ denotes matrix elements of the kinetic energy term while $\tilde{v}^{(2)}$ and $\tilde{v}^{(3)}$, ...
correspond to fully anti-symmetric two-, three-, ... interaction matrix elements respectively.  

The quantum description of such a system requires a priori the knowledge of its $N$-Body 
wave function $\Psi^*( \{ {\mathbf r}_i \} ,t)$ or more generally its $N$-body density 
matrix denoted by $D(\{{\mathbf r}_i \},\{{\mathbf r'}_i \},t)$. $\{{\mathbf r}_i \}$ is a short-hand 
notation for the particles coordinates $({\mathbf r}_1, \cdots, {\mathbf r}_N  )$. 
The complexity of the Many-body problem comes from the number of degrees of freedom to consider. 
Except for very small  number of particles, 
the total number of degrees of freedom to treat becomes prohibitory to get 
the exact ground state or the evolution of such a complex system.
Therefore, we are forced to seek simplifications where much less relevant degrees
of freedom are considered. The most common strategy is to assume a hierarchy 
between those degrees of freedom depending on their complexity. The starting point 
of the hierarchy consists in focusing on one-body degrees of freedom only. 
At the second level, one- and two-body degrees of freedom are incorporated 
simultaneously and so on and so forth up to the exact description.
 
The aim of the present section is to consider the first level. 
Such an approach is motivated  
first by the fact that most of the observations 
generally made on an interacting system are related to one-body quantities: deformation, 
collective motion... Since any one-body operator writes  
$O^{(1)} = \sum_{ij} \left\langle i |O|j  \right\rangle a^\dagger_i a_j$, 
all the information on one-body properties is contained in the one-body density matrix defined as
\begin{eqnarray}
\rho^{(1)}_{ji}(t) & = &  Tr( a^\dagger_i a_j D(t)) \equiv \langle a^\dagger_i a_j  \rangle.
\end{eqnarray} 
Some properties of the one-body density as well as its connection with higher order densities are discussed 
in appendix \ref{app:dens}. Large effort is devoted to provide the best approximation on the one-body density only 
without solving the full problem. The main difficulty comes from the fact that the one-body density 
could not be fully isolated from other more complex degrees of freedom. 
Therefore by reducing the information on a closed system into a small set of variables, we are left 
with an open quantum system problem where this subset is coupled to the surrounding sets of irrelevant degrees of freedom.

In this chapter, variational principles are used as a starting point to discuss the reduction of information in 
many-body systems. 
Concepts like relevant/irrelevant observables, effective Hamiltonian dynamics, projections are first introduced 
from a rather general point of view. These concepts are then illustrated in the specific case 
of interacting particles.   


\subsection{Variational principles in closed systems}
\label{sec:varia}

Variational principles are powerful tools to provide approximate solutions 
for static or dynamical properties of a system when few degrees of freedom 
are expected to contain the major part of the information
\cite{Ker76,Bla86,Dro86,Fel00}.  
For time dependent problem, the Rayleigh-Ritz 
variational principle generalizes as  
\begin{eqnarray}
S = \int_{t_0}^{t_1} dt \left<  \Psi(t) \right|  i\hbar \partial_t - H  \left|  \Psi(t)  \right>,
\label{eq:varia}
\end{eqnarray} 
where $S$ denotes the action. The action should be minimized, i.e. $\delta S = 0$ under fixed boundary 
conditions $\left|  \delta \Psi(t_0)  \right> =0$ and 
$\left< \delta \Psi(t_1) \right|=0$.  The variation has to be made on all components of the wave-function. Denoting by 
$\Psi_i$ these components in a specific basis with $\Psi_i (t) = \langle i | \Psi(t)\rangle$, $S$ becomes\footnote{Note that, last expression can very easily be transformed
into a more symmetric and more natural form:
\begin{eqnarray}
S &=& \int_{t_0}^{t_1} dt {\cal L}
[\Psi^*,\Psi, \dot \Psi, \dot \Psi^*].
\end{eqnarray}
}
\begin{eqnarray}
S &=& \int_{t_0}^{t_1} dt \sum_i \Big\{ i\hbar \Psi^*_i (t) \partial_t  \Psi_i (t) - 
\sum_j \Psi^*_i (t) H_{ij} \Psi_j(t) \Big\} \label{eq:vargen1} \\
&\equiv& \int_{t_0}^{t_1} dt \Big\{ i\hbar \Psi^*  \partial_t  \Psi  - {\cal H} \left[ \Psi,\Psi^*
\right]  \Big\}, \nonumber \\
&=& \int_{t_0}^{t_1} dt {\cal L}[\Psi^*,\Psi, \dot \Psi] 
, \label{eq:vargen2}
\end{eqnarray}
where ${\cal H}[\Psi^*,\Psi ]$ and ${\cal L}[\Psi^*,\Psi ]$ stands for the time-dependent Hamiltonian and Lagrangian respectively 
written in a functional form.
In expression (\ref{eq:vargen2}), a discrete basis is used. The generalization to continuous basis is straightforward. If 
the states $\{ i\}$ do form a complete basis of the full Hilbert space relevant for the considered problem, then 
the minimization procedure leads to
\begin{eqnarray}
\left\{
\begin{array} {ccc}
i\hbar \partial_t | \Psi(t) \rangle &=& H  | \Psi(t) \rangle , \\
&& \\
-i\hbar \partial_t \langle \Psi(t) | &=&  \langle \Psi(t) | H  ,
\end{array}
\right.
\end{eqnarray} 
which is nothing but the standard Schr\"odinger equation and its adjoint. 
Note that the second equation has been obtained by making variations with respect to the components $\Psi_i$ after integrating by
parts and underlines the crucial role of boundary conditions. The connection to classical equation of motion can be made 
using the functional form and introducing the field $\Phi$ and momenta $\Pi$ coordinate such that $\Psi = (\Phi + i \Pi) /\sqrt{2}$, leading 
to \cite{Ker76}
\begin{eqnarray}
\frac{\partial \Phi}{\partial t} &=& \frac{\partial {\cal H}}{\partial \Pi}, ~~~  
\frac{\partial \Pi}{\partial t} = - \frac{\partial {\cal H}}{\partial \Phi},
\end{eqnarray}
which are nothing but Hamilton's equations for the conjugate variables $(\Phi,\Pi)$. 

\subsection{Selection of specific degrees of freedom and Ehrenfest theorem}
\label{sec:ehrenfest}
The interest of variational principle is obviously not to recover the Schr\"odinger equation but stems from the possibility to 
restrict the variation to a smaller sub-space of the full Hilbert space and/or to a specific class of wave-functions.
Then, the dynamics is not exact anymore but will be the best approximation within the selected space or trial states 
class.   

We will consider here the important case where specific local transformations exist between any of the trial state $| \Psi \rangle$ and surrounding 
states. Explicitly, we consider the case:
\begin{eqnarray}
| \Psi + \delta \Psi \rangle = e^{\sum_\alpha \delta q_\alpha A_\alpha} | \Psi \rangle ,
\label{eq:variastate}
\end{eqnarray}  
where $\{\delta q_\alpha \}$ and $A_\alpha$ denotes respectively a set of parameters and operators. In most cases, 
the set of trial states is written as \cite{Cho96} 
\begin{eqnarray}
| \Psi ({\mathbf Q}) \rangle &=& R({\mathbf Q})  | \Psi(0) \rangle = e^{{\mathbf Q}.{\mathbf A}} | \Psi(0) \rangle 
\label{eq:transfogen}
\end{eqnarray} 
$R({\mathbf Q})$ is an element of the Lie Group constructed from a parameters set ${\mathbf Q} \equiv \{ q_{\alpha} \}$ and 
from its generators ${\mathbf A} \equiv \{A_\alpha \}$. 
Most often $| \Psi(0) \rangle$, is a state of the irreducible representation of the group. 
The most common examples are coherent states, independent particle states or quasi-particles 
vacuum (the two latter cases will be illustrated below). States written as in eq. (\ref{eq:transfogen}) are implicit 
functionals of ${\mathbf Q}$, in the following, the simple notation $| {\mathbf Q} \rangle \equiv | \Psi ({\mathbf Q}) \rangle$
is used. Variations with respect to the wave-function are now replaced by variations with respect to the parameters ${\mathbf Q}$
with:
\begin{eqnarray}
\left\{
\begin{array} {ccc}
| \delta {\mathbf Q} \rangle &=& \sum_\alpha \delta q_\alpha \left(\frac{\partial }{\partial q_\alpha}| {\mathbf Q} \rangle  \right) \\
\\
\left< \delta {\mathbf Q} \right| &=& \sum_\alpha \delta q^*_\alpha(t) \left( \frac{\partial }{\partial  q^*_\alpha} \left< {\mathbf Q} \right| \right) 
\end{array}, 
\right.
\label{eq:varia1}
\end{eqnarray}
or using the transformation (\ref{eq:transfogen}) between trial states: 
\begin{eqnarray}
| \delta {\mathbf Q} \rangle &=& \sum_\alpha \delta q_\alpha A_\alpha | {\mathbf Q} \rangle 
~{\rm and}~
\left< \delta {\mathbf Q} \right| = \left\langle {\mathbf Q} \right| \sum_\alpha \delta q^*_\alpha (t)  A_\alpha.
\label{eq:varia2} 
\end{eqnarray}
Using expressions (\ref{eq:varia1}) in the minimization, leads to the classical 
Euler-Lagrange equation of motion for the parameters \cite{Fel00}:
\begin{eqnarray}
\frac{d}{dt} \frac{\partial {\cal L}}{\partial \dot q_\alpha} = \frac{\partial {\cal L}}{\partial q_\alpha}, \hspace*{1.cm} 
\frac{d}{dt} \frac{\partial {\cal L}}{\partial \dot q^*_\alpha} = \frac{\partial {\cal L}}{\partial q^*_\alpha}.
\end{eqnarray} 
If instead, expressions (\ref{eq:varia2}) are used, the following two equations of motion, corresponding respectively to the 
variations $\delta q^*_\alpha$ and $\delta q_\alpha$, are obtained:
\begin{eqnarray}
\left\{
\begin{array}{ccc}
i\hbar \langle  {\mathbf Q}  | A_\alpha  | \dot {\mathbf Q}  \rangle 
&=& \langle  {\mathbf Q}  | A_\alpha H | {\mathbf Q}  \rangle , \\
&& \\
i\hbar \langle  \dot {\mathbf Q}  | A_\alpha  | {\mathbf Q}  \rangle  &=& -\langle  {\mathbf Q}  | H A_\alpha | {\mathbf Q}  \rangle , \\
\end{array}
\right.
\label{eq:evolq}
\end{eqnarray} 
which combined together gives the evolution 
\begin{eqnarray}
i\hbar \frac{d  \langle A_\alpha  \rangle}{dt} = \langle [A_\alpha , H ]  \rangle.\label{eq:ehr}
\end{eqnarray} 
We recognize here nothing but the Ehrenfest theorem, giving the evolution 
of any operator $ \{ A_\alpha \}$ with the Hamiltonian $H$. Therefore, starting from a density $D(t_0) = \left|  {\mathbf Q}\right>\left<  {\mathbf Q}\right|$, 
for one time step the $\{ \langle A_\alpha \rangle\}$ evolutions identify to 
the exact evolution although the state is constrained to remains in 
a sub-class of trial states. 

\subsection{General aspects and validity of the mean-field theory}

\begin{figure}
\resizebox{0.45\textwidth}{!}{
  \includegraphics{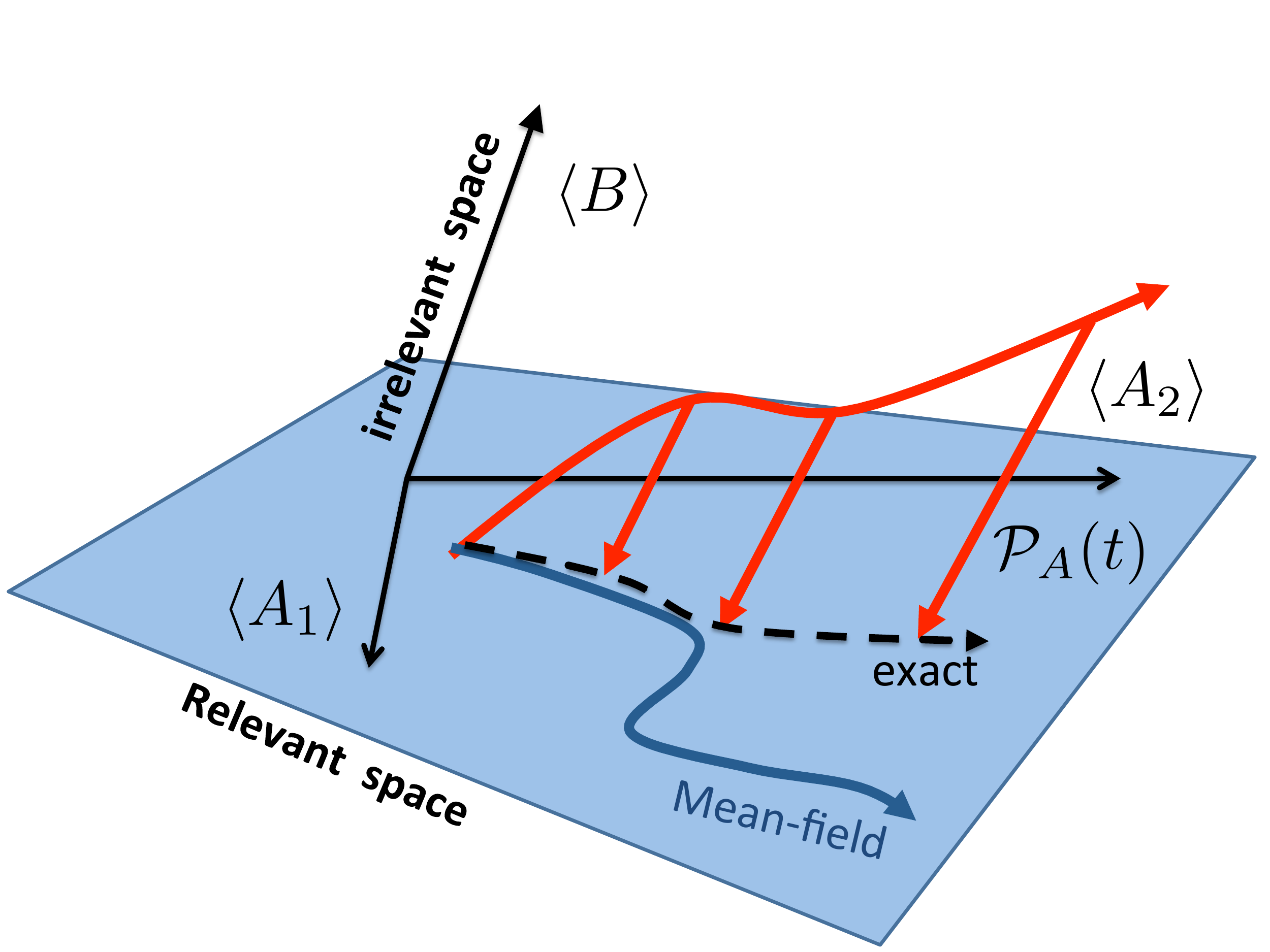}
}
\caption{(Color online) Schematic representation of the mean-field evolution. 
The space associated to relevant DOF, the $\{ \langle A_\alpha  \rangle \}$  is displayed 
by the shaded blue area, while irrelevant degrees of freedom are labelled generically by 
$ \{ \langle B \rangle \}$. Assuming, that the state is initially properly described in the relevant space, the exact 
dynamics escape from this space due to 
the coupling with irrelevant DOFs. The mean-field evolution can be seen as an approximation 
to the exact dynamics projected on the relevant space. This projection is expected to be 
perfect for short time, but will deviate from the exact projection for long time.   }
\label{fig:proj}      
\end{figure}

The mean-field concept is inherently connected to the selection of some degrees of freedom (DOF) that 
are assumed to be of particular relevance for the specific problem under interest. Starting from a simple trial 
state, the use of a variational principle insures that the $\langle A_\alpha  \rangle$ dynamics is good 
over short time. Due to the specific properties of the variational space, it also lead to a closed set of the 
equation of motion, i.e.
\begin{eqnarray}
\frac{d  \langle A_\alpha  \rangle}{dt} = {\cal F} \left(  \{ \langle A_\beta \rangle \} \right).\label{eq:allobs}
\end{eqnarray}
Therefore, the knowledge of the expectation values $\{ \langle A_\alpha  \rangle \}$ at initial time is sufficient 
to perform the long time dynamics.    
The functional $ {\cal F} (\cdot)$ is generally a rather complex, most often non-linear function of the relevant DOF. 
The mean-field dynamics can be schematically represented as in Fig. \ref{fig:proj} and corresponds to an approximation
to the exact evolution 
projected on the selected DOF. 
Besides the schematic picture, a projector ${\cal P}_A(t)$ can be explicitly 
introduced (see appendix \ref{sec:acp} and \ref{sec:projecteddyn}) associated to the information 
carried out by the observation of relevant DOF at a given time. Then, the Hamiltonian can be separated 
as:
\begin{eqnarray}
H &=& \underbrace{{\cal P}_A(t) H}_{H_{\rm MF}(t)} + \underbrace{(1-{\cal P}_A (t)) H}_{V_{\rm res}(t)}. \nonumber
\end{eqnarray} 
Then, the mean-field approximation consists in neglecting the residual interaction  $V_{\rm res}(t)$. 
The validity of the mean-field approximation strongly depends on the information carried out by 
the observables $\{ A_\alpha \}$ and on the effects of the residual part of the Hamiltonian.         

If the system is allowed to explore more general variational states, the evolution a priori will not be closed 
in terms of the relevant degrees of freedom and will be given by:   
\begin{eqnarray}
\frac{d  \langle A_\alpha  \rangle}{dt} = {\cal H} \left(  \{ \langle A_\beta \rangle \}, \{\langle B_i \rangle \}\right)\label{eq:allobs2}
\end{eqnarray}
where $\langle B_i \rangle$ is a generic notation for all degrees of freedom that are not in the $\langle A_\alpha \rangle$ space. 
Due to the coupling between the selected DOF and other DOF, the mean-field evolution is not expected to be valid for the long time evolution. To overcome this
difficulty, it is necessary to go beyond mean-field by treating explicitly or implicitly the effect of 
irrelevant DOF on the set of $\{ A_\alpha \}$. The end of this section illustrates applications 
of the mean-field theory in the context of interacting fermions while the main goal of 
this review is essentially devoted to general  methods going Beyond Mean-Field.  

\subsection{Application of mean-field theory to the Many-Body problem}

Notions introduced in previous section are illustrated here for the problem of $N$ interacting particles. Our starting point is 
the variational principle written in this case as 
\begin{eqnarray}
S &=& \int_{t_0}^{t_1} dt  \int_{\{ {\mathbf r}_i \}}
 \left[ \prod_i d^3\mathbf r_i \right]  \nonumber \\
&& \Psi^*( \{ {\mathbf r}_i \} ,t)  \Big\{ i\hbar \partial_t - H( \{{\mathbf r}_i \} ) \Big\}  \Psi^*( \{ {\mathbf r}_i \} ,t)
\label{eq:varianbody} 
\end{eqnarray} 
where $\Psi$ denotes a N-body wave-packets functional 
of the particles positions $\{{\mathbf r}_i \} \equiv ({\mathbf r}_1, \cdots, {\mathbf r}_N  )$. $H$ corresponds to a Many-Body Hamiltonian given by (\ref{eq:hamil}). For the sake of simplicity, we will consider here two-body 
Hamiltonian only.  

In the following, the so-called "independent 
particle" or "mean-field" approximation is first presented for fermions.  
As illustrated previously, different strategies can be employed to introduce mean-field, namely variational
principle, Ehrenfest theorem, or projection leading to effective Hamiltonian. Although, strong connections 
exist between them, important features might be completely missed by using only one of them.
For instance, the reduction of information is best seen using variational principle while
missing pieces appear more clearly using the Ehrenfest theorem and/or a direct 
separation  of the Hamiltonian into a mean-field and residual part. For this reason, different approaches are discussed below.     

\subsubsection{The independent particle approach for fermionic systems}

If the two-body interaction is neglected in the Hamiltonian, then, the exact many-body wave-function is exactly known and 
reduces to an anti-symmetric product (Slater determinant) of single-particle orthogonal states, denoted by $\{ \varphi_\alpha \}$     
\begin{eqnarray}
\Psi(\{ {\mathbf r}_i \}) & = & {\cal A}\left(\varphi_1({\mathbf r}_1) \cdots \varphi_N({\mathbf r}_N) \right) \label{eq:slater} 
\end{eqnarray}
where ${\cal A}( . )$ denotes the anti-symmetrization operator. The associated 
one-body density matrix $\rho$ reads 
\begin{eqnarray}
\rho &=& \sum_{\alpha=1,N} | \varphi_\alpha \rangle \langle \varphi_\alpha |.
\end{eqnarray}
It can be easily 
checked that $Tr(\rho) = N$ and $\rho^2 = \rho$ underlining 
that it has exactly $N$  occupation numbers equal to one while the others equal to zero. In the following, 
we will use the greek notation  $\alpha$
for occupied (hole) states.
For independent particle states, correlations matrices vanish 
at any order and all the information on the system is contained in the one-body density matrix. This is illustrated 
by the fact that, 
for any order $k$, the $k$-body density matrix is given by  an anti-symmetric product of the one-body density matrix 
(see appendix A):
\begin{eqnarray}
\rho_{12} &=& \rho _{1}\rho _{2}\left( 1-P_{12}\right) , \nonumber \\
\rho _{123} &=& \rho _{1}\rho _{2}\rho_{3}\left( 1-P_{12}\right) \left( 1-P_{13}-P_{23}\right), \nonumber \\
&& \cdots \nonumber
\end{eqnarray}     
Here, notations of refs. \cite{Lac04,Sim08}) are used, where 
the indices refer to the particle on which the operator is applied.
Therefore, all observables including the energy becomes a functional of the one-body density matrix components, that 
plays the role of the $\{ \langle A_\alpha \rangle \}$ introduced previously
 
\subsubsection{ Mean-field from variational principle} 

When the two- (and more) body interaction is plugged in, the wave-function cannot be written in the simple form (\ref{eq:slater}). However, one 
could still find an approximate solution by restricting the trial wave-function to a Slater determinant, this is the so-called Hartree-Fock 
or Time-Dependent Hartree-Fock approximation first proposed in refs. \cite{Har28,Foc30} and \cite{Dir30}. 
In that case, the action reduces to \cite{Ker76}
\begin{eqnarray}
S &=& \int_{t_0}^{t_1} ds  \sum_\alpha 
\int_{\mathbf r} d^3{\mathbf r} \Big\{ i\hbar 
 \varphi^*_\alpha({\mathbf r},s)   \dot \varphi^*_\alpha({\mathbf r},s)  \nonumber \\
 && \hspace*{2.5cm} - {\cal H} [\{ \varphi_\alpha \}, 
\{ \varphi^*_\alpha \} ] \Big\},
\label{eq:variamf} 
\end{eqnarray} 
where ${\cal H} [\{ \varphi_\alpha \}, \{ \varphi^*_\alpha] \} ] $ is given by 
\begin{eqnarray}
{\cal H} [\{ \varphi_\alpha \}, \{ \varphi^*_\alpha \} ] &=& \sum_\alpha  \langle \varphi_\alpha | T | \varphi_\alpha \rangle \nonumber \\
&+& \frac{1}{4} \sum_{\alpha,\beta} 
 \langle \varphi_\alpha \varphi_\beta | \tilde v_{12} | \varphi_\alpha  \varphi_\beta \rangle . 
\end{eqnarray}
Such a state can grasp part of the two-body effects through the introduction of a self-consistent mean-field. Variation of the action 
(\ref{eq:variamf}) have to be made with respect to the components of the single-particle basis $\{ \varphi^*_\alpha ({\mathbf r}) \}$ or 
its complex conjugate $\{ \varphi_\alpha ({\mathbf r}) \}$ leading to
\begin{eqnarray}
i\hbar \frac{\partial }{\partial t} \varphi_\alpha ({\mathbf r})
&=& 
\displaystyle 
\frac{\delta {\cal H}}{\delta  \varphi^*_\alpha ({\mathbf r}) } ~{\rm and}~
i\hbar \frac{\partial }{\partial t} \varphi^*_\alpha ({\mathbf r}) = 
\displaystyle - \frac{\delta {\cal H}}{\delta  \varphi_\alpha ({\mathbf r}) } .  \label{eq:spmf}
\end{eqnarray}    
Above equations of motion are generally written in terms of the mean-field Hamiltonian defined through
\begin{eqnarray}
\frac{\delta {\cal H}}{\delta  \varphi^*_\alpha } &\equiv& h[\rho] \varphi_\alpha  ~{\rm with} ~
h[\rho] = T + {\rm Tr}_2 \tilde v_{12} \rho_2.
\label{eq:potmf}
\end{eqnarray}
Here, $T$ denotes matrix elements of the kinetic part while the second term corresponds to the average  
potential created by the $N$ particles. In equation (\ref{eq:potmf}), ${\rm Tr}_{2}(.)$  is the partial trace on the second 
particle (for instance $\left\langle i| {\rm Tr}_2 
\tilde v_{12} \rho_2  | j \right\rangle = \sum_{kl} \left\langle i k | \tilde v_{12}| jl   \right\rangle \left\langle l |\rho |  k \right\rangle$). Finally, the N-body wave-function reduces to a set of coupled Schr\"oedinger equation for occupied states (using the short-hand notation $| \varphi_\alpha \rangle = | \alpha \rangle$): 
\begin{eqnarray}
i\hbar \frac{d | \alpha \rangle}{dt} &=& h[\rho] | \alpha \rangle. \label{eq:mfstandard}
\end{eqnarray} \\

\noindent {\bf One-body density evolution:}
From the single-particle state evolution (\ref{eq:spmf}), we deduce that 
\begin{eqnarray}
i\hbar \partial_t \rho &=& (i\hbar \partial_t | \alpha \rangle ) \langle \alpha | + 
i\hbar \partial_t | \alpha \rangle  ( i\hbar \partial_t \langle \alpha |) \nonumber \\
&=& [h[\rho] , \rho ] . \label{eq:mfrho}
\end{eqnarray}
This equation of motion, called mean-field approximation or TDHF, 
represents the optimal path in the space of one-body observables for short time evolutions. Indeed,
Slater determinants correspond to a specific class of trial states discussed in section \ref{sec:ehrenfest}. According 
to the Thouless theorem \cite{Tho60}, any local transformation of a Slater determinant $| \Psi \rangle$ into another Slater determinant writes
\begin{eqnarray}
| \Psi+ \delta \Psi \rangle = e^{\sum_{i j} \delta Z_{ij} a^\dagger_i a_j} | \Psi  \rangle. 
\end{eqnarray}  
Said differently, the set of one-body operators $\{a^\dagger_i a_j \}$ are generators of the transformation between Slater determinants.
Accordingly, the variational principle automatically ensures that 
\begin{eqnarray}
i\hbar \frac{d}{dt} \langle  a^\dagger_i a_j \rangle &=& \langle [a^\dagger_i a_j , H ] \rangle
\end{eqnarray}  
along the path. It could indeed be checked 
that the evolution of one-body observables estimated through the Ehrenfest theorem using a Slater determinant 
gives the mean-field evolution (\ref{eq:mfrho}).  

\subsubsection{Mean-field dynamics from Thouless Theorem} 
\label{section:meanfieldThouless}

Mean-field evolution corresponds to a projected dynamic onto the space of relevant one-body
degrees of freedom where the coupling to irrelevant degrees of freedom (correlation) is neglected.
A projected Hamiltonian could be explicitly constructed using the projection technique introduced 
in section \ref{sec:projecteddyn}. Here, a more direct method is used first to directly separate the Hamiltonian 
into a mean-field part and residual part and second to illustrate that mean-field
could be obtained even without the variational principle. 
   
To precise the missing part, we write the Slater determinant in second quantization form  
$| \Psi \rangle = \Pi_\alpha a^\dagger_\alpha | - \rangle$ and complete the occupied states by a set (possibly infinite) of unoccupied single-particle
states (also called particle states) labeled by $\bar \alpha$ and associated to 
the creation/annihilation $a^\dagger_{\bar \alpha}$ and  $a_{\bar \alpha}$. 
The completed basis verifies
\begin{eqnarray}
\sum_\alpha \left| \alpha \right\rangle \left\langle \alpha \right| + 
\sum_{\bar \alpha} \left| \bar \alpha \right\rangle \left\langle \bar \alpha \right| \equiv \rho + (1-\rho) = {1}.
\end{eqnarray} 
From this closure relation, any creation operator associated to a single-particle states 
$\left| i \right\rangle$ decomposes 
as 
\begin{eqnarray}
a^\dagger_i &=& \sum_\alpha a^\dagger_\alpha \left\langle  \alpha \left.  \right| i \right\rangle + 
\sum_{\bar \alpha} a^\dagger_{\bar \alpha} \left\langle  \bar \alpha \left.  \right| i \right\rangle.
\label{eq:oadiph}
\end{eqnarray}
For instance, 
restarting from the general expression of $H$ and expressing the different single-particle states $(i,j,k,l)$ in the 
particle-hole basis gives:
\begin{eqnarray}
H \left| \Psi \right\rangle &=& \Big\{ -\frac{1}{2} Tr(\tilde v_{12} \rho_1 \rho_2 )  \nonumber \\
&&  \nonumber \\
&&+ \sum_{\alpha,\beta} \left\langle \beta \left| \rho h[\rho] \right| \alpha \right\rangle  a_{{\beta}}^{\dagger}  a_{\alpha}
+ \sum_{\bar \alpha \alpha} \left\langle  \bar \alpha \left| h[\rho] \right| \alpha \right\rangle 
a_{\bar{\alpha}}^{\dagger} a_{\alpha} \nonumber
\nonumber \\
&&+ \frac{1}{4} \sum_{\bar \alpha  \bar \beta \alpha \beta }  
\left\langle \bar \alpha \bar \beta \left| \tilde v \right| \alpha \beta \right\rangle 
a^\dagger_{\bar \alpha} a^\dagger_{\bar \beta} a_{\beta} a_{\alpha} \Big\}\left| \Psi \right\rangle
\label{eq:hphi}
\end{eqnarray}
where commutations have been performed in such a way that all creation operators are on the left and where 
$a^\dagger_\alpha | \Psi \rangle =  a_{\bar \alpha} | \Psi \rangle =0$ has been used.  The three terms above are 
denoted hereafter respectively by $E_0[\rho]$, $H_{MF}[\rho]$ and $V_{res}[\rho] $.

This expression is helpful 
to understand the approximation made at the mean-field level.
In previous section, 
we have shown that mean-field provides the best approximation for one-body degrees of freedom using the Ehrenfest theorem. Here, we will show that 
the mean-field evolution is equivalent to an effective Hamiltonian dynamics where $ V_{res}$ is neglected.   

Assuming that only $E_0[\rho]$ and $ H_{MF}[\rho]$ contribute to the evolution. Then, over an infinitesimal time step $dt$, 
the new state is approximated by  
\begin{eqnarray}
\left| \Psi (t+dt) \right\rangle \simeq \exp\left(\frac{dt}{i\hbar} E_0[\rho]\right) \exp \left(\frac{dt}{i\hbar} 
H_{MF}[\rho] \right) \left| \Phi \right\rangle.
\end{eqnarray} 
The first exponential is simply a global phase factor and will not contribute to observable evolution. 
The second contribution corresponds to an exponential of a one-body operator, which 
according to the Thouless Theorem 
\cite{Tho60} 
transforms a Slater Determinant into another Slater determinant. 

Indeed, using fermionic commutation rules, gives (see appendix \ref{app:mfthouless})
\begin{eqnarray}
\exp \left(\frac{dt}{i\hbar} 
 H_{MF}(\rho) \right) 
\left| \Phi \right\rangle &=& \Pi_{\alpha}  a^{\dagger}_{\alpha + d\alpha} \left| - \right\rangle,
\end{eqnarray}  
where the states $\left| \alpha + d \alpha \right\rangle$ are the new single-particle states deduced 
from the $\left| \alpha \right\rangle$ through the mean-field evolution, Eq. (\ref{eq:mfstandard}).
Besides the fact 
that the mean-field is directly recovered, 
another interest of the present approach is to provide 
an effective Hamiltonian that is directly separated into a relevant and irrelevant part. As shown in next chapter, 
an explicit expression of $\hat V_{res}$ is useful to 
discuss the departure from a mean-field dynamics.

\subsection{Mean-field with pairing correlations}
\label{sec:tdhfb}
   
Mean-field theory is sometime restricted to the approximation were the many-body wave-function is replaced by a Slater 
determinant state. Here, mean-field will be more generally referred to the approximation where the trial state is a 
quasi-particle vacuum. Slater determinants is a sub-class of quasi-particle states with occupation $1$ and $0$. Applying the 
same technique as above, leads to the Time-Dependent Hartree-Fock Bogoliubov (TDHFB) where pairing correlations 
can be included.    
In that case, the generator of transformations 
between quasi-particle vaccua are the set of 
operators $\{a^\dagger_i a_j , a^\dagger_i a^\dagger_j , a_i a_j \}$. \\


\noindent {\bf  Quasi-particle vacuum: }       
We now consider a quasi-particle vacuum written as  
\begin{equation}
| \Psi \rangle \sim \prod_{\alpha} \beta_\alpha | - \rangle,
\label{vacuum}
\end{equation}
where the $\{ \beta_\alpha \}$ denotes a complete set of quasi-particle annihilation operators. This form automatically insures 
$\beta_\alpha | \Psi \rangle=0$ for any $\alpha$. The new quasi-particle states are defined 
through a specific linear combination (Bogoliubov transformation) 
of single-particle creation/annihilation operators 
$\{ a^\dagger_i, a_i \}$ \cite{Rin80}
\begin{eqnarray}
\left\{
\begin{array} {cc}
\beta_\alpha   = & \sum_{i} U^*_{i \alpha} a_i + V^*_{i\alpha} a^\dagger_i \\
\beta^\dagger_\alpha = & \sum_{i} U_{i\alpha } a^\dagger_i + V_{i\alpha } a_i .
\end{array}
\right.
\label{transfoBogo}
\end{eqnarray}
where matrices $U$ et $V$ have specific properties to insure 
that new operators $\{ \beta_\alpha , \beta^\dagger_\alpha \}$
verify fermionic anti-commutation rules. 

The information on the system is not anymore contained only in the normal density. Indeed, one should 
introduce the anomalous density whose 
matrix elements are defined by  
$\kappa_{i j} = \langle a_j a_i \rangle$ 
(which also implies $\kappa^*_{i j } = \langle a^\dagger_i a^\dagger_j \rangle$). 
Latter contractions cancel out for independent particle systems. 
The Bogoliubov  transformation (Eq.~(\ref{transfoBogo}))can be inverted to express the $a^\dagger$ and $a$ operators  
in terms of quasi-particles operators :
\begin{eqnarray}
\left\{
\begin{array} {cc}
a_i   = & \sum_{\alpha} U_{ i\alpha} \beta_\alpha + V^*_{i \alpha } \beta^\dagger_\alpha \\
a^\dagger_i   = & \sum_{\alpha} V_{ i \alpha} \beta_\alpha + U^*_{i \alpha } \beta^\dagger_\alpha .
\end{array}
\right.
\end{eqnarray}
Using these expressions in $\rho$ et $\kappa$, we deduce
\begin{equation}
\rho_{i j} = \sum_\alpha V_{j \alpha}V^*_{i \alpha} = 
\left(V^* V^T\right)_{i j},\hspace*{0.5cm} \kappa_{i j}= \left(V^* U^T\right)_{i j}.
\end{equation}
These contractions are generally presented as a generalized density matrix defined as 
\begin{eqnarray}
{\cal R}  = \left( 
\begin{array} {cc}
\left( \langle a^\dagger_j a_i \rangle \right) & \left(\langle a_j^{ } a_i \rangle \right)\\
&\\
\left(\langle a^\dagger_j a^\dagger_i \rangle \right) &  \left(\langle a_j a^\dagger_i \rangle   \right)
\end{array} 
\right)
 = \left( 
\begin{array} {cc}
\rho & \kappa \\
- \kappa^* & 1-\rho^*  
\end{array} 
\right).
\label{matriceR}
\end{eqnarray}
The new contractions make possible to treat a certain class of correlations that were neglected previously.
Using the Wick theorem \cite{Rin80,Bla86}, components of the associated two-body correlation matrix now read
\begin{eqnarray}
\rho^{(2)}_{ijkl}=\langle ij | \rho_{12} | kl \rangle &=&
\langle a^\dagger_k a^\dagger_l a_j a_i \rangle \nonumber \\
&=& \overline{ a^\dagger_k a_i }~\overline{ a^\dagger_l a_j } 
- \overline{a^\dagger_k a_j}~\overline{ a^\dagger_l a_i} + \overline{ a^\dagger_k a^\dagger_l} ~\overline{ a_j a_i}\nonumber \\
&=& \rho_{ik} \rho_{jl} - \rho_{il} \rho_{jk} + \kappa_{ij} \kappa^*_{kl} .
\label{Wick2corps}
\end{eqnarray}
On opposite to Slater determinants, the correlation matrix denoted by  
 $C_{12}$ does not a priori vanish. We further see that the HFB theory 
leads to separable form of the two body correlation matrix elements:
\begin{eqnarray}
C_{ijkl} &=& \kappa_{ij} \kappa^*_{kl}.
\label{eq:ckk}
\end{eqnarray}
In turn, the HFB is more complex than the HF one. For instance, 
the state is not anymore an eigenstate of the particle number operator. 
We say that the particle number symmetry $U(1)$  is explicitly broken.
Fluctuations associated to the particle number $N = \sum_\alpha a^\dagger_\alpha a_\alpha$ now write
\begin{eqnarray}
\langle N^2 \rangle - \langle N \rangle^2 = 2 \, {\rm Tr} (\kappa \kappa^\dagger) = 2 \, {\rm Tr}(\rho-\rho^2).
\end{eqnarray} 
In general, this quantity is non-zero for a quasi-particle vacuum.
This implies for instance that at least the particle number should be constrained in average 
in nuclear structure studies (this is generally done by adding a specific Lagrange multiplier to the variational principle).  
It is also worth to mention that in TDHFB, the expectation value $\langle N \rangle $ 
is a constant of motion. 
Therefore, no specific care of particle number is necessary in the dynamical case if it has been properly adjusted at initial time.  \\

\noindent {\bf TDHFB Equations:} Since the generators of transformation between quasi-particle states now include 
the $\{a_i a_j \}$ and their hermitian conjugate, minimization of the action is now equivalent to optimize associated 
equations of motion given by the equation of motion :  
\begin{eqnarray}
i \hbar \frac{d}{dt} \rho_{ji}   &=& i \hbar \frac{d}{dt}\langle a^\dagger_{i}a_{j} \rangle 
= \langle [a^\dagger_{i}a_{j},\hat{H} ]\rangle, \label{OBDM}\\
i \hbar \frac{d}{dt} \kappa_{ji} &=& i \hbar \frac{d}{dt}\langle a_{i}a_{j} \rangle 
= \langle [a_{i}a_{j},\hat{H} ]\rangle.\label{OBPT}
\end{eqnarray}
Using the Wick theorem, the set of TDHFB coupled equations are obtained \cite{Sim08}:
\begin{eqnarray}
i \hbar \frac{d}{d t}\rho &=& \left[h,\rho \right] + \kappa \Delta^\ast  - \Delta \kappa^\ast , \label{eq:TDHFBrho} 
\end{eqnarray}
and
\begin{equation} \label{kap}
i \hbar \frac{d}{d t}\kappa = h \kappa + \kappa h^\ast  - \rho \Delta - \Delta \rho^\ast  + \Delta, 
\end{equation}
where $\Delta$ denotes the pairing field:
\begin{eqnarray}
\Delta_{ij}=\frac{1}{2}\sum_{kl} \tilde {v}_{ijkl} \kappa_{kl}. 
\label{eq:Delta}     
\end{eqnarray} 
Finally, using the generalized density matrix $\mathcal{R}$ and generalized HFB Hamiltonian $\mathcal{H}$, defined as
\begin{eqnarray}
{\cal H}  \equiv \left( 
\begin{array} {cc}
h & \Delta \\
- \Delta^* & - h^*
\end{array} 
\right),
\end{eqnarray}
equations on $\rho$ and $\kappa$ can be written, in a more convenient form, as  
\begin{eqnarray}
i \hbar \frac{d{\cal R}}{dt} = \left[{\cal H} , {\cal R} \right] .
\label{eq:tdhfbR}
\end{eqnarray}
The 
TDHFB equation generalizes the TDHF case
(Eq.~(\ref{eq:mfrho})) by accounting for pairing effects in the dynamical evolution. 
There is nowadays a clear effort to extent state of the art mean-field codes by including pairing correlations
\cite{Has07,Ave08,Eba10,Ste11,Sca12}. 
Note finally that, similarly to 
the TDHF case, one can also directly obtain the mean-field equation with pairing by splitting the 
Hamiltonian into a HFB part and a residual interaction part (not shown here) \cite{Sca12}.    

\subsection{Summary and discussion}

In this section, basic ingredients of mean-field theory have been presented starting from a variational principle. 
Variational principles are very helpful to understand to what extend mean-field approximation provides an 
optimal description of selected degrees of freedom and can be understood as a projection of the exact dynamics 
on a subspace of observables. 

The independent particle approximation has the great advantage to replace the exact many-body problem 
by a much simpler one-body problem that can most often be treated numerically. However, it is rarely used 
in the form presented here, i.e. starting from an Hamiltonian and performing the Hartree-Fock or the Hartree-Fock 
Bogoliubov approximation.  The first reason is that
correlations called "beyond mean-field" play an important role: direct two-body effects, pairing, quantum zero point motion in collective 
space. TDHFB, presented above, corresponds to one of the possible extension of mean-field able to account for pairing effects. 
In the next section, a description of recent advances in quantum transport theory beyond mean-field is made.

A second and more subtle difficulty is that Hartree-Fock approximation starting from the bare interaction, for instance in condensed matter 
or in nuclear physics, does not 
provide a sufficiently good approximation to serve as a starting point for the nuclear many-body problem. 
To overcome this difficulty, the independent particle picture is still used but in a functional spirit 
within the Density Functional Theory (condensed matter) or Energy Density Functional (nuclear physics) framework.



%
%

\section{Dynamical Theories Beyond mean-field: a survey on deterministic approaches}
\label{sec:beyond}

\begin{figure}
\resizebox{0.45\textwidth}{!}{
  \includegraphics{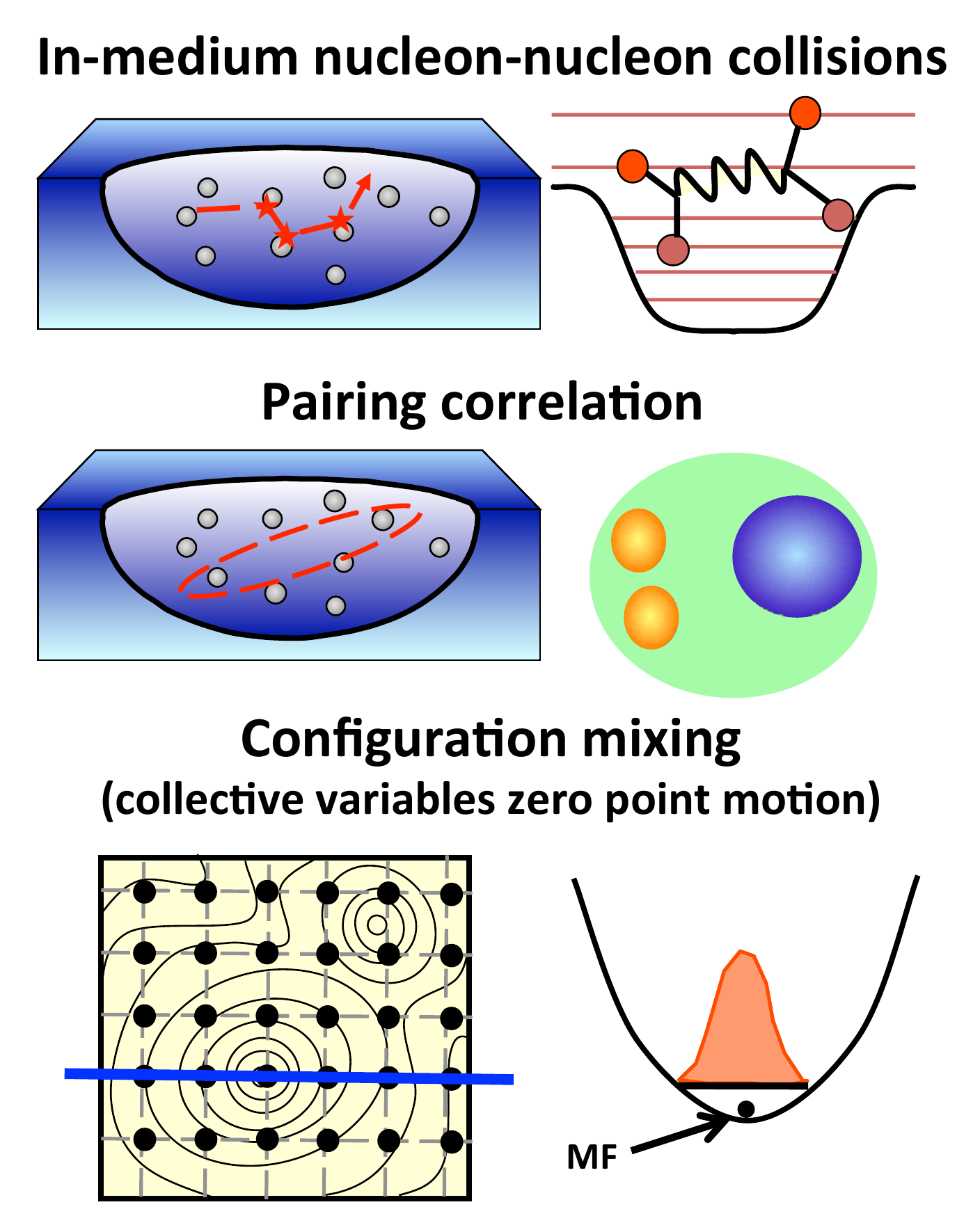}
}
\caption{Schematic illustration of the different types of correlation 
beyond mean-field. 
From top to bottom, direct in-medium nucleon-nucleon collisions, pairing and correlations associated to configuration mixing 
are respectively shown. Assuming that a system is properly described by a Slater determinant, 
direct nucleon-nucleon collisions is the first source of departure from the independent particle picture and is the physical 
process at the origin of thermalization. 
However, at low 
internal excitation energies, this effect is strongly hindered due to Pauli effect induced by surrounding nucleons and other 
correlations 
dominate. Pairing affects nuclear structure properties like masses, collective motion, pair transfer, ...
Configuration mixing, generally incorporated through the Generator Coordinate method, tell us that nuclei 
could not a priori be simply described by a single Slater determinant. While the latter 
misses fluctuations collective space, configuration mixing incorporates it properly. }
\label{fig:beyond}      
\end{figure}

In previous sections, the independent particle approximation to the N-body problem has been introduced. 
This approximation has played and continues to play a major role for our understanding of interacting systems. 
While the gross features of most nuclei are properly  
accounted for by replacing the complex many-body wave-function 
by a Slater determinant and an effective Hamiltonian (EDF), most often, physical processes 
reveal correlations beyond mean-field \cite{Ben03}. The complexity of nuclei stems from the many facets 
of correlations (see Figure \ref{fig:beyond}).
For instance short and long range correlations in static nuclei could
only be accounted for by a proper treatment of pairing effects and configuration mixing. 
Conjointly, as collision energies between two nuclei increase,  
the Pauli principle becomes less effective to block direct nucleon-nucleon collisions. 
Then, two-body correlations should explicitly be 
 accounted for. 
During the past decades, several approaches have been 
introduced to treat correlations beyond 
mean-field in a quantum theory. 
The development of such a theory has been strongly 
influenced by concepts developed for open quantum systems \cite{Bre02}. In that case, one-body 
degrees of freedom are the relevant observables and play the role of a system coupled to the surrounding 
environment of more complex observables. Recent advances in theories treating correlations beyond mean-field are 
presented here. A comprehensive list of theories introduced in this section is given in table 
\ref{tab:approches}, in each case the associated acronym and key observables are given. 

\begin{table*}[th]
    \begin{center}
{ 
\begin{tabular}{|l|l|l|l|} \hline
Name & approximation & Quantities & associated observables \\
 &  &  evolved & \\
\hline
&&&\\
TDHF & mean-field (m.-f.) & $\rho = \sum_\alpha | \varphi_\alpha \rangle \langle \varphi_\alpha |$ & one-body \\ 
&&&\\
TDHF-Bogoliubov  & m.-f. + pairing  & $\rho$, $\kappa$  & generalized one-body \\
(TDHFB) &&&\\
&&&\\
\hline
\hline \\
{\bf Beyond-Mean-Field } \\
{\bf Deterministic} \\
\hline
\hline \\
Extended-TDHF  & m.-f. + NN collision  & $\rho = \sum_\alpha | \varphi_\alpha \rangle n_\alpha \langle \varphi_\alpha |$  & one-body  \\ 
(ETDHF) & (dissipation) &&\\

&&&\\
Time Dept. Density Mat.  & m.f. + two-body correlations & $\rho$, $C_{12}$ & one- and two-body\\ 
(TDDM) &&&\\
&&&\\
TDDM$^P$ & m.f. + two-body correlations & $\rho$, $C_{12}$ & one- and two-body\\ 
 & (approximation of TDDM & & \\
 & focused on pairing) & & \\ 
&&&\\
\hline
\hline \\
{\bf Beyond-Mean-Field } \\
{\bf Stochastic} \\
\hline
\hline \\
Stochastic mean-field& m.-f. + initial fluctuation  & $D= \overline{| \Psi \rangle \langle \Psi |}$ & conf. mixing   \\ 
 (SMF)  & &Random Initial Value &\\
&&& \\
Stochastic-TDHF& m.-f. + NN collision  & $D= \overline{| \Psi \rangle \langle \Psi |}$  & one-body   \\ 
 (STDHF) &(dissipation+fluctuations)& Quant. Jump between SD& \\
&&&\\
Quantum Monte-Carlo & Exact (within stat. errors) & $D= \overline{| \Psi_1 \rangle \langle \Psi_2 |}$ & all   \\ 
 (QMC) & Quantum Jump &&\\
\hline
\end{tabular}
}
\end{center}
\caption{Summary of microscopic approaches presented in this document.}
\label{tab:approches}
\end{table*}

\subsection{Limitation of the mean-field theory.}
\label{sec:depart_tdhf} 

Mean-field theories is attractive due to its simplicity compared to 
the exact treatment of a many-body problem. While this approach  can grasp
many phenomena, it also suffers for some drawbacks that are discussed below.   

\subsubsection{Quantum fluctuations in collective space}
\label{sec:qf}

The TDHF or TDHFB approaches, discussed in previous section describe the quantal 
evolution of single-particles or quasi-particles. While quantal fluctuations in collective space are not strictly zero, 
it is clear from the Wick theorem that mean-field theory, relies on a quasi-classical approximation for quantum fluctuations in the space 
of relevant degrees of freedom, i.e.:
\begin{eqnarray}
\frac{1}{2}  \left( \langle A_\alpha A_\beta \rangle + \langle A_\beta A_\alpha \rangle\right)\approx  \langle A_\alpha \rangle \langle A_\beta \rangle . 
\end{eqnarray}  
This classical approximation does not necessarily implies that mean-field alone cannot be predictive 
for fluctuations. Unfortunately, for nuclear systems, fluctuations are severely underestimated 
compared to experimental observations. This problem is rooted in the description of the system in terms 
of a single independent particle or quasi-particle pure state. In nuclear structure studies, zero point collective fluctuations 
are generally incorporated through the mixing of different configurations, this is the so-called Generator coordinate method. In transport models, as we will illustrate below, one can eventually cure this problem by incorporating 
DOF associated to fluctuations directly in the description.

\subsubsection{departure from the single-particle (or quasi-particle) picture}

Even, if the system is properly described by a pure independent particle state at initial time, 
it is expected that correlations beyond the mean-field will be built up in time leading to a failure 
of mean-field approximation for long time evolution (see also Fig. \ref{fig:proj}.). 
In expression (\ref{eq:hphi}), a clear separation is made between 
what is properly treated at the mean-field level ($E_0[\rho]$ and 
$H_{MF}[\rho]$) and what is neglected, i.e. $V_{res}[\rho]$. At this point several comments are in order:
\begin{itemize}
\item[$\bullet$] The validity of the mean-field approximation depends on the intensity of the residual interaction
which itself depends on the SD state $\left| \Phi \right\rangle$ and therefore will significantly depend
on the physical situation. Using simple arguments \cite{Lic76}, 
the time $\tau_{SD}$ over which the Slater determinant picture breaks down could be expressed as:
\begin{eqnarray}
\tau_{SD} &=& \frac{\hbar}{2}  \Big(\frac{1}{N}
\sum_{\bar \alpha \bar \beta \alpha \beta} 
|\left\langle \bar \alpha \bar \beta \left| \tilde v \right| \alpha \beta \right\rangle|^2 \Big)^{-1/2}.
\end{eqnarray} 
In nuclear physics, typical values of the residual interaction leads to $\tau_{SD} \simeq 100-200$ fm/c.
Therefore, even if the starting point is given by an independent particle wave-packet, the exact 
evolution will deviate rather fast from the mean-field dynamics. This gives strong arguments in favor
of theories 
beyond TDHF. 
\item[$\bullet$] An alternative expression of the residual interaction which is valid in any basis, is
\begin{eqnarray}
V_{res}[\rho] &=& \frac{1}{4} \sum_{ijkl} \left\langle i j \left|( 1-\rho_1 )( 1-\rho_2 ) \tilde v_{12} \rho_1 
\rho_2 \right| k l \right\rangle a^\dagger_i a^\dagger_j a_l a_k. \nonumber
\label{eq:vres}
\end{eqnarray}  
This expression illustrates that the residual interaction associated to a Slater determinants could be seen as 
a "dressed" interaction which properly account for Pauli principle. Physically, the residual interaction corresponds
to direct nucleon-nucleon collisions between occupied states (2 holes) which could only scatter toward unoccupied
states (2 particles) due to Pauli blocking. 
We say sometimes that the residual interaction has a 2 particles-2 holes (2p-2h) nature.      
\end{itemize}    
Due to the residual interaction, the exact many-body state will become a more and more complex 
superposition of Slater determinants during the time evolution.
 As stressed in the introduction, due to the complexity 
of the nuclear many-body problem, the exact dynamic is rarely accessible. In the following section, methods 
to include correlations beyond mean-field, like direct nucleon-nucleon collisions, are 
discussed.  

\subsubsection{Strategy for Beyond mean-field approach to dynamics}

We have seen in previous section that the mean-field evolution can be obtained by 
selecting few relevant degrees of freedom. Then, these DOF are assumed to contain all the information on the system 
and any irrelevant DOF is assumed to be a functional of the $\{ \langle A_\alpha \rangle \}$.  Accordingly, the system evolution reduces to the $ \langle A_\alpha  \rangle$ evolution solely (see section \ref{sec:meanfield}):
\begin{eqnarray}
\frac{d  \langle A_\alpha  \rangle}{dt} = {\cal H} \left(  \{ \langle A_\beta \rangle \}, \{ \langle B_i \rangle \} \right) \rightarrow {\cal F} (  \{ \langle A_\beta \} \rangle ). \nonumber 
\end{eqnarray}
The most natural way to extend mean-field dynamics is  to explicitly treat the DOF $\{ \langle B_i \rangle \}$ that 
were neglected at the mean-field level (see Fig. \ref{fig:proj}). Except in few specific cases, one cannot consider the complete space 
of DOF and the set of irrelevant DOF should itself be truncated. This is what is done below where for instance only one- and two-body degrees of freedom are considered.  Then, Eq. (\ref{eq:allobs}) is complemented by a second set of equations of motion:
\begin{eqnarray}
\frac{d  \langle B_i  \rangle}{dt} = {\cal G} \left( \{  \langle A_\beta \rangle \}, \{ \langle B_i \rangle \} \right). \label{eq:irrel}
\end{eqnarray}
The main difficulty is the number of degrees of freedom to follow in time that becomes rapidly prohibitive for practical 
implementation. Noting that the knowledge of expectation values of the irrelevant DOF are rarely necessary, it is sometimes 
possible to reduce the complexity by not considering these DOF explicitly but by treating their effect 
on the relevant DOF space. Then, an improved evolution of the $\{\langle A_\alpha  \rangle \}$ can be found that 
takes the form:
\begin{eqnarray}
 \frac{d  \langle A_\alpha  \rangle}{dt} ={\cal F} ( \{  \langle A_\beta \rangle\} ) + {\cal K}  ( \{ \langle A_\beta \rangle \}) +
\delta K (\{ \langle A_\beta \rangle \} ). \label{eq:EMF}
\end{eqnarray}
where ${\cal F} (\cdot)$ is the previous mean-field functional. ${\cal K}  (  \langle A_\beta \rangle )$ denotes the effect of coupling to irrelevant DOF inducing 
departure from the mean-field path, while $\delta K (\langle A_\beta \rangle  )$ treats possible effect of the component 
$\{ \langle B_i \rangle \}$ that are not properly treated in the relevant space (like initial quantum fluctuations) 
and that propagates within the mean-field. Eq. (\ref{eq:EMF}) are generally rather complex and are obtained by first integrating Eq. (\ref{eq:irrel}) in time and then by projecting onto the space of relevant degrees of freedom.  In particular, contrary to the original mean-field, Eq. (\ref{eq:EMF}) might be non-local in time, i.e. the evolution at time $t$
depends on the whole system history.

\subsection{General correlated dynamics: the BBGKY hierarchy}
\label{sec:bbgky}
 
The use of the Ehrenfest theorem (section \ref{sec:ehrenfest}), underlines that the mean-field theory is particularly 
suited to describe one-body degrees of freedom. A natural extension of mean-field consists in following 
explicitly two-body degrees of freedom. 
Considering now the Ehrenfest theorem for the one and two-body degrees of freedom 
leads to two coupled
equations for the one and two-body density matrix components 
$\rho^{(1)}_{ij} = 
\langle a^\dagger_j a_i \rangle$ and $\rho^{(2)}_{ij, kl} = 
\langle a^\dagger_k a^\dagger_l a_j a_i \rangle$ 
\begin{equation}
\left\{ 
\begin{array}{cl}
i\hbar \frac{\partial }{\partial t}\rho _{1}
=&\left[ t_1,\rho _{1}\right] + \frac{1}{2}{\rm Tr}_{2}\left[ \tilde v_{12},\rho_{12}\right]  \\ 
&  \\ 
i \hbar \frac{\partial }{\partial t}\rho_{12} =& [t_1 + t_2 + \frac{1}{2} \tilde v_{12}, \rho_{12}] 
+ \frac{1}{2} Tr_3 \left[ \tilde v_{13} + \tilde v_{23} , \rho_{123} \right] \\ 
\end{array}
\right.  \label{eq:BBGKY}
.
\end{equation}
Above equations are the first two equations of a hierarchy equations, known 
as the Bogolyubov-Born-Green-Kirkwood-Yvon (BBGKY) hierarchy\cite{Bog46,Bor46,Kir46}
where the three-body density evolution is also coupled to the four body density evolution and so on and so 
forth. Here, we will restrict to the equations on $\rho_{1}$ and $\rho_{12}$ which have often served as
the starting point to develop transport theories beyond mean-field \cite{Cas90,Rei94,Abe96,Lac04}. 

\subsection{The Time-Dependent Density-Matrix Theory}
\label{sec:tddm}

Mean-field approximation neglects two-body and higher correlations ($C_{12}=0$). 
In that case, the equations on $\rho_{1}$ reduces to TDHF. A natural extension corresponds to neglecting three-body and 
higher order correlations ($C_{123} = 0$) \footnote{Introducing 
the permutation operator $P_{12}$ between two particles, defined as $P_{12} \left| ij \right\rangle = 
\left| ji \right\rangle$. The two-body correlation matrix is given by:
\begin{equation}
C_{12}=\rho _{12}-\rho _{1}\rho _{2}(1-P_{12})  \label{eq:c2}
\end{equation}
while the three-body correlations $C_{123}$ reads 
\begin{equation}
\begin{array}{ll}
C_{123}= & \rho _{123}-\rho _{1}C_{23}\left( 1-P_{12}-P_{13}\right) \nonumber \\
& -\rho_{2}C_{13}\left( 1-P_{21}-P_{23}\right) \\ 
& -\rho _{3}C_{12}\left( 1-P_{31}-P_{32}\right) \\
& -\rho _{1}\rho _{2}\rho
_{3}\left( 1-P_{13}\right) \left( 1-P_{12}-P_{23}\right).
\end{array}
\label{eq:c3}
\end{equation}}.  
The resulting theory where coupled equations between the one-body density $\rho_1$ and the two-body correlation
$C_{12}$ are followed in time are generally 
called Time-Dependent Density-Matrix (TDDM) theory (see for instance \cite{Cas90}):
\begin{eqnarray}
i\hbar \frac{\partial }{\partial t}\rho _{1} &=&\left[ h_{1}[\rho],\rho _{1}\right] + \frac{1}{2}{\rm Tr}_{2}\left[ \tilde v_{12},C_{12}\right] 
\end{eqnarray} 
together with
\begin{eqnarray}
i \hbar \frac{dC_{12}}{dt} &=& [ h[\rho]_1 + h[\rho]_2 , C_{12}] \nonumber \\
&+&  (1- \rho_1)(1- \rho_2) \tilde v_{12} \rho_{1}  \rho_{2}  - \rho_{1}  \rho_{2} \tilde v_{12} (1- \rho_1)(1 - \rho_2)  \nonumber \\
&+& \frac{1}{2} (1- \rho_1 - \rho_2) \tilde v_{12} C_{12} -  \frac{1}{2} C_{12} \tilde v_{12} (1- \rho_1 - \rho_2)   \nonumber \\
&+& \frac{1}{2} {\rm Tr}_3 \left[ (\tilde v_{13} + \tilde v_{23}) ,  \rho _{1}C_{23}\left( 1-P_{12}-P_{13}\right) \right] \nonumber \\
&+& \frac{1}{2} {\rm Tr}_3 \left[ (\tilde v_{13} + \tilde v_{23}) ,  \rho_{2} C_{13}\left( 1-P_{21}-P_{23}\right)  \right]. 
\label{eq:tddm} 
\end{eqnarray}
The second term in the evolution of $C_{12}$, denoted by $B_{12}$, is
called the Born term. It contains the physics of direct in-medium nucleon-nucleon 
collisions. Comparing $B_{12}$ and expression ($\ref{eq:vres})$, we see that it is directly proportional 
to the residual interaction. Indeed, starting from a Slater determinant 
($C_{12}(t_0) = 0$), this is the only term that does not cancel out in the evolution of $C_{12}$
over short time. In particular, it will be responsible for the departure from an independent 
particle picture.
The third term, denoted by ${P}_{12}$ has a less  straightforward interpretation. For instance, it has been 
shown that ${P}_{12}$ could be connected to pairing correlations  \cite{Toh04}. Finally the last two terms 
contains higher order p-p and h-h correlations. It is finally worth mentioning that the last term could eventually be 
modified to better account for conservation laws (see discussion in \cite{Pet94}).

Applications of the TDDM theory faces two major difficulties. First, since 
two-body degrees of freedom are explicitly considered, huge matrices have to be 
treated numerically and appropriate truncation schemes should be performed.
In addition, to make realistic applications to nuclei imply the use of
contact interactions (Skyrme like).
These interactions, which are zero range in $r$-space are thus of infinite range in momentum
space. This unphysical behavior of the interaction is critical in practice, since during
nucleon-nucleon collisions, particles will scatter to too high momentum. No clear solution
to this problem exists so far in the TDDM theory \cite{Lac04}.    
Due to these difficulties, only a few applications have been carried out so far for
collective vibrations \cite{Deb92,Luo99,Toh01,Toh02b}, and very recently for nuclear
collisions \cite{Toh02}. Guided by the BCS approach to pairing, a simplified version of 
TDDM, called TDDM$^P$ has also been proposed to account approximately for both pairing 
and direct nucleon-nucleon collisions in Ref. \cite{Ass09}.

\subsection{Direct in-medium two-body collisions and Extended TDHF theory}
\label{sec:extended}

Pairing correlations become less important when the internal excitation 
of the system increases. Conjointly, Pauli principle is less effective to 
block direct nucleon-nucleon collisions. Two-body collisions are included in the 
Born term $B_{12}$ in eq. (\ref{eq:tddm}). In the following, we only account for 
this term in the evolution of $C_{12}$\cite{Won78,Won79,Dan84,Bot90,Ayi80} leading to
\begin{equation}
i\hbar \frac{\partial }{\partial t}C_{12}-\left[ h_{1}[\rho] + h_{2}[\rho] ,C_{12}\right]
=B_{12}  \label{eq:c12f12}.
\end{equation}

The standard strategy to include collisions is closely related to the theory of 
open quantum systems \cite{Bre02}. 
Two-body correlations are interpreted as an environment for one-body degrees of freedom.
To account for two-body effects without dealing directly with two-body matrices, a projection 
technique "a la Nakajima-Zwanzig" \cite{Bre02} is used. First, the correlation equation 
of motion is integrated from the initial time to $t_{0}$ to time $t$ as
\begin{equation}
C_{12}(t)=-\frac{i}{\hbar }ds\int_{t_{0}}^{t}U_{12}\left( t,s\right)
B_{12}\left( s\right) U_{12}^{\dagger} \left( t,s\right) + \delta C_{12}(t),
\label{eq:c12t}
\end{equation}
where $U_{12}(t,s)$ represents the independent particle propagation of two
particles, $U_{12}=U_{1}\otimes U_{2}$ with 
\begin{eqnarray}
\displaystyle U(t,s)=\exp
\left( -\frac{i}{\hbar }\int_{s}^{t}h[\rho (t^{\prime })]dt^{\prime }\right).
\end{eqnarray} 
In expression (\ref{eq:c12t}), the first term represents correlations
due to the residual interaction during the time interval. The second
term describes propagation of the initial correlations $C_{12}(t_{0})$ from $
t_{0}$ to $t$, i.e. 
\begin{eqnarray}
\delta C_{12}(t)=U_{12}(t,t_{0})C_{12}(t_{0})U_{12}^{\dag }(t,t_{0}).
\end{eqnarray}
Reporting this expression in the evolution of $\rho_1$, a generalization of TDHF 
theory is obtained (where we omit the indice "1" in $\rho_1$)
\begin{equation}
i\hbar \frac{\partial }{\partial t} \rho = [h [\rho],\rho] + K[\rho] + \delta K(t).
\label{eq:ESTDHF}
\end{equation}
Two additional terms appear compared to the original mean-field transport theory.
$K[\rho]$, called collision term, reads 
\begin{eqnarray}
K[\rho] &=& -\frac{i}{\hbar }\int_{t_{0}}^{t}ds {\rm Tr}_{2}[v_{12},U_{12}(t,s)B_{12}(s)U_{12}^{\dagger }(t,s)] ,
\label{eq:k}
\end{eqnarray}
and contains the effect of direct in medium collisions on the evolution of one-body degrees of freedom.
As we will see below, this term is anticipated to be responsible for dissipative aspects leading eventually to 
the onset of thermalization in the interacting system.

The second term corresponds to the effect of initial correlations propagated through the mean-field and 
that will induce deviations from the mean-field picture. It can be written in a compact form as 
$\delta K(t)$ and is given by:
\begin{eqnarray}
\delta K(t) =& Tr_{2}[v_{12},\delta C_{12}(t)] .
\label{eq:dk} 
\end{eqnarray}
This term depends on the initial conditions considered and will be the subject of the next chapter.
We should note that Eq.(\ref{eq:ESTDHF}) has been introduced in the semi-classical limit in Ref.\cite{Ayi88}.
In this equation the initial correlation term $\delta K(t)$ is treated as the stochastic part of the collision term.
This approach is referred to as the Boltzmann-Langevin model. For further details please see Ref.\cite{Abe96,Ran90}.

\subsubsection{Irreversible process and Extended TDHF}

Let us first illustrate the advantages 
of the introduction of collision term on top of the mean-field dynamics and neglect initial correlations, i.e. 
$\delta K[\rho] = 0$. The resulting theory 
is called Extended TDHF with a non-Markovian collision term (or with "memory effects"). 
The terminology "non-Markovian" (in opposition to "Markovian") comes from the fact that the system 
at time $t$ depends not only on the density at time $t$ but also on its full history 
due to the presence of a time integral in Eq. (\ref{eq:k}).

Extended TDHF has  rarely been
directly applied because of the numerical effort required. In order to
illustrate these difficulties, let us introduce the single-particle basis $%
\left| \alpha \left( t\right) \right\rangle $ that diagonalizes the one body
density $\rho_1 (t)$ at a given time: 
\begin{equation}
\rho (t)=\sum \left| \alpha (t)\right\rangle \;n_{\alpha }(t)\;\left\langle
\alpha (t)\right| .  \label{eq:rhodiag}
\end{equation}
This basis explicitly depends on time and will be called "natural" basis 
or "canonical" basis hereafter. 
As we do expect from nucleon-nucleon collisions, the 
collision term induces a mixing of single-particle degrees of freedom during
time evolution. 
Using the weak coupling approximation in combination with the first order
perturbation theory, the ETDHF equation can be transformed into a generalized
master equation for occupation numbers which account for the Pauli principle : 
\begin{equation}
\frac{d}{dt}n_{\alpha }(t)=\int_{t_{0}}^{t}ds\left\{ \left( 1-n_{\alpha
}\left( s\right) \right) {\mathcal{W}_{\alpha }^{+}}\left( t,s\right)
-n_{\alpha }\left( s\right) {\mathcal{W}_{\alpha }^{-}}\left( t,s\right)
\right\}. 
\label{eq:master}
\end{equation}
Here,  the explicit form of the gain $\mathcal{W}_{\lambda }^{+}$ and loss $\mathcal{W}_{\lambda
}^{-}$ kernels could be found in ref. \cite{Lac98}. Therefore, in contrast to TDHF where occupation
numbers are constant during the time evolution, in ETDHF the $n_\alpha$ evolve and can
eventually relax toward equilibrium. Such a relaxation is the only way to properly account 
for the thermalization process in nuclei. In ref. \cite{Lac98}, the inclusion of correlation with 
Extended TDHF has been successfully tested in the simple case of two interacting nucleons in one dimension
where the exact solution can also be obtained. 

\subsection{Summary on deterministic approaches}

The aim of the present section was to give an overview of deterministic approaches that goes beyond 
the mean-field approximation. Among the approaches, some of them are explicitly introducing additional degrees of freedoms to follow in time, like in the TDDM case while others try to projected the effect of correlations on relevant degrees of freedom keeping only the information of the one-body density. In all cases, the level of complexity is significantly enhanced compared to mean-field and most often, application to realistic situations becomes very difficult if not impossible. In the rest of this review article, we present alternative methods based on stochastic quantum mechanics to describe either initial correlations effect or correlations that built up in time.


%
%

\section{Mean-field with initial quantum fluctuations}
\label{sec:stochmf}
TDHF or Extended TDHF provide approximate solutions of the N-body problem
starting from a well-defined initial state and leading to a
unique final state. These approaches are appropriate to describe mean values 
of one-body observables but generally misses fluctuations in collective space.  
To treat the quantum zero point motion 
in collective space, one can for instance account for configuration mixing through 
the so-called Time-Dependent Generator Coordinate Method (TDGCM) \cite{Rin80,Goe80,Goe81,Rei87}.
Such approach that keeps the full quantum coherence in collective space, 
is however rather involved numerically \cite{Gou05} and is nowadays restricted 
to system close to the adiabatic limit suited for rather small internal excitation.
To approximately treat  both quantal zero-point fluctuations and possible thermal statistical 
fluctuations, a stochastic scheme, called hereafter Stochastic Mean-Field (SMF) theory, has been proposed in ref. \cite{Ayi08}. 
  
\subsection{General strategy}
The Stochastic Mean-Field approach starts from the hypothesis that a quantum 
dynamical problem can be sometimes replaced by a superposition of classical evolutions with properly chosen initial conditions. Such a replacement can even be exact in some cases \cite{Her84,Kay94}.  To focus on the main hypothesis 
of the SMF approach, let us consider 
a quantum system described by a set of degrees of freedom. To make connection with section \ref{sec:meanfield}, we assume that we are interested in a subset of DOF 
associated with the set of operators $\{ A_\alpha \}$. Knowing the wave-function or more generally the density matrix $D(t_0)$, at initial time, one can estimates the expectation values of the DOF as well as associated quantal fluctuations:
\begin{eqnarray}
\langle A_\alpha \rangle (t_0) &=& {\rm Tr}(A_\alpha D (t_0)), \nonumber \\
\sigma_{\alpha \beta} (t_0) &=& \frac{1}{2} ( \langle A_\alpha A_{\beta}  + A_{\beta} A_\alpha  \rangle ) - \langle A_\alpha \rangle \langle  A_{\beta} \rangle . \nonumber 
\end{eqnarray} 
As discussed previously,  the exact evolution of the system requires a priori to solve the exact Liouville-von Neumann
equation for $D(t)$ and can lead to rather complex coupling between relevant and irrelevant degrees 
of freedom. The mean-field approximation provides a simple way to focus on the mean-values $\{\langle A_\alpha \rangle \}$ but generally fails to reproduce quantum fluctuations due to its quasi-classical nature. 

The Stochastic Mean-Field (SMF) approach provides an approximate way to also treat fluctuations keeping the simplicity of mean-field equation of motion. In this approach, a statistical ensemble of 
initial values for the $\{ A^{(n)}_\alpha(t_0) \}$ are considered. Here, the label $(n)$ refers to a sample of the statistical ensemble. This variables are now considered as classical variables whose initial statistical properties are chosen 
in such a way to reproduce quantum mean-values and fluctuations, i.e.:
\begin{eqnarray}
\overline{ A^{(n)}_\alpha(t_0)} &=& \langle A_\alpha \rangle (t_0) , \nonumber  \\
\overline{ A^{(n)}_\alpha(t_0) A^{(n)}_\beta(t_0)} - \overline{ A^{(n)}_\alpha(t_0) }~ \overline{A^{(n)}_\beta(t_0)}&=&\sigma_{\alpha \beta} (t_0),
\end{eqnarray}    
where the average $\overline{X^{(n)}}$ corresponds to the classical average over different samples. Then, each initial 
configuration is evolved according to the classical equation of motion:
\begin{eqnarray}
\frac{d   A^{(n)}_\alpha }{dt} = {\cal F} \left( \{ A^{(n)}_\beta \} \right).\label{eq:smfobs}
\end{eqnarray}
In a many-body problem, the mean-field equation can be regarded as the quasi-classical approximation. As we will see, using the SMF technique in combination with the mean-field 
equation of motion is a powerful tool not only to improve the description of fluctuations but also to treat the effect of fluctuations on one-body DOF. This 
approach, contrary to other stochastic methods, have the specificity that fluctuations are introduced at initial time only. Then, each initial condition 
is propagated with its own mean-field.   
      
\subsection{ Stochastic Mean-Field in many-body systems}
\label{sec:smf}

Let us consider the situation where the initial state is properly described either by a Slater determinant or a statistical ensemble of 
independent particles with density of the form: 
\begin{eqnarray}
\hat D = \frac{1}{z}\exp \left(\sum \lambda_i a^\dagger_i a_i \right). \label{eq:mbdensity}
\end{eqnarray}
where $z$ is a normalization factor while $(a^\dagger_i , a_i)$ are the creation/annihilation operators associated 
to the canonical basis $|\Phi_i \rangle $. 
Accordingly, its initial one-body density matrix is given by 
\begin{eqnarray}
\rho (t_0)= \sum_i |\Phi_i (t_0) \rangle n_i \langle \Phi_i (t_0)|. \label{eq:densmf}
\end{eqnarray}
As we have discussed in section \ref{sec:meanfield}, the mean-field approximation focus on the evolution 
of the one-body DOF only, Eq.  (\ref{eq:mfrho}) that could be simulated using the set of single-particle states Schroedinger 
equation:
\begin{eqnarray}
\label{eqmfstate}
i\hbar \frac{\partial }{\partial t}| \Phi _{i} (t) \rangle 
= h(\rho) |\Phi _{i} (t) \rangle,
\end{eqnarray}
while keeping the occupation number $n_i$ constant in time. 

The mean-field theory is a quantal approach and even if it usually underestimates fluctuations of collective 
observables in the nuclear physics context, these fluctuations are non-zero. Within mean-field theory, the expectation value of
an observable $\hat A$ is obtained through $\langle \hat A \rangle = {\rm Tr} (\hat A \hat D)$ where $\hat D$ has the 
form (\ref{eq:mbdensity}). Accordingly, the quantal average and fluctuation of a one-body observable $\hat A$ along the 
mean-field trajectory are given by:
\begin{eqnarray}
\langle \hat A \rangle &=& \sum_i \langle \Phi_i(t) |\hat A | \Phi_i(t) \rangle n_i \label{eq:avermf}
\end{eqnarray}
and 
\begin{eqnarray}
\sigma^2_A(t) &=&\langle \hat A^2 \rangle - \langle \hat A \rangle^2 \nonumber \\
&=& \sum_{ij} | \langle \Phi_i(t) |\hat A | \Phi_j(t) \rangle|^2 n_i (1-n_j). \label{eq:flucmf} 
\end{eqnarray}

Let us now apply the general SMF strategy described above. We are interested here in describing one-body DOF that we will treat as classical objects with initial fluctuations. Since the knowledge of one-body DOF is equivalent to the knowledge of the density matrix $\rho$, we can directly consider the density matrix itself 
classically. Therefore a set of initial density matrix, labelled by $\rho^{(n)}$ are considered.
The initial fluctuations should be chosen in such a way that the average properties and fluctuations of one-body observables 
identify with the quantal expectations given by Eqs. (\ref{eq:avermf}) and (\ref{eq:flucmf}) at $t=t_0$. It has been shown in ref. \cite{Ayi08}
that the matrix elements $\rho _{ij}^{(n)}$ in the complete basis $\{ \Phi _{i} (t_0) \}$ can be  
taken as uncorrelated Gaussian numbers with mean values equal to
\begin{eqnarray}
\overline {\rho _{ij}^{(n)}}=\delta _{ij}n_j \label{eq:meanrho}
\end{eqnarray}
while their variances are determined as,
\begin{eqnarray}
\label{eq4}
\overline {\rho _{ij}^{(n)} \rho _{{j}'{i}'}^{(n)} } 
&=& \frac{1}{2}
\delta_{j{j}'}\delta _{i{i}'}\left[n_i(1 - n_j) + n_j(1 - n_i) \right].  \label{eq:flucrho}
\nonumber\\
\end{eqnarray}
Considering the observable $\hat A$, within SMF, for a given event, its value at initial time is a fluctuating quantity given by:
\begin{eqnarray}
A^{(n)}(t_0) &=& \sum_{ij} \langle \Phi_i (t_0) |\hat A| \Phi_j (t_0) \rangle \rho^{(n)}_{ji}. 
\end{eqnarray} 
An important aspect in SMF, already mentioned previously, is that both mean-values and fluctuations are obtained by performing
classical average on the initial sampling. This gives:
\begin{eqnarray}
\overline{A^{(n)}(t_0)} &=& \sum_{ij} \langle \Phi_i (t_0) |\hat A| \Phi_j (t_0) \rangle \overline{ \rho^{(n)}_{ji}} \nonumber \\ 
&=& \sum_{i} \langle \Phi_i (t_0) |\hat A| \Phi_i (t_0) \rangle n_i(t_0). \nonumber
\end{eqnarray} 
 Introducing $\delta A^{(n)} = A^{(n)} - \overline{A^{(n)}}$, we also have (omitting $t_0$ in the first line):
 \begin{eqnarray}
\overline{\delta A^{(n)}(t_0)\delta A^{(n)}(t_0)} &=& \sum_{ijkl} \langle \Phi_i |\hat A| \Phi_j \rangle \langle \Phi_k |\hat A| \Phi_l  \rangle 
\overline{  \rho^{(n)}_{ji}\rho^{(n)}_{lk}}, \nonumber \\
&=&  \sum_{ij} | \langle \Phi_i(t_0) |\hat A | \Phi_j(t_0) \rangle|^2 n_i (1-n_j), \nonumber
\end{eqnarray}
proving that the classical average  identifies with the quantal average, Eq. (\ref{eq:flucmf}) at $t_0$ 
 
 In SMF, each initial condition is evolved in time using its self-consistent mean-field equation of motion that is obtained simply
 by replacing $\rho$ by $\rho^{(n)}$ in Eq.  (\ref{eq:mfrho}). For a given event, the equation of motion then reads 
 \begin{eqnarray}
\label{eq55}
i\hbar \frac {\partial}{\partial t} \rho^{(n)}(t) =[h(\rho^{(n)}), \rho^{(n)}(t)]
\end{eqnarray}
with the initial condition $\rho^{(n)}(t_0) = \sum_{ij} |\Phi_i (t_0) \rangle \rho^{(n)}_{ij} \langle \Phi_j (t_0)|$. 
Note that, trajectories are independent from each others. 

It can be shown without any difficulty that the density along each path can be written as 
\begin{eqnarray}
\label{eq3}
\rho ^{(n)} (t) = \sum\limits_{ij} | 
\Phi_{i}^\ast(t;n) \rangle \rho_{ij}^{(n)} 
\langle \Phi_{j} (t;n ) |,
\end{eqnarray}
 where the matrix elements $\rho_{ij}^{(n)} $ are kept fixed in time while 
 the single-particle wave-functions evolve according to the Schroedinger equation 
 \begin{eqnarray}
\label{eq5}
i\hbar \frac{\partial }{\partial t}| \Phi _{i} (t;n ) \rangle
= h(\rho^{(n)} ) | \Phi _{i} (t;n ) \rangle,
\end{eqnarray}
with the constraint  $ | \Phi _{i} (t_0;n ) \rangle =  | \Phi _{i} (t_0) \rangle$. Note that, even if the single-particle 
states are identical at initial time from one event to the other, this will not be the case for $t>t_0$ due to the 
self-consistency of the mean-field.  

\begin{figure}
\resizebox{0.45\textwidth}{!}{%
  \includegraphics{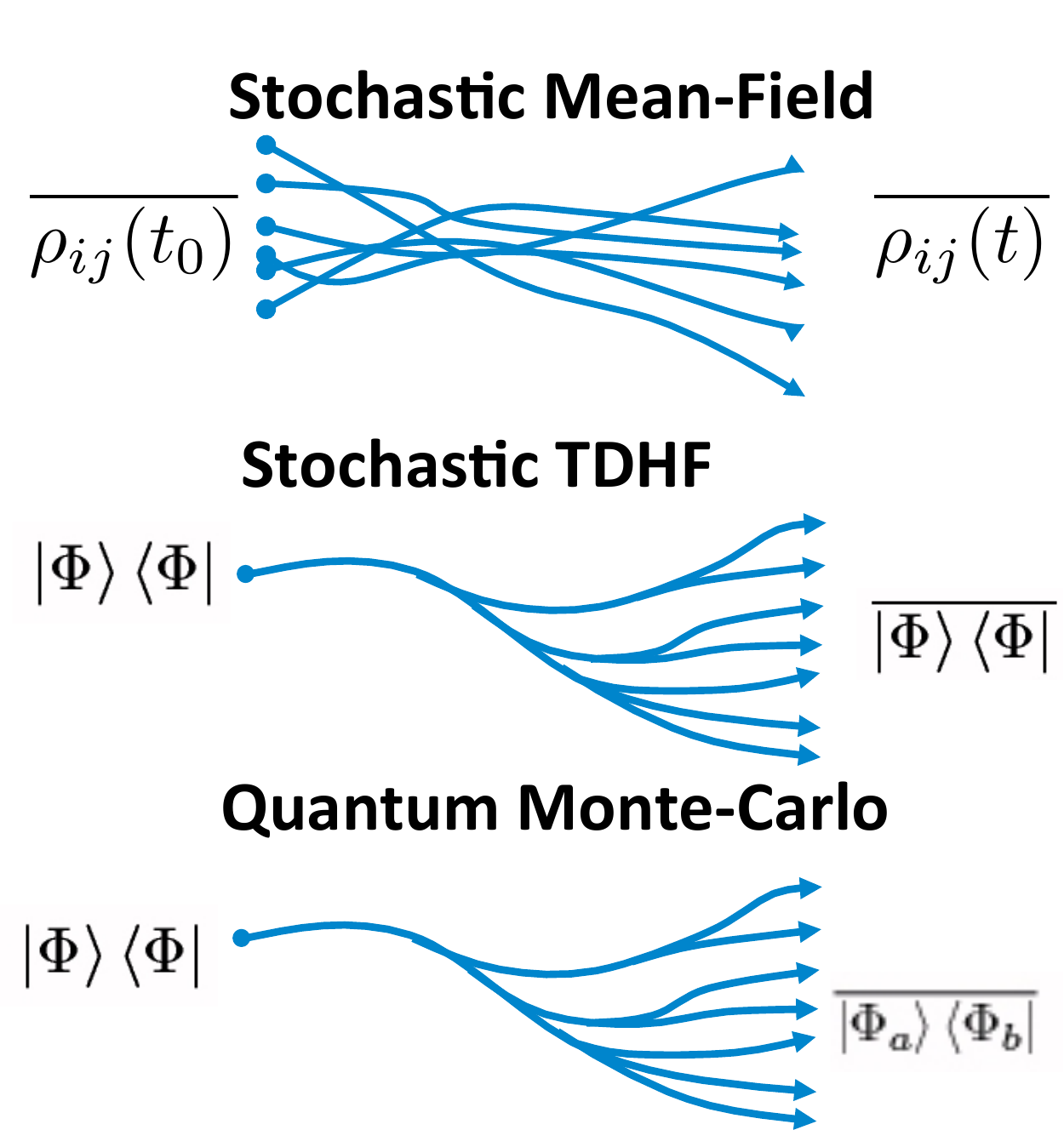}
}
\caption{Illustration of the different types of stochastic theories introduced in this section.
Top: The Stochastic Mean-Field theory is introduced to account for the effect of possible initial correlation and treat fluctuations beyond the mean-field approximation.
In that case, a statistical ensemble of initial densities is chosen and each set of initial values evolves independently from the others according to the quasi-classical mean-field equation.
Middle:  In the Stochastic TDHF theory, a random noise is introduced at each time step. This noise induces 
quantum jumps between densities of pure Slater determinants states. By averaging over different trajectories, 
the effect of two-body collisions on one-body density evolution is incorporated similarly to the Extended TDHF 
theory.
Bottom: In the Quantum Monte-Carlo approach, all two-body correlations are a priori treated. In that case, starting from 
an initial independent particle state, the evolution is replaced by an ensemble of stochastic density evolutions each being
written as a dyadic of Slater determinants 
$D = | \Phi_a \rangle \langle \Phi_b |$. 
The exact evolution is expected to be recovered 
by averaging over the dyadic. }
\label{fig:stochastic}       
\end{figure}   

\subsection{Dispersion of one-body observable}

As a first demonstration, we illustrate that for small amplitude fluctuations the SMF approach 
gives the same result for dispersion of one-body
observable as the one deduced by Balian-V\'en\'eroni (BV) using a variational formulation. 
The fluctuating part of a one body observable $\hat{Q}$ in an event is determined by, 
\begin{eqnarray}
\label{Q}
\delta Q^{(n)}(t)=\sum_{ij}  \langle \Phi_j(t;n)|\hat{Q}|\Phi_i(t;n)\rangle  \delta \rho^{(n)}_{ij}(t)
\end{eqnarray}
where $\delta \rho^{(n)}(t)$ denotes the single-particle density matrix, $\rho^{(n)}(t)= \rho(t)+ \delta \rho^{(n)}(t)$ fluctuations.
In order to calculate the variance of a one-body observable, we consider small amplitude fluctuations. Small fluctuations of the density matrix is 
determined by the time-dependent RPA equation, which is obtained by linearizing the stochastic TDHF Eq. (\ref{eq55}) around the average evolution.
As a result, the expectation value of the one-body observable can be expressed as
\begin{eqnarray}
\delta Q^{(n)}(t_1)= \sum_{ij} \langle \Phi_i(t_0)|B(t_0)|\Phi_j(t_0) \rangle \delta \rho^{(n)}_{ji}(t_0)
\end{eqnarray}
where $t_1$ represents the final time at which the observation is made and $t_0$ is the initial time. The one-body 
operator $B(t)$ is defined according to
\begin{eqnarray}
B_{ji}(t)  = \sum_{kl}Q_{lk}~\left\langle kj \left| \exp\left[-\frac{i}{\hbar}\int_{t}^{t_1}R(s)ds\right] \right|li \right\rangle .
\end{eqnarray}
where $R(t)$ is the linearized Liouville matrix obtained from the linearized TDHF equation.
It is easy to show that time evolution of the one-body operator $B(t)$ is determined by the dual of the time-dependent RPA according to \cite{Ayi08},
\begin{eqnarray}
i\hbar \frac {\partial}{\partial t} B(t) = [h(\rho), B(t)]+
{\rm Tr} \left(\frac{\partial h}{\partial \rho}\right)\cdot [B(t),\rho]. \nonumber 
\end{eqnarray}
The solution is determined by backward evolution with the boundary condition $B(t_1)=Q$. 
Then, the variance of the observable is calculated as,
\begin{eqnarray}
\sigma^2_Q(t_1)= \sum |\langle \Phi_i(t_0)|B(t_0)|\Phi_j(t_0)\rangle|^2n_i(1-n_j).
\end{eqnarray}
This result is identical with the formula derived from the Balian-V\'en\'eroni principle assuming that the deviation 
from the mean-field trajectory is small \cite{Bal85}.

\subsection{Adiabatic projection on collective path}

In order to illustrate the fact that the SMF equation describes the dynamics of fluctuations in accordance with the one-body dissipation mechanism, we give another example in this section. We consider that the collective motion is slow and can be described by a few relevant collective variables. For example, in induced fission, collective variables
maybe taken as the relative distance of fragments, mass-asymmetry and deformation parameters. Here, we consider a single collective variable $q(t)$, and introduce the quasi-static single-particle representation,
\begin{eqnarray}
h(q) \Psi_j(\vec{r};q)=\epsilon_j(q) \Psi_j(\vec{r};q),
\end{eqnarray}
where $h(q)=h[\rho(q)]$ denotes the mean-field Hamiltonian, in which the time dependence of the local density  $\rho(\vec{r},\vec{r};t)$ is parameterized in terms of the collective variable in a suitable manner. We expand the
single-particle density in terms of the single-particle representation,
\begin{eqnarray}
\label{QPE}
\rho(\vec{r}, \vec{r}^{\prime};t)=
\sum_{kl} \Psi_k^{*}(\vec{r};q) \rho_{kl}(t)\Psi_l(\vec{r}^{\prime};q).
\end{eqnarray}
Both the elements of density matrix $\rho_{kl}(t)$ and collective variable $q(t)$ are fluctuating quantities. In this section for clarity of notation, we ignore the event label $n$ on these quantities. We determine the matrix elements of density in the lowest order perturbation theory in dynamical coupling $ \langle \Psi_k|\partial \Psi_l/\partial q \rangle \dot{q}(t)$. It is preferable that the wave functions are close to diabatic structure. Since in the diabatic representation, dynamical coupling is expected to be small, hence, it can be treated in the weak-coupling approximation. Diabatic
single-particle representation can approximately be constructed by ignoring small symmetry breaking terms in the mean-field potential \cite{Norenberg81,Ayik82}.

In order to determine the temporal evolution of collective variable, we use the total energy conservation,
\begin{eqnarray}
E &=& \sum_{lk} \langle \Psi_k(q)|T|\Psi_l(q)\rangle  \rho_{lk}(t)+\nonumber \\
& &\frac{1}{2} \sum_{ijlk} \rho_{ji}(t) \langle \Psi_k(q)\Psi_i(q)|V|\Psi_l(q)\Psi_j(q) \rangle \rho_{lk}(t). \nonumber 
\end{eqnarray}
In the many-body Hamiltonian, for simplicity we take an effective two-body interaction potential energy $V$. The total energy depends on time implicitly via collective variable $q(t)$ and explicitly via matrix elements $\rho_{lk}(t)$. Energy conservation requires,
\begin{eqnarray}
\frac{dE}{dt}=\dot{q}\frac{\partial E}{\partial q}+
\frac{\partial E}{\partial t}=0 
\end{eqnarray}
In this expression $-\partial E/\partial q$ represents a dynamical force acting on the collective variable. The force depends on time, and it evolves from an initial diabatic form accompanied with deformation of Fermi surface towards an adiabatic limit associated with the adiabatic potential energy. The second term represents the rate of change of the energy due to explicit time dependence,
\begin{eqnarray}
\frac{\partial E}{\partial t} &=& \sum_{lk} \langle \Psi_k(q)|h(\rho)|\Psi_l(q) \rangle 
\frac {\partial}{\partial t} \rho_{lk}(t)\nonumber\\
&=& \sum_{k} \epsilon_k(q)
\frac{\partial}{\partial t} n_{k}(t). \label{Edot}
\end{eqnarray}
Here, $n_{k}(t)=\rho_{kk}(t)$ represents the occupation factors of the quasi-static single-particle states. It is possible to derive a master equation for the occupation factors by substituting the expansion Eq. (\ref{QPE}) into Eq. (\ref{eq55}). The collision term in the master equation involves memory effects. 
In weak-coupling limit, the memory time of the collision kernel is determined by the correlation time of the coupling matrix elements defined as
$\tau_c=\hbar/\Delta$, where $\Delta$ is the energy range of the form factor of the coupling matrix elements. 
For slow collective motion, we can neglect the memory effects and  carry out an expansion in powers of $t-t_1$, as it was done in the linear 
response treatment of ref. \cite{Hofmann76}. On the other hand, in a parabolic potential approximation of collective potential energy, it is possible
to take approximately the memory effect into account by incorporating harmonic propagation of collective motion during
short time intervals. In Eq. (\ref{Edot}) factoring out $\dot{q}(t)$ from each term, it is possible to deduce a generalized Langevin equation of 
motion for the collective variable \cite{Takigawa04,Ayik05},
\begin{eqnarray}
\label{LE}
M \ddot{q} + \frac{1}{2} \frac {dM}{dq} \dot{q}^2+ \frac{\partial
E}{\partial q} = -\gamma~ \dot{q} +\xi(t) \end{eqnarray}
where $M$, $\gamma$ and $\xi$ denotes the inertia, the friction
coefficient and the stochastic force, respectively. Expressions for these quantities is given in \cite{Ayi08}.
As a result of stochastic properties of the initial correlations, stochastic force has a 
Gaussian distribution with zero mean $\overline{\xi}(t)=0$, and a second moment which can be expressed as, 
\begin{eqnarray}
\label{correlation}
\overline{\xi(t) \xi(t_1)}=
\int_{-\infty}^{+\infty}\frac{d\omega}{2\pi} e^{-i
\omega(t-t_1)}~\hbar \omega~ \coth \frac{\hbar
\omega}{2T}~\gamma(\omega)
\end{eqnarray}
Here $\gamma(\omega)$ denotes frequency dependent friction  coefficient. This result represents the quantum 
fluctuation-dissipation relation  associated with the one-body dissipation mechanism and it naturally emerges 
from the stochastic approach  presented here \cite{Gardiner91,Weiss99}. 

\begin{figure}[htbp] 
\begin{center}
\resizebox{0.75\columnwidth}{!}{
  \includegraphics{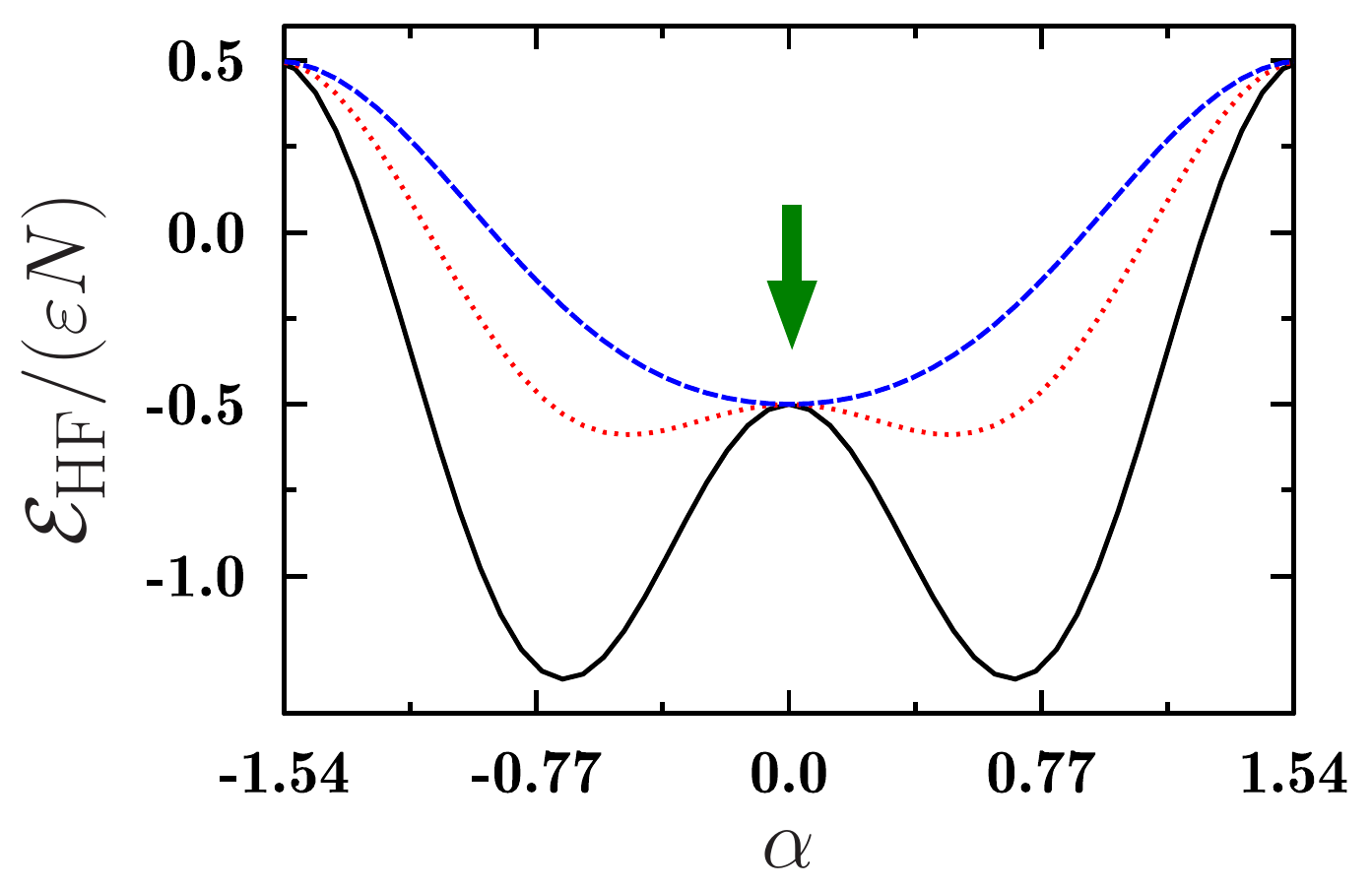}} 
  \end{center}
\caption{(color online). Evolution of the Hartree-Fock energy ${\cal E}_{\rm HF}$ as a function of $\alpha$ for $\chi=0.5$ (dashed line), 
$\chi=1.8$ (doted line) and $\chi=5$ (solid line) for $N=40$ particles.
The arrow indicates the initial condition used in the SMF dynamics (From ref. \cite{Lac12}).} 
\label{fig2:lacroix} 
\end{figure} 

\subsection{Dynamics near a saddle point: spontaneous symmetry breaking}

As mentioned in the introduction, the mean-field theory alone cannot break a symmetry 
by itself. The symmetry breaking can often be regarded as the presence of a saddle point 
in a collective space while the absence of symmetry breaking in mean-field just means that the system will stay 
at the top of the saddle point if it is there initially. Such situation is well illustrated in the Lipkin-Meshkov-Glick
model.  This model consists of $N$ particles 
distributed in two N-fold degenerated single-particle states separated by an energy $\varepsilon$. The associated Hamiltonian 
is given by (taking $\hbar=1$),
\begin{eqnarray}
H = \varepsilon J_z - V(J^2_x  -  J_y^2) , 
\label{eq:hamillipkin}
\end{eqnarray}
where $V$ denotes the interaction strength while $J_i$ ($i=x$, $y$, $z$), are the quasi-spin operators defined as
\begin{eqnarray} 
J_z &=& \frac{1}{2} \sum_{p=1}^{N} \left(c^\dagger_{+,p}c_{+,p} - c^\dagger_{-,p}c_{-,p}\right) , \nonumber \\
J_x &=& \frac{1}{2} (J_+ + J_-), ~~~J_y = \frac{1}{2i} (J_+ - J_-)
\end{eqnarray}
with $J_+ = \sum_{p=1}^{N} c^\dagger_{+,p}c_{-,p}$, $J_- = J_+^\dagger$ and where 
$c^\dagger_{+,p}$ and $c^\dagger_{-,p}$ are creation operators associated with the upper and lower single-particle levels.
In the following, energies and times are given in $\varepsilon$ and $\hbar/\varepsilon$ units respectively.
\begin{figure}[htbp] 
\begin{center}
\resizebox{0.8\columnwidth}{!}{
  \includegraphics{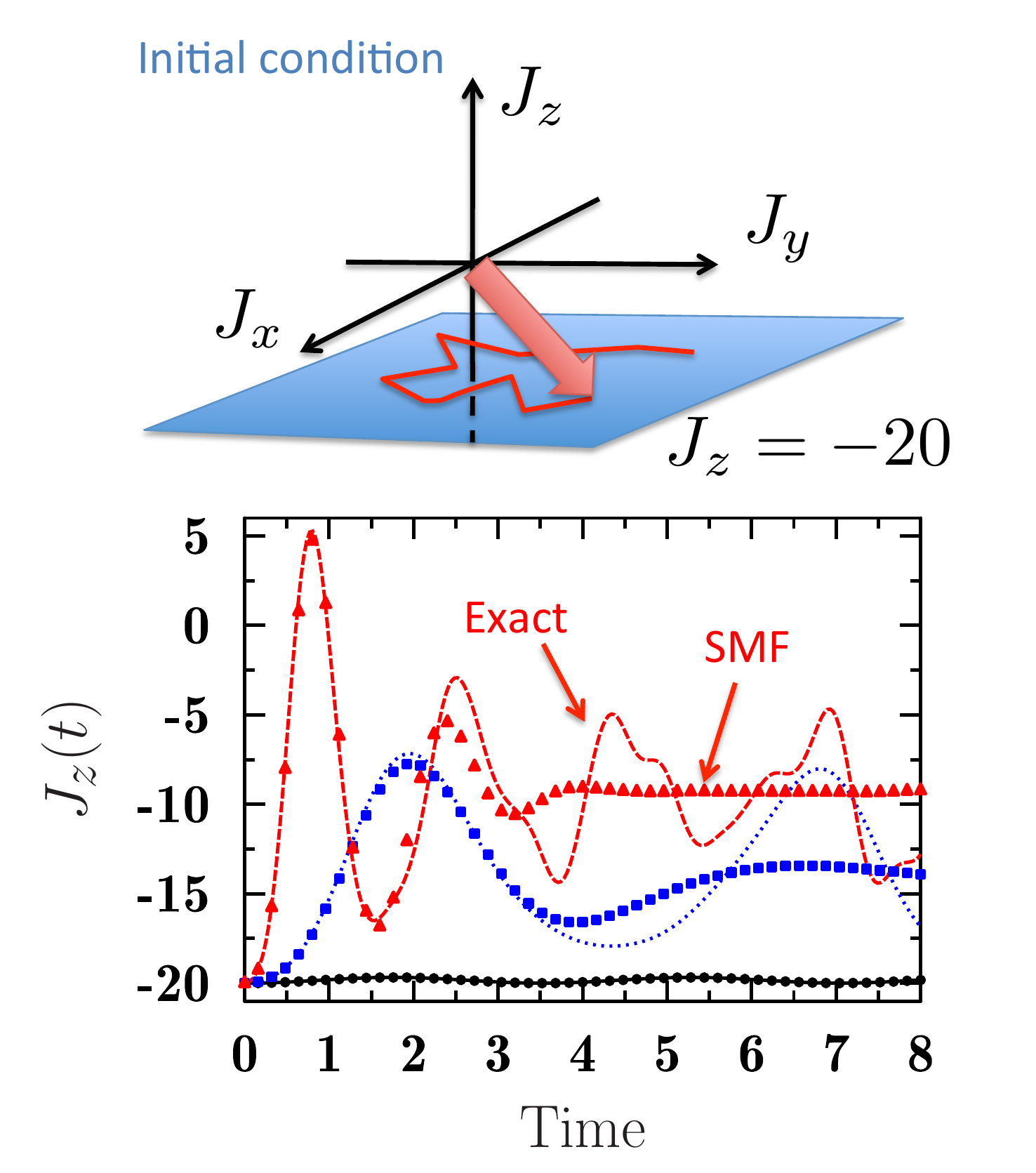}} 
  \end{center}
\caption{(color online) Top: illustration of the initial sampling used for the SMF theory in the collective space of quasi-spins.
Bottom:  Exact evolution of the $z$ quasi-spin component obtained when the initial state is $|j,-j\rangle$ for three 
different values of $\chi$: $\chi = 0.5$ (solid line), $\chi=1.8$ (dotted line) and $\chi=5.0$ (dashed line) for $N=40$ particles. The corresponding 
results obtained with the SMF simulations are shown with circles, squares and triangles respectively (adapted from \cite{Lac12}).} 
\label{fig3:lacroix} 
\end{figure}

It can be shown that the TDHF dynamic can be recast as a set of coupled equations between 
the expectation values of the quasi-spin operators $j_i \equiv  \langle J_i \rangle/N$ (for $i=x$,  $y$ and $z$) given by:
\begin{eqnarray}
\frac{d}{dt} 
\left(
\begin{array} {c}
 j_x     \\
  j_y  \\
 j_z 
\end{array}
\right)
&=& \varepsilon 
\left(
\begin{array} {ccc}
 0   & -1 + \chi  j_z  &  \chi   j_y  \\
 1+ \chi   j_z    &  0 & \chi  j_x  \\
 -2 \chi   j_y & -2 \chi   j_x  & 0
\end{array}
\right)
\left(
\begin{array} {c}
 j_x    \\
 j_y \\
 j_z 
\end{array}
\right) \label{eq:tdhf}
\end{eqnarray}
where $\chi = V(N-1) / \varepsilon$. Note that, this equation of motion is nothing but a special case 
of eq. (\ref{eq:mfrho}) where the information is contained in the three quasi-spin components. 
 To illustrate the symmetry breaking in this model it is convenient 
to display the Hartree-Fock energy ${\cal E}_{\rm HF} $ as a function of the $j_z$ component (Fig. \ref{fig2:lacroix}). 
Note that, here the order parameter $\alpha = \frac{1}{2} {\rm arccos}(-2j_z)$ is used for convenience. 
When the strength parameter is larger than  a critical value ($\chi > 1$), the parity symmetry is broken in $\alpha$ direction. 
For $\chi > 1$,  if the system is initially at the position indicated by the arrow in Fig. \ref{fig2:lacroix}, with TDHF it will remain at this point, i.e. 
this initial condition is a stationary solution of Eq.  (\ref{eq:tdhf}). 
\begin{figure}[htbp] 
\begin{center}
\resizebox{0.6\columnwidth}{!}{
  \includegraphics{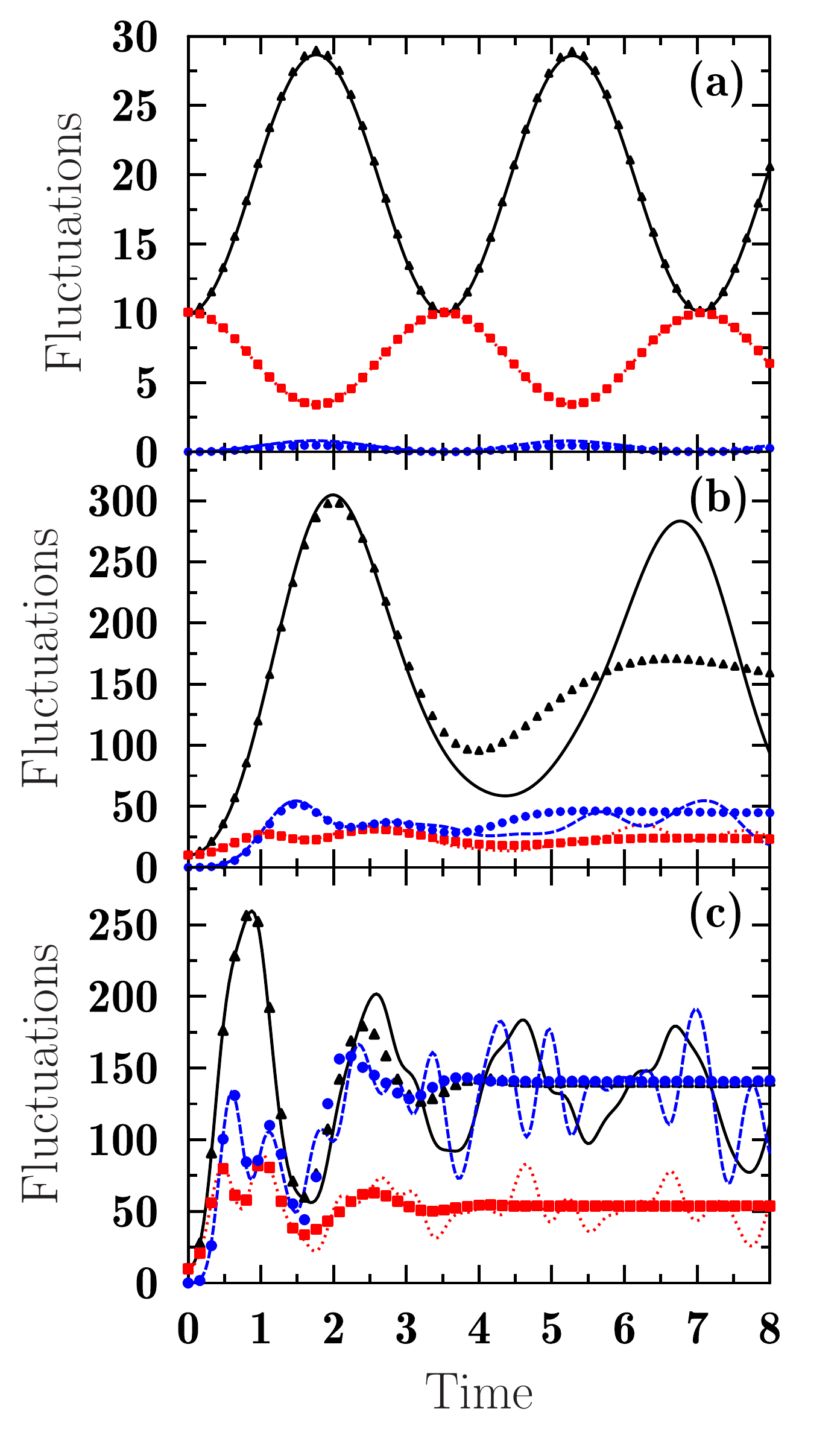}} 
  \end{center}
\caption{(color online)  Exact evolution of dispersions of quasi-spin operators obtained when the initial state is $|j,-j\rangle$ for three 
different values of $\chi$, from top to bottom $\chi = 0.5$ (a), $\chi=1.8$ (b) and $\chi=5.0$ (c) are shown. In each case, 
solid, dashed and dotted lines correspond to fluctuations of each quasi-spin projection 
$\sigma^2_x (t)$, $\sigma^2_y (t)$ and $\sigma^2_z (t)$, respectively.
In each case, results of the SMF simulations are shown with triangles ($\sigma^2_x$), squares ($\sigma^2_y$) and 
circles ($\sigma^2_z$). (taken from \cite{Lac12}).} 
\label{fig4:lacroix} 
\end{figure} 

Following the strategy discussed above, a SMF approach can be directly formulated in collective space where initial random conditions
for the spin components are taken. Starting from the statistical properties (\ref{eq:meanrho}) and (\ref{eq:flucrho}), it can be shown that the
quasi-spins should be  initially  sampled according to Gaussian probabilities with 
first moments given by \cite{Lac12}:
\begin{eqnarray}
\overline{j^\lambda_x(t_0)}=\overline{j^\lambda_y(t_0)} = 0,
\end{eqnarray}
and second moments determined by, 
\begin{eqnarray}
\overline{j^\lambda_x(t_0)j^\lambda_x(t_0)} = \overline{j^\lambda_y(t_0) j^\lambda_y(t_0)} = \frac{1}{4N}. 
\end{eqnarray}
while the $z$ component is a non fluctuating quantity.
  
An illustration of the initial sampling (top) and of  results obtained by 	
averaging mean-field trajectories with different initial conditions is shown in Fig. \ref{fig3:lacroix} 
and compared to the exact dynamic.
As we can see from the figure, while the original mean-field gives constant quasi-spins as a function of time, the SMF approach greatly improves 
the dynamics and follows the exact evolution up to a certain time that depends on the interaction strength. As shown in Fig. \ref{fig4:lacroix}, the stochastic approach not only improves the description of the mean-value of one-body observables but also the fluctuations. It is finally worth to mention that the BV variational principle has been applied to this model in ref. \cite{Bon85}.  The BV gives very good improvement 
beyond mean-field in the small $\chi$ limit where a perturbative treatment is valid but failed to describe the large $\chi$ case where 
symmetry breaking is important. 

\subsection{Stochastic Mean-Field with pairing correlations}

As we have shown in section \ref{sec:tdhfb}, to describe superfluid systems, it is advantageous to break 
explicitly the symmetry associated to particle number, the $U(1)$ symmetry.  Then, the one-body density $\rho$ 
is replaced by the generalized density ${\cal R}$ that evolves at the mean-field level according to Eq. (\ref{eq:tdhfbR}).  
It has been shown recently that the SMF can be generalized to treat superfluid systems also by taking advantage
of the $U(1)$ symmetry breaking. Then, an ensemble of initial conditions ${\cal R}^{(n)}(t_0)$ followed by TDHFB 
evolutions \cite{Lac13}:
\begin{eqnarray}
i \hbar \frac{d }{dt}{\cal R}^{(n)}(t)
&=& \left[{\cal H}\left[{\cal R}^{(n)}(t)\right], {\cal R}^{(n)}(t)  \right], \label{eq:smftdhfb}
\end{eqnarray}  
are considered.
The statistical properties of the elements of density matrix 
${\cal R}^{(n)}_{\alpha \beta}(t_0)$ are specified in terms of statistical properties of the normal $\rho^{(n)}_{\alpha\beta}(t_0)$ and 
anomalous $\kappa^{(n)}_{\alpha\beta}(t_0)$ density matrices. 
Elements of the normal and the anomalous density matrices are uncorrelated 
Gaussian random numbers with the mean values,   
\begin{eqnarray}
\overline{\rho^{(n)}_{\alpha \beta} }&=& \delta_{\alpha\beta} f_\alpha, ~~~\overline{\kappa^{(n)}_{\alpha\beta}} = 0, \label{eq:flucrk}
\end{eqnarray} 
and the second moments defined by,
\begin{eqnarray}
\overline{ \delta \rho^{(n)}_{\alpha \beta} ~  \delta \rho^{(n)*}_{\alpha' \beta'}}  =  
\frac{1}{2}  \delta_{\alpha \alpha'} \delta_{\beta \beta'} \left[ f_\alpha (1 - f_\beta)  + f_\beta (1-f_\alpha) \right], \label{eq:flucrhopair} \\
\overline{ \delta \kappa^{(n)}_{\alpha \beta} ~  \delta \kappa^{(n)*}_{\alpha' \beta'} }  = 
\frac{1}{2}  \delta_{\alpha \alpha'} \delta_{\beta \beta'} \left[ f_\alpha f_\beta + (1-f_\alpha) (1-f_\beta) \right]. \label{eq:fluckappa}
\end{eqnarray}
Note that here $\alpha$ is a label referring to the quasi-particle basis and $f_\alpha$ denotes the initial quasi-particle occupancy. 

Such a generalized description is adequate for the description of pairing correlations and allows for the treatment of of effect beyond
the independent quasi-particle picture. Similarly to the previous case where pairing was neglected, 
It can be shown in particular that the above statistical properties properly reproduces 
the initial quantal fluctuations of the mean-values and fluctuations of generalized "one-body" operators written as (see appendix A of ref. \cite{Lac12}):
\begin{eqnarray}
\hat Q &=& \sum_{ij} Q^{11}_{ij} a^\dagger_i a_j + \sum_{ij}  \left( Q^{20}_{ij} a_j a_i + Q^{20*}_{ji}  a^\dagger_i a^\dagger_j \right).
\end{eqnarray} 

The method has been benchmarked for a system described by a pairing Hamiltonian (picked fence model)  \cite{Ric64}
\begin{eqnarray}
H & = &  \sum_{i=1}^{K}  \varepsilon_i \hat N_i + 
\sum_{i j}^K  G _{ij} \hat S^+_i   \hat S^-_j,   \label{eq:rich}
\end{eqnarray}  
where the different operators $\hat N_i$, $S^+_i$ and $S^-_i$ are now related respectively to the occupation, pair creation and pair annihilation 
operators of the level $i$. As in the LMG model, the TDHFB equation can be written in terms of the quasi-spin operators expectation values:
\begin{eqnarray}
\frac{d}{dt} 
\left( 
\begin{array} {c}
   S^x_i(t)  \\
   S^y_i(t)  \\
   S^z_i(t)
\end{array} 
\right) =  \left(
\begin{array} {c c c}
0 &  -2  \tilde \varepsilon_i (t)  &  + 2 \Delta^y_i  \\
2 \tilde \varepsilon_i (t) & 0 &  -2 \Delta^x_i  \\
- 2 \Delta^y_i  & 2 \Delta^x_i & 0
\end{array}
\right) 
\left( 
\begin{array} {c}
   S^x_i(t) \\
   S^y_i(t) \\
   S^z_i(t)
\end{array} 
\right) \label{eq:tdhfbspin}
\end{eqnarray} 
where $S^x_i =  ( S^+_i +  S^-_i )/2$, $S^y_i = ( S^+_i - S^-_i )/2i$
and $ S^z_i = ( N_i - \Omega_i)/2$ denote the expectation values of the corresponding operators(see ref. \cite{Lac12} for more details). 
Connection with the standard 
TDHFB theory can be made by noting that $ S^z_i(t)$ is directly linked to the normal density while $S^x_i(t) $
and $S^y_i(t)$ corresponds to the real and imaginary part of the anomalous density $\kappa$. 

\begin{figure}[htbp] 
\begin{center}
\resizebox{0.8\columnwidth}{!}{
  \includegraphics{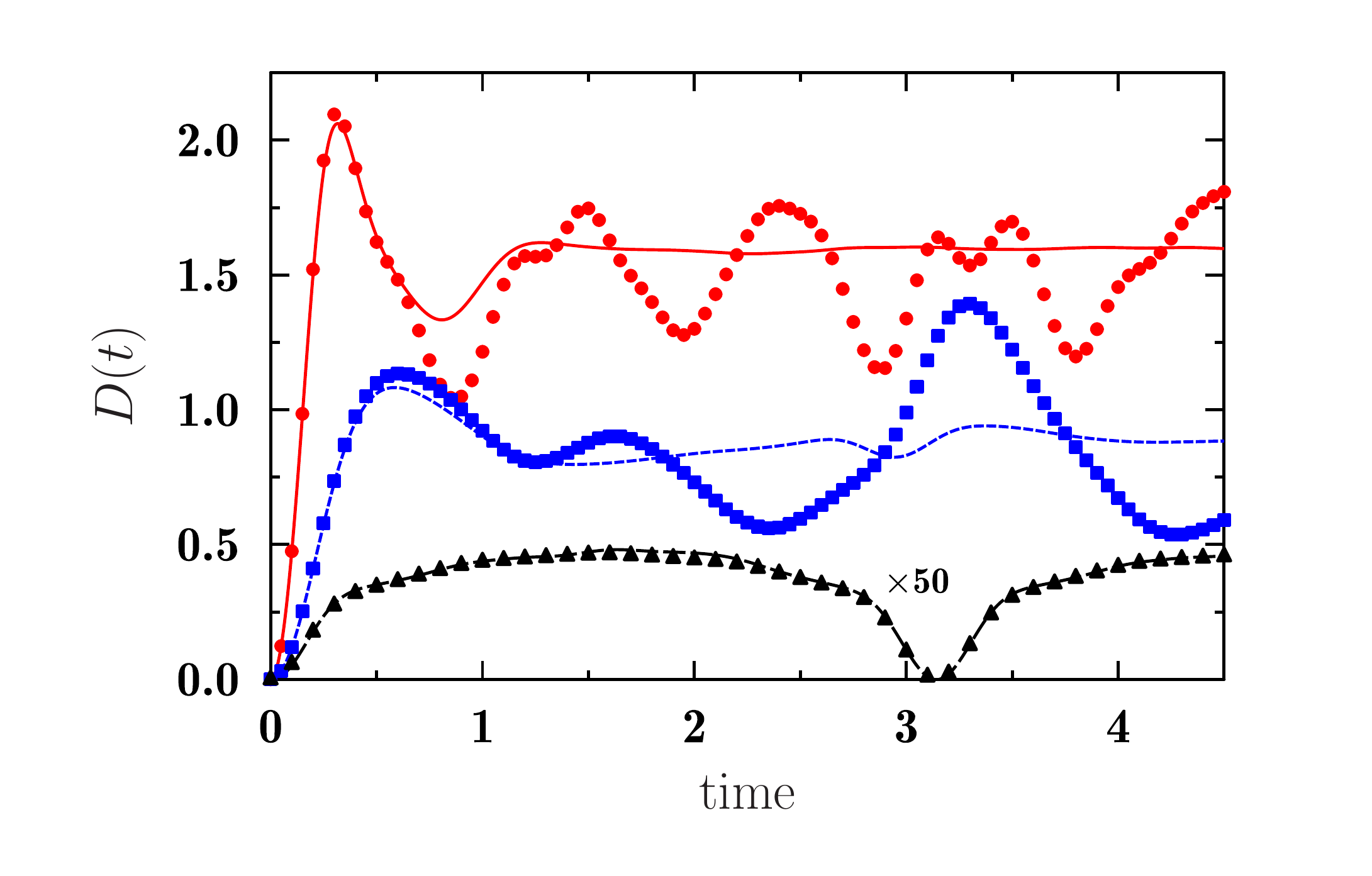}} 
  \end{center}
\caption{ (color online) Exact evolution of the single-particle fragmentation ($D(t)$) for different pairing interaction strength:
weak (black triangles), intermediate (blue squares) and strong (red circles) pairing. 
The results obtained by averaging over TDHFB trajectories are shown respectively by black long dashed line, 
blue short dashed line and red solid line (taken from Ref. \cite{Lac13}).
} 
\label{fig:dt} 
\end{figure}

In the SMF approach built on TDHFB, the three initial quasi-spin become fluctuating quantities. For each initial condition, the 
TDHFB equation is then solved in time to improve mean-field.  The specific case where the system is assumed to be initially in 
a non-superfluid phase has been considered was used in Ref. \cite{Lac12}. This situation is similar to the one shown in Fig. \ref{fig2:lacroix}
where the order parameter can for instance be replaced by the pairing gap. The system has initially a vanishing pairing and therefore is at the 
position located by the arrow in the figure. Since in a pure mean-field description, the system cannot break the symmetry it will stay at the saddle point. It could be indeed shown that, if the initial pairing correlation is zero, the TDHFB equation reduces  to the TDHF evolution and that a Slater determinant is stationary.  When SMF is used, non-vanishing pairing can occur event-by-event leading to non-trivial dynamics.   As an illustration of the SMF with pairing theory predictive power, the quantity $D(t) =  2 \sum_i n_i(t) (1-n_i(t)) $ that measures the fragmentation 
of the single-particle states around the Fermi energy, is shown in Fig. \ref{fig:dt} and compared to the exact evolution starting from a Slater determinant. Note that the TDHF approach without fluctuations would lead simply to $D(t) =0$.
Similarly to the case presented in previous section, the SMF approach is able to provide rather reasonable evolution 
where TDHF and TDHFB would simply fail. 

\subsection{A survey of recent applications to nuclear collisions}

In recent works, dissipation mechanism \cite{Ayi09} and nucleon exchange \cite{Was09b,Yilmaz11a} are investigated in the special case of central collisions of heavy-ions near barrier energies. Also, some applications of the 
SMF approach have been carried out for analyzing early development of spinodal instabilities in nuclear matter \cite{Ayik08b,Ayik09b,Ayik11,Yilmaz11b,Yilmaz13}.  In this section, we review recent investigations on dissipative mechanism in central heavy-ion collisions.

\subsubsection{Macroscopic reduction of information in mean-field dynamics}

\begin{figure}[hpt]
\begin{center}
\resizebox{0.8\columnwidth}{!}{
 \includegraphics{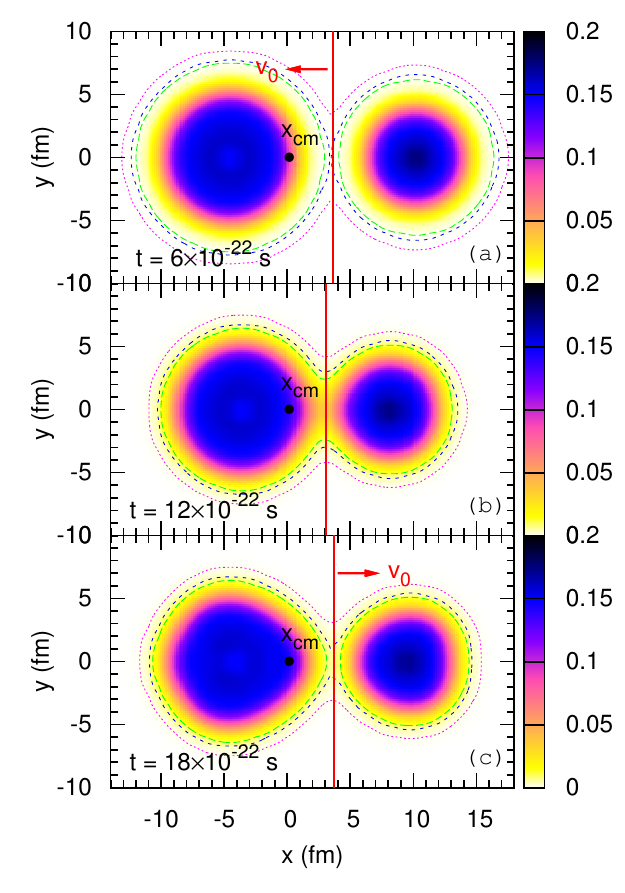}
} 
  \end{center}
\caption{Nucleon density profiles at the reaction plane, $\rho(x,y,z=0)$, are indicated by contour plots for the central collision
of $^{40}$Ca + $^{90}$Zr system at $E_{\rm cm}=97$ MeV in units of fm$^{-3}$. The black dot is the center of mass point. 
The red lines indicate the positions of the window $x_0$ and $v_0=dx_0/dt$ denotes velocity of the window. The three times correspond
respectively to before (top), during (middle) and after (bottom) contact (From \cite{Yilmaz11a}).}
\label{fig1:collision}
\end{figure}

For not too heavy systems, in central heavy-ions collisions above the Coulomb barrier lead to fusion, 
and at below the barrier energies, colliding nuclei exchange a few nucleons and re-separate. 
Such reactions are often investigated using time-dependent mean-field theories in three dimensions.
 As an example, Fig. \ref{fig1:collision} shows the density profiles at the reaction plane, $\rho(x,y,z=0,t)$ 
 in the $^{40}$Ca + $^{90}$Zr collisions at center-of-mass energy $E_{\rm cm}=97$ MeV for three different times. 
 The collision energy is below the Coulomb barrier energy and the system re-separate after contact

An important aspect in heavy-ion collisions is the knowledge of transport properties, related to dissipative aspects, during the approaching 
phase. In particular, the sharing of energy between internal single-particle and collective DOF is essential to understand how the collective 
phase-space is populated in time. When the system can still be considered as binary, one can get information on dissipation from the microscopic mean-field by focusing on specific macroscopic observables.  For instance the evolution of the relative distance gives insight 
in the energy loss while the number of particles in each nucleus provides direct information on the nucleon exchange process.  

A geometric projection method was proposed  by introducing the window between projectile-like and target-like nuclei according to the 
procedure outlined in \cite{Was08,Ayi09,Was09b}. Let us assume as a convention that the target is on the right while the projectile 
is on the left. Any local observable, denoted generically by $A$ of the target or projectile can be computed using 
 \begin{eqnarray}
 \label{qproj}
A_T(t)&=&\int d^3 {\mathbf r} Q( {\mathbf r} ) \Theta(x_0-x)\rho({\mathbf r} ,{\mathbf r} ), \nonumber \\
A_P(t)&=&\int d^3 {\mathbf r} Q( {\mathbf r} )[ 1 - \Theta(x_0-x)]\rho({\mathbf r} ,{\mathbf r} ),
\end{eqnarray} 
where $\Theta$ is the step function. The most common choice for $A$ are either equal to one (for the mass), to the center of mass position, to momentum or to angular momentum. With proper combinations of these quantities, one can construct a set of specific observables, denoted generically by $\{ A_\lambda \}$ associated to the relative motion or relative particle content. At the mean-field level, one can anticipates that the equation of motion will reduce to a set of equations:
\begin{eqnarray}
\frac{d}{dt} A_\lambda (t) &=& {\cal F}(A_\lambda (t)) + {\cal V}(A_\lambda (t), t).  
\end{eqnarray}
The two terms respectively stand for a driving force stemming from a potential energy surface in collective space and a dissipative kernel.
Note that dissipation of one-body type is automatically contained in TDHF. 

It is anticipated that one-body dissipation is rather well described by mean-field. However, fluctuations leading to the dispersion of one-body DOF are strongly underestimated. When initial fluctuations are included in the SMF framework,  the macroscopic 
equation of motion will be approximately transformed as:
 \begin{eqnarray}
\frac{d}{dt} A^{(n)}_\lambda (t) &=& {\cal F}(A^{(n)}_\lambda (t)) + {\cal V}(A^{(n)}_\lambda (t), t) + \delta \xi^{^{(n)}}_{\lambda} (t), 
\end{eqnarray}
where a new fluctuating term $\delta \xi^{^{(n)}}_{\lambda} (t)$ appears. The properties of the fluctuating quantity will depend on the specific 
collective observables and will in general be rather complex and non-markovian. In the markovian limit,  as we will see in examples below, 
the macroscopic evolution will be similar to a Brownian motion 
where fluctuations are related to the set of second moment $\overline { \delta \xi^{^{(n)}}_{\lambda} (t) \delta \xi^{^{(n)}}_{\lambda'} (t')}$.  

\subsubsection{Energy dissipation}

\begin{figure}[hbtp]
\begin{center}
\resizebox{1.0\columnwidth}{!}{
 \includegraphics{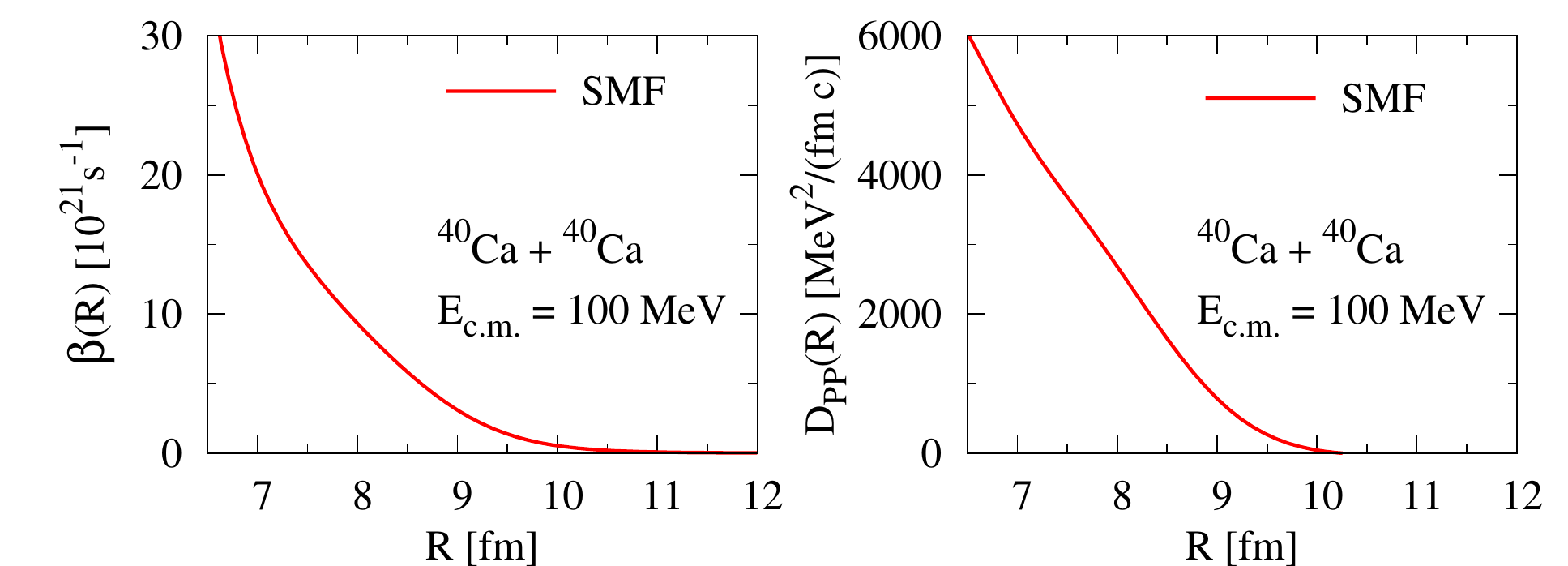}
} 
  \end{center}
  \caption{(Color online) Evolution of reduced friction (left) and diffusion coefficient (right) as a function of the relative 
distance for the head-on $^{40}$Ca+$^{40}$Ca  collision at center of mass energy $E_{\rm c.m.} = 100$ MeV.}
\label{fig2dubna}
\end{figure}

The powerfulness and applicability of the SMF have been recently 
illustrated in fusion reactions by extending the work of ref. \cite{Was08,Was09c}. Using a macroscopic reduction of the stochastic 
mean-field evolution, central collisions leading to fusion have been mapped to a 
one-dimensional macroscopic Langevin evolution on the relative distance $R$ between the two nuclei
given by \cite{Ayi09}:
\begin{eqnarray}
\dot P^{(n)} = - \partial_R U(R^{(n)} ) 
- \gamma [R^{(n)}] \dot {R}^{(n)} + \xi_P^{(n)} (t). \label{eq:langevin} 
\end{eqnarray}
$U(R^{(n)} )$ and $\gamma (R^{(n)} )$ denote the nuclear+coulomb potential and dissipation 
associated to one-body friction respectively and are already present at the mean-field level \cite{Was08,Was09c}. $\xi _P^{(n)} (t)$ is a 
Gaussian random force acting on the relative
motion reflecting stochasticity in the initial value. This fluctuating part leads to diffusion in collective space 
which can be approximated by 
\begin{eqnarray}
\overline{\xi _P^{(n)} (t)\xi _P^{(n)}({t}')} &\simeq& 2\delta(t-{t}')D_{PP}(R) , \nonumber  
\end{eqnarray}
where $D_{PP}(R)$ denotes 
the momentum diffusion coefficient. The latter term is nothing but the one that is missing in the original theory and is of primer 
importance to properly describe observables fluctuations. An example of reduced friction $\beta(R)\equiv \gamma(R)/\mu (R)$ and diffusion 
coefficients $D_{PP}(R)$ estimated from the macroscopic reduction of SMF is given in Fig. \ref{fig2dubna}
for the head-on $^{40}$Ca+$^{40}$Ca  collision.   

\subsubsection{Nucleon exchange}

The possibility to estimate transport coefficients associated to fluctuation and dissipation from a fully 
microscopic quantum transport theory is a major breakthrough. However, to be really convincing, one should 
in addition prove that the increase of fluctuations is consistent with experimental observations. To prove that 
SMF can be a predictive framework, we have recently considered transfer reactions \cite{Was09b}. 
Fragment mass distributions deduced from Heavy-Ion reactions have been extensively studied.
It is seen that the dispersion in mass scales approximately with the average number of 
exchanged nucleons. While mean-field properly describes the latter, it miserably fails to account for the 
dispersion. This phenomena is rather well understood in macroscopic models but has not been yet reproduced 
microscopically. To address this issue, we have considered head-on collisions below the Coulomb barrier. 
In that case, nuclei approach, exchange some nucleons and then re-separate. Similarly to the 
relative distance case, a macroscopic reduction onto the projectile (resp. target) mass, denoted by $A^{(n)}_P$ (resp. 
$A^{(n)}_T$) can be introduced.
For small fluctuations, the ensemble average quantities are equivalent to the results obtained by the standard mean-field approximation. As a result, the Langevin equation for the nucleon exchange becomes,
\begin{eqnarray}
\label{eq:langevin-mass}
\frac{d}{dt}A_T^{(n)}(t) = v_A(t)+\left( \frac{\partial v_A(t)}{\partial A_T} \right)\delta A_T^{(n)}(t) + \xi_A^{(n)}(t),
\end{eqnarray}
where $v_A$ is the drift coefficient for nucleon exchange. The quantity
$\xi_A^{(n)}(t)$ denotes the fluctuating part of the nucleon flux. 
Similarly to the momentum case, a simplified markovian assumption leads to:
\begin{equation}
\overline{\xi_A^{(n)}(t)\xi_A^{(n)}(t')}= 2\delta(t-t')D_{AA}(t),
\end{equation}
where $D_{AA}(t)$ is the diffusion coefficient for nucleon exchange. This result establishes the
connection with the nucleon exchange picture developed in the 80's has been made \cite{Ayi09,Was09b}. 
In particular, computable expressions of the drift and diffusion coefficients were proposed allowing for quantitative description.
\begin{figure}
\begin{center}
\resizebox{0.8\columnwidth}{!}{
  \includegraphics{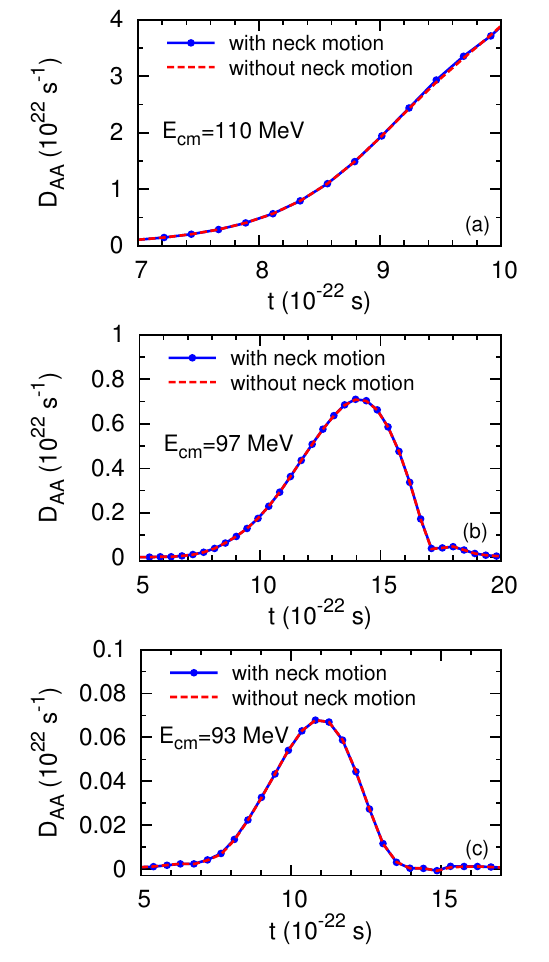}} 
  \end{center}
\caption{(color online) Nucleon diffusion coefficients are plotted versus time in central collisions of $^{40}$Ca + $^{90}$Zr system at three different center-of-mass energies. (for more details see \cite{Yilmaz11a})}
\label{figure7collision}
\end{figure}
Fig. \ref{figure7collision} illustrates the diffusion coefficients obtained in the central collisions of $^{40}$Ca + $^{90}$Zr system 
as a function of time and at center-of-mass energies below the barrier, 
$E_{\rm cm}=93$ MeV (c) and $E_{\rm cm}=97$ MeV (b), and above the barrier energy $E_{\rm cm}=110$ MeV (a).

Employing the Langevin Eq. (\ref{eq:langevin-mass}) we can calculate the variance  
$\sigma^2_{AA}(t)=\langle (A^\lambda_{T})^2\rangle-\langle A^\lambda_{T}\rangle^2$  of fragment mass distribution.  
It follows  that the variance is determined by
\begin{equation}
 \frac{d}{dt}\sigma^2_{AA}(t) = 2\alpha(t)\sigma^2_{AA}(t)+ 2D_{AA}(t),
 \label{variance} 
\end{equation} 
where $\alpha(t)=\partial v_A(t)/\partial A_T$. Because of very small value of the mean nucleon transfer, 
we can neglect the contribution from drift term and solve the variance equation 
(\ref{variance}) to find, 
\begin{equation}
\sigma^2_{AA}(t) = 2 \int_0^t D_{AA}(s)ds.
\label{eq:sigma}
\end{equation} 

\begin{figure}
\begin{center}
\resizebox{0.8\columnwidth}{!}{
 \includegraphics{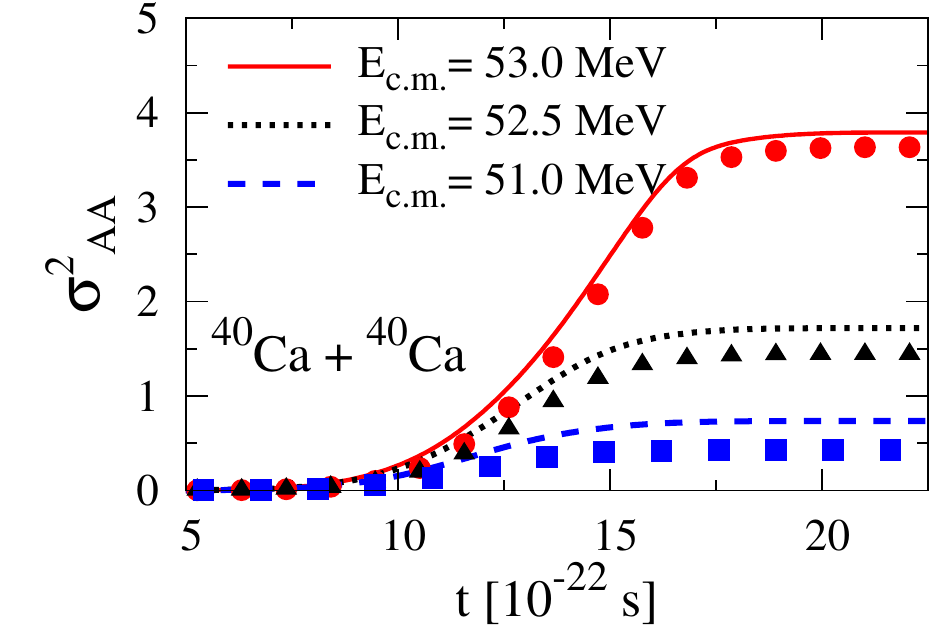}
} 
  \end{center}
\caption{(Color online) Evolution of $\sigma_{AA}^2$ calculated in SMF approach for
 $^{40}$Ca${}+^{40}$Ca (top) at different center of mass energies.
Number of exchanged particles is superimposed
by the filled-circles, filled-squares, and filled-triangles
from high to low energies. (Adapted from \cite{Was09b})
\label{fig3dubnabis}}
\end{figure}

An illustration of $\sigma^2_{AA}(t)$ for  $^{40}$Ca${}+^{40}$Ca  reactions is given in Fig. \ref{fig3dubnabis}
and compared to the number of exchanged nucleons, denoted by $N_{\rm ex}$. In all cases, both quantities 
are very close from each other and lead to much higher dispersion than the original mean-field.
Indeed, with mean-field, the estimated asymptotic values in the latter case are $0.004$, $0.008$ and $0.008$ from low to high energy
and are much less than the final number of exchanged nucleons that are equal to $0.43$, $1.44$ and $3.63$ respectively.
On opposite, the predicted asymptotic mass dispersions are equal to $0.73$, $1.72$ and $3.79$ and is much closer 
to $N_{ex}$ (see also figure \ref{fig3dubnabis}).
This numerical test provides a strong support for the validity of the stochastic mean-field approach.

From the investigation made on heavy-ion collisions, it has been shown that the SMF theories extend he usual TDHF approach and provide
a useful tool to describe not only dissipative aspects but also fluctuations. In particular, it might provide a rather simple approach 
to overcome most of the shortcoming of TDHF. At present, quantitative aspects have been mainly obtained using a simplified semiclassical 
approximation. We have observed that, below the Coulomb barrier energies, the semiclassical expression underestimates the nucleon drift deduced from the standard mean-field description with TDHF equations. In the near future, a fully quantal approach avoiding the semi-classical 
limitations will be required.

\subsection{Summary on the SMF theory}

The SMF theory has important aspects that make it very attractive (see discussion below).
On the theoretical side \cite{Ayi08}, for small amplitude fluctuations, this model
gives a result for the dispersion of a one-body observable that is identical to
the one obtained using the Balian-V\'en\'eroni (BV) variational approach~\cite{Bal85}.
It is also shown that, when the SMF is projected on a collective variable, it gives rise to a
generalized Langevin equation~\cite{Mor65} that incorporates one-body dissipation and
one-body fluctuation mechanisms in accordance with quantal
dissipation-fluctuation relation. These connections give a strong
support that the SMF approach provides a consistent microscopic description
for dynamics of density fluctuations in low energy nuclear reactions.

From the practical point of view, this approach 
is much simpler than the TDGCM. Indeed, by neglecting the interferences between trajectories, each evolution can be made independently
from the others. In addition, on contrary to the Stochastic TDHF case, that is discussed 
in section \ref{eq:stdhf}, randomness appears only at the initial time and should not 
a priori face the difficulty of a statistical explosion of the trajectory number. SMF framework has been recently applied to fusion \cite{Ayi09}
and transfer reactions \cite{Was09b}. The latter study has in particular pointed out that fluctuations of one-body observables are largely 
increased as compared to the TDHF and seem consistent with experimental observations. This issue is a long standing problem  
that was unsolved until now in a fully quantum microscopic approach. Finally recent tests in cases where spontaneous symmetry 
breaking might be particularly important are very promising. 





%
%

\section{Stochastic Schroedinger equation for N-body problems}
\label{sec:jump}

In previous section, we have seen that quantum fluctuations beyond mean-field
can eventually be incorporated approximately by introducing fluctuations at the initial instant. 
Here, we are interested in the stochastic treatment of correlations that are not present initially, but that built up in time such as those included in the Extended TDHF or TDDM theories presented in section \ref{sec:beyond}. It is shown 
below that 
these correlations can be treated too by adding noise to the self-consistent mean-field, replacing then the initial 
problem by a set of simpler evolutions of Slater determinants with the great difference, compared to SMF, that the noise is continuously added during the time-evolution.

\subsection{Stochastic process in Slater Determinant space} 

Before describing the specific case of ETDHF, let us understand in a simple manner  
how a stochastic process can be introduced.
Starting from a simple Slater determinant state $\left| \Psi(t = 0) \right\rangle  = \left| \Phi(t_0) \right\rangle$, 
correlations will develop in time and we do expect that the exact Many-Body state writes:
\begin{eqnarray}
\left| \Psi(t) \right\rangle  = \sum_k c_k (t) \left| \Phi_k(t) \right\rangle,
\end{eqnarray}   
where $\left| \Phi_k \right\rangle$ denotes a complete (eventually time-dependent) 
basis of Slater-Determinant states. Accordingly, the many-body density writes 
\begin{eqnarray}
D(t) &=& \sum_{k,k'} c_k (t)c_{k'} (t) \left| \Phi_k(t) \right\rangle \left\langle \Phi_{k'}(t) \right|.
\label{eq:exact}
\end{eqnarray} 
The extended and stochastic version of TDHF that will be presented below, 
implicitly assume that the many-body density can be 
properly approximated by its diagonal components \cite{Rei92a,Lac06a}  
\begin{eqnarray}
D(t) &\simeq& \sum_{k} P_k  \left| \Phi_k(t) \right\rangle \left\langle \Phi_k(t) \right|,
\end{eqnarray} 
where $P_k = |c_k (t)|^2$. The probability $P_k$ obeys a master equation that eventually 
could be simulated using quantum jumps. The resulting density is obtained through the average
over different stochastic paths, i.e.
\begin{eqnarray}
D(t) & \simeq & \overline{\left| \Phi_k(t) \right\rangle \left\langle \Phi_k(t) \right|}
\end{eqnarray}      
Physically, this can be understood as follows. The irrelevant degrees
of freedom (complex internal degrees of freedom) interacts with the relevant degrees of freedom
(single-particle degrees of freedom) and induce a fast decay towards zero for the 
off-diagonal matrix elements. 
This phenomenon is know as a decoherence process \cite{Kue73,Joo03}.

\subsubsection{Extended TDHF in the short memory approximation}
\label{eq:stdhf}

The Extended TDHF can eventually be interpreted as an average over quantum jumps between Slater determinants, a theory
generally called Stochastic TDHF (STDHF) \cite{Rei92a,Rei92b,Bal81,Lac06a}.   
Let us first introduce ETDHF using a different technique than the truncation of the BBGKY
hierarchy. This theory is expected to be valid in the weak coupling limit, i.e. when the residual 
interaction introduced in section \ref{section:meanfieldThouless} is small. 
Such a theory can indeed be obtained using time-dependent perturbation theory. Starting from an initial density $D(t_0)$, the evolution
is given at second order in perturbation theory by 
\begin{eqnarray}
i\hbar \frac{dD(t)}{dt} &=& [H_{\rm MF}(t), D(t)] \nonumber \\
&-& \frac{ 1 }{2 \hbar^2 } {\rm T} 
\left( \int_{t_0}^t \int_{t_0}^t 
\left[ 
V_{\rm res} (s') ,\left[ V_{\rm res}(s) , D(s) \right] \right]
 ds' ds \right) , \nonumber
\end{eqnarray}
where ${\rm T}(.)$ denotes the time-ordering operator and where $ V_{\rm res}(s)$ denotes the residual interaction 
written in the interaction picture using the mean-field propagator. The expression above is non-local in time, showing 
that the evolution between a time $t$ and an initial time  time $t_0$ depends not only of the system at time $t$ but also on the former time. 

The second term in the evolution of $D(t)$ essentially involves two different characteristic
times. The first one is the correlation time $\tau _{cor}$, that is defined as 
\begin{equation}
\overline{\overline{ V_{res}\left( t\right) V_{res}\left( s
\right) }} \propto e^{-\left| t-s\right|
/\tau _{cor}}  , \label{c4_Vmem}
\end{equation}
where the average $\overline{\overline{ ~\cdot ~}}$ 
denotes an average over all possible single-particle states combinations. 
This time, characteristic of
the residual interaction, is directly related to the mean energy $\Delta$
exchanged during nucleon-nucleon collisions through the relation $\tau_{cor}=\hbar /\Delta $ \cite{Wei80}. 
The second characteristic time, 
called relaxation time $\tau _{rel}$, corresponds to the time-scale
associated to the reorganization of single-particle states. 

Here, we consider the limit  
$\tau _{cor} \langle \langle \tau _{rel}$ that is valid for a sufficiently dilute system
when the binary collisions are well separated in time and use 
\begin{eqnarray}
\overline{\overline{V_{\rm res} (t)  V_{\rm res}(s)}} \propto  V^2_{\rm res}(t) F\left(\frac{|t-s|}{\tau_{\rm cor}} \right).
\label{eq:short}
\end{eqnarray}   
where $F$ is a function that tends to zero over a time-scale ${\tau_{\rm cor}}$ much smaller than the typical
time associated to the reorganization of one-body degrees of freedom.
In that limit, the density in the integral can be approximated by $D(s)\simeq D(t)$ leading finally to 
\begin{eqnarray}
i\hbar \frac{dD(t)}{dt} &=& [H_{\rm MF}(t), D(t)] \nonumber \\
&-&\frac{g}{2} 
\Big\{ V_{\rm res}(t)  V_{\rm res}(t)  D(t) +  D(t) V_{\rm res}(t) V_{\rm res}(t) \nonumber \\
&-& 2 V_{\rm res}(t)  D(t) V_{\rm res} (t) 
\Big\}
 \label{eq:lindbladnbody}
\end{eqnarray}
where the constant 
\begin{eqnarray}
g \equiv \frac{1}{\hbar^2} \iint F(|s-s'| / \tau_{\rm cor}) ds ds'
\end{eqnarray} 
is introduced. The approximation above leads to an equation of motion for the density $D(t)$ that is local in time, 
and therefore memory effects have disappeared. 

Eq. (\ref{eq:lindbladnbody}) equation is nothing but a Lindblad equation that is generally found in open quantum 
systems \cite{Lin75,Lin75a,Bre02}. Therefore, starting from 
second-order perturbation theory and assuming the short memory approximation leads naturally to an Open Quantum System
equation of motion. 

\subsubsection{Dissipation in one-body space}
Eq. (\ref{eq:lindbladnbody}) is rather complicated and 
involves complex many-body 
operators. Here, we are mainly interested in one-body degrees of freedom. Starting from Eq. (\ref{eq:lindbladnbody}), the 
one-body density matrix evolution reads \cite{Lac06a}:
\begin{eqnarray}
\frac{d \rho}{dt} &=& \frac{1}{i\hbar}\left[ h_{MF}(\rho),\rho \right] -  \frac{g}{2} {\cal D}(\rho).
\label{eq:drho2}
\end{eqnarray}
${\cal D}(\rho)$, called "dissipator" hereafter, corresponds to the average effect of the
residual interaction and reads
\begin{eqnarray}
\left< j \left| {\cal D} \right| i \right> &=& {\rm Tr}\left( D
\left[\left[a^+_i a_j , V_{\rm res}  \right], V_{\rm res} \right]  \right).
\label{eq:diss}
\end{eqnarray}
Assuming that the system is initially in a pure state described by a Slater determinant $\left| \Phi (t_0) \right>$
formed of $N$ orthonormal single particle states denoted by $\left| \alpha  \right>$, the
associated initial one-body density matrix reads $\rho = \sum_{\alpha } \left| \alpha  \right>
\left< \alpha  \right|$. 
Using the residual interaction expression, Eq. (\ref{eq:hphi}), 
${\cal D}(\rho)$ can finally be recast as:  
\begin{eqnarray}
{\cal D}(\rho) =  Tr_2 \left[ \tilde v_{12}, B_{12} \right],
\label{eq:dissrho}
\end{eqnarray}
where $B_{12}$ is nothing but the Born term appearing in the Extended TDHF theory. Indeed, a similar 
expression could have been directly obtained starting from the ETDHF theory in the Markovian limit.
Equation (\ref{eq:drho2}) is a master equation for the one-body density. It could also be put into 
a Lindblad form using the fact that the residual interaction can always be decomposed as (see for instance \cite{Koo97,Jui02})
\begin{eqnarray}
V_{\rm res} = -\frac{1}{4} \sum_n \lambda_n {\cal O}^2_n,
\label{eq:dvoo}
\end{eqnarray}  
where $\lambda_n$ are real and where the ${\cal O}_n$ correspond to a set of commuting Hermitian one-body operators written as
${\cal O}_n = \sum_{\bar{\alpha}\alpha} \left< \bar{\alpha} \left| O_n \right| \alpha \right>a^\dagger_{\tilde{\alpha}} a_\alpha$.
Reporting in eq. (\ref{eq:dissrho}), ${\cal D}(\rho)$ can be recast as
\begin{eqnarray}
{\cal D}(\rho)  = \sum_{mn} \Gamma_{mn} \left[O_n O_m ~\rho +  \rho ~O_n O_m-  2 O_m ~ \rho ~ O_n
\right].
\label{eq:disslast}
\end{eqnarray}
The coefficient $ \Gamma_{mn}$ are given by  
\begin{eqnarray}
\Gamma_{mn} = \frac{1}{2} \lambda_m \lambda_n Tr(O_m (1-\rho) O_n \rho).
\end{eqnarray}
We recognize in this expression, the quantum covariance between the operator ${\cal O}_n$ and ${\cal O}_m$ , i.e.
$Tr(O_m (1-\rho) O_n \rho) = \left< {\cal O}_m {\cal O}_n \right> - \left< {\cal O}_m\right>\left< {\cal O}_n\right>$.
Expression (\ref{eq:disslast}) has the form of the dissipator appearing usually in the Lindblad equation\cite{Bre02}.
Therefore, the evolution of one-body degrees of freedom associated to equation (\ref{eq:lindbladnbody})
identifies with a Markovian quantum master equation generally obtained in quantum open systems. A large amount of work
is devoted to the simulation of such master equation by quantum jump methods (see for instance
\cite{Dio86,Car93,Rig96,Ple98,Bre02}) and one can take advantage of the most recent advances in this field.
This aspect has however rarely been discussed in the context of self-interacting system. 

\subsubsection{Stochastic process in one-body space}

Following ref. \cite{Bre02}, we introduce the Hermitian matrix $\Gamma$ with components $\Gamma_{mn}$.
An economical method to introduce quantum jumps
is to use the unitary transformation $u$ that diagonalizes $\Gamma$, i.e. $\Gamma  = u^{-1}\gamma u$,
where $\gamma$ is the diagonal matrix of the eigenvalues of $\Gamma$.
New operators $A_k$ can be defined
by the transformation $A_k = \sum_n u^{-1}_{k n} O_n$. The dissipator is then recast as  
\begin{eqnarray}
{\cal D}(\rho)  = \sum_{k} \gamma_{k} \left[A^2_k \rho +  \rho A^2_k -  2 A_k \rho A_k
\right].
\label{eq:dissverylast}
\end{eqnarray}
Last expression can be simulated using the average over the stochastic mean-field dynamics:
\begin{eqnarray}
d \rho &=& \frac{dt}{i\hbar}\left[ h_{MF}(\rho),\rho \right] -  g\frac{ dt }{2} {\cal D}(\rho) + dB_{sto},
\label{eq:drhosto}
\end{eqnarray}
where $dB_{sto}$ is a stochastic one-body operator which,  using Ito rules \cite{Gar85} (see also appendix \ref{sec:ito}), reads
\begin{eqnarray}
dB_{sto} &=&\sum_k  \left\{ dW_k (1 - \rho)A_k \rho + dW^*_k \rho A_k (1 - \rho) \right\}.
\end{eqnarray}
Here $dW_k$ denotes stochastic variables given by $dW_k  = -i d\xi_k \sqrt{g \gamma_k}$,
where $\{ d \xi_k \}$ correspond to a set of real gaussian stochastic variables with mean zero and
$\overline{d\xi_k d\xi_{k'}} = \delta_{kk'}dt$.

\subsubsection{Quantum jump for single-particle states}

It is worth noticing that the proposed dissipative equation and its stochastic counterpart are
only well defined if the density is initially prepared as a pure Slater-determinant state. We now turn to
the essential properties of equation (\ref{eq:drhosto}). First, it preserves the number of particles $Tr(d\rho) = 0$.
In addition, if initially $\rho^2 = \rho$, then
\begin{eqnarray}
d\rho d\rho -g\frac{dt}{2} \left[\rho {\cal D}(\rho) + {\cal D}(\rho)\rho  \right] = -g\frac{dt}{2} {\cal D}(\rho)
\end{eqnarray}
which is obtained using Ito stochastic rules and retaining only terms linear in $dt$. Last expression demonstrates
that $(\rho+d\rho)^2 = \rho+ d\rho$. Thus, $\rho$ remains a projector along the stochastic path. As a consequence,
the pure state nature of the many-body density matrix is preserved along the stochastic
path, i.e. $D=\left| \Phi(t)  \right> \left< \Phi (t) \right|$ where $\left| \Phi  \right>$ is a normalized
Slater determinant at all time. The associated stochastic Schroedinger equation for single-particle states reads
\begin{eqnarray}
d\left| \alpha  \right> &=& \left\{ \frac{dt}{i\hbar} h_{MF}(\rho) + \sum_k dW_k (1-\rho) A_k \right. \nonumber \\
&-&  \left. g\frac{dt}{2} \sum_k \gamma_k \left[ A^2_k \rho + \rho A_k \rho A_k -2 A_k \rho A_k \right]
   \right\}
\left| \alpha  \right>. \nonumber 
\end{eqnarray}
This last expression can be directly used for practical applications.
 
In this section, we have shown that the effect of residual interaction at second order in perturbation 
and projected on one-body degrees of freedom gives the Extended TDHF approximation in the short-memory time (Markovian) 
approximation. In such a limit, starting from a pure Slater Determinant state, 
the dissipative dynamics can be replaced by a quantum jump process where the N-body state remains a SD 
along each stochastic trajectory. The possibility to account for the effect of correlation on top of a mean-field 
dynamics has been discussed extensively in the early 80's. For instance, it has been  
proposed to treat each direct nucleon-nucleon collisions as a random process \cite{Bal81,Gra81}. Alternatively, following 
a similar strategy as the one presented in this section and starting from perturbation theory \cite{Rei92a,Rei92b}, 
the Fermi golden rule has been used to introduce a Stochastic TDHF theory. The main difficulty is to avoid the explosion 
of the number of trajectories and therefore find physical criteria to only follow relevant trajectories. 
The approach presented here makes more transparent the connection of a many-body system where specific degrees of freedom
are of interest and the theory of Open Quantum Systems. In addition, the stochastic evolution of single-particle states 
are directly the equations that should be implemented in practice. It should however be noted that the possible explosion 
of trajectories is not the only reason that may limit the application of Stochastic TDHF. Indeed, such a theory 
is well defined if we start from a Hamiltonian but is less clear in the context of density functional theory that most often is 
not directly linked to the underlying many-body hamiltonian.

\subsection{Exact Quantum Monte-Carlo from functional integrals method}

Approximations to the N-body problem such as ETDHF, STDHF or SMF focus on one-body degrees of freedom. 
In these framework some many-body effects such as interferences
between different channels are lost. In particular, we do expect that most of the extensions of TDHF presented above  
will not be able to describe two-body or more complex degrees of freedom. Mean-field 
theories by projecting out the evolution onto a specific class of degrees of freedom can then be regarded 
as a system open to the surrounding more complex observables  
(see for instance discussion in \ref{eq:stdhf}). From the Open Quantum System point of view, the introduction 
of Extended TDHF and then Stochastic TDHF can be considered as a rather standard way to introduce dissipation using first 
the Nakajima-Zwanzig approach, second the Markovian approximation and then the stochastic unraveling.
Less conventional approaches based on quantum Monte-Carlo can be used 
to treat exactly the dynamics of a system coupled to an environment \cite{Lac05b,Lac08}. A similar exact reformulation also exists in the case of 
interacting particles using the functional integral method.

Functional integrals techniques have often been used to replace the exact Many-Body problem 
by an average over different "effective" one-body problem \cite{Lev80-a,Lev80-b,Neg88}.
In ref. \cite{Koo97}, the general strategy to obtain ground state properties
of a many-body system using Monte-Carlo methods, the so called Shell-Model Monte-Carlo, is described. 
Recently, this technique has been combined with mean-field theory to obtain Stochastic TDHF equations 
which in average lead to the exact evolution \cite{Car01,Jui02}. 
The goal of the present section is to demonstrate that one could always treat exactly the problem of interacting particle with density 
given by Eq. (\ref{eq:exact}) by an appropriate stochastic process between Slater determinants. The exact density will then be obtained by an average 
\begin{eqnarray}
D(t) & \simeq & \overline{\left| \Phi_k(t) \right\rangle \left\langle \Phi_k'(t) \right|}
\end{eqnarray}  
where states in the left differ from states on the right.
 
\subsubsection{ Functional integrals for schematic residual interaction:}

We again consider that, at a given time, the Many-Body state is a Slater Determinant  
$\left| \Psi(t)  \right\rangle= \left| \Phi \right\rangle$.  For short time step
$\Delta t$, we have 
\begin{eqnarray}
\left| \Psi(t+ \Delta t) \right\rangle &=& \exp \left( \frac{\Delta t}{i\hbar}  H \right)
\left| \Phi (t) \right\rangle \nonumber \\
&\simeq & \Big(1+ \frac{\Delta t}{i\hbar}  H + O(\Delta t) \Big) \left| \Phi(t) \right\rangle.
\end{eqnarray} 
Due to the presence of a two-body interaction in $ H$, the state $\left| \Psi(t+ \Delta t) \right\rangle$ differs from
a Slater Determinant. However, it is proved here that it could be replaced exactly by an average 
over quantum jumps between SD states.

At any time, the Hamiltonian can be decomposed as a mean-field and a residual part. For simplicity, it is  
first assumed that 
\begin{eqnarray}
V_{\rm res} =  A^2,
\end{eqnarray}
$A$ being a one-body operators. A Gaussian probability $G(x)$ with mean zero and variance $1$ is introduced and
the complex number $\Delta \omega \equiv \sqrt{\frac{2\Delta t}{i\hbar}}$ is defined as well as 
the one-body operator $  S(\Delta t , x)$ with
\begin{eqnarray}
 S(\Delta t , x) \equiv \frac{\Delta t}{i\hbar}  H_{\rm MF} + x \Delta \omega  A .
\end{eqnarray}
Considering the average value of $ S(\Delta t , x)$ and keeping 
only terms up to $\Delta t$, we obtain:
\begin{eqnarray}
\int_{-\infty}^{+\infty} e^{ S(\Delta t , x)} G(x) dx &=& 1 + \frac{\Delta t}{i\hbar}  H_{\rm MF} 
+ \overline{x} ~\Delta \omega  A \nonumber \\
&+& \overline{x^2} ~(\Delta \omega)^2  A^2  + O(\Delta t) \nonumber \\
&=& 1 + \frac{\Delta t}{i\hbar}  H + O(\Delta t).
\end{eqnarray}
By averaging over the different realization of $x$, we recover the exact propagator over short time step. Note 
that more general relations could be found using the Hubbard-Stratonovish transformation 
(see for instance \cite{{Koo97}}).
Using the above relation, we see that
\begin{eqnarray}
\exp \left( \frac{\Delta t}{i\hbar}  H \right) \left| \Phi \right\rangle &=& 
\int_{-\infty}^{+\infty} dxG(x) e^{ S(\Delta t , x)}\left| \Phi(t) \right\rangle \nonumber \\
&\equiv& 
\int_{-\infty}^{+\infty} dxG(x) \left| \Phi_x (t+ \Delta t) \right\rangle
\end{eqnarray} 
Due to the one-body nature of $ S$, each $\left| \Phi_x (t+ \Delta t) \right\rangle$ is a Slater determinant. 
Therefore, 
we have demonstrated that the evolution of the exact state could be replaced by an ensemble of Slater determinants. 
The technique could be iterated for each $\left| \Phi_x (t+ \Delta t) \right\rangle$ to obtain the exact long time dynamics as an average 
over Slater determinant states. 
In the continuous time 
$\Delta t \rightarrow dt$, we will introduce the notation 
\begin{eqnarray}
\left| d \Phi \right\rangle &=& \left\{ \frac{dt}{i\hbar}  H_{\rm MF} + x d\omega A \right\}\left| \Phi \right\rangle  
\label{eq:SSEl} 
\end{eqnarray} 
which will be called Stochastic Schroedinger Equation and describe the quantum jump process between Slater determinants.
Several comments are in order:
\begin{itemize}
\item[$\bullet$] Since $ S(\Delta t, x)$ is not a priori Hermitian, the dynamics does not preserves the orthogonality 
of the single-particle wave-function. Such a non-orthogonality should properly be treated during the 
time evolution \cite{Jui02,Lac05}.
\item[$\bullet$] Starting from a Many-Body density written as $D(t) = \left| \Phi \right\rangle \left\langle \Phi \right|$, at an 
intermediate time, the average density writes 
\begin{eqnarray}
D(t) = \overline{\left| \Phi_1 (t) \right\rangle \left\langle \Phi_2 (t) \right|} ,
\end{eqnarray}  
where $\left| \Phi_1 \right\rangle$ evolves according to Eq. (\ref{eq:SSEl}) while $\left\langle \Phi_2 \right|$
evolves according to
\begin{eqnarray}
\left\langle \Phi_2(t+ \Delta t) \right| &=& 
\left\langle \Phi_2(t) \right| \exp\left\{ -\frac{\Delta t}{i\hbar}  H_{\rm MF} + y \Delta \omega^*  A \right\}. \nonumber
\end{eqnarray} 
$ y $ is a noise independent of $x$, with mean zero and $\overline{y y} = 1$. Since the 
evolution is exact, any one-, two- or k-body observable $ Q$ estimated through $\langle   Q \rangle \equiv 
Tr(D(t) Q)$ will follow the exact dynamics \cite{Lac05}.
\end{itemize}

\subsubsection{\bf General Many-Body  Hamiltonian:}  

The functional integral method has been introduced above using a 
schematic separable residual interaction. For a general two-body Hamiltonian, one can take advantage of the 
decomposition of the residual interaction according to Eq. (\ref{eq:dvoo}).
Therefore, for realistic interactions one should introduce as many stochastic Gaussian independent 
variables as the number of operators entering in the sum. In practice, this number defines the 
numerical effort which in general is very large. For this reasons only few applications to the dynamics 
of rather simple systems exist so far.   
Last, the extension of above stochastic theories to HFB state has been given in ref. \cite{Lac06b}.

\subsection{Quantum Monte-Carlo method for closed systems from optimal observables evolution}

Using the functional integral method, it has been shown above that the exact evolution of particles interacting 
through a two-body Hamiltonian can be  
replaced by a set of stochastic evolutions of densities written as $D = | \Phi_a \rangle \langle \Phi_b |$ where 
both $| \Phi_a \rangle $ and $| \Phi_b \rangle $ are independent particle states. More generally, several studies \cite{Car01,Jui02,Jui04,Bre04a,Bre04b,Lac05b} have 
shown that the exact dynamics of a many-body system can be replaced by the average over "densities" of the form  
\begin{eqnarray}
D(t) = \left| Q_a \right> \left< Q_b \right|,
\label{eq:dab}
\end{eqnarray}  
where $\left| Q_a \right> $ and $\left| Q_b \right>$ belong to a specific class of trial states 
introduced in section \ref{sec:ehrenfest}.
One of the disadvantage of the functional integral approach is that the link with observable evolution is highly non-trivial. 
Here, a different 
strategy proposed in ref. \cite{Lac07} is introduced to design the quantum Monte-Carlo process. 
The method is not specifically dedicated to the 
N-body problem. Therefore, it is presented starting from any class of trial states.  
The basic idea is to directly
use observables evolution  to deduce the stochastic contribution.    
In section  \ref{sec:ehrenfest}, it is shown 
that mean-field approximation can be regarded as the optimal path for the expectation values of the observables $
\{\langle  A_\alpha \rangle \}$ that generate transformations between trial states. Accordingly, mean-field dynamic
insures that the exact Ehrenfest evolution is obtained for these observables over short time. Here, we consider 
evolution within the class of trial states given by 
\begin{eqnarray}
\left| Q_a + \delta Q_a \right>  &=& e^{\sum_\alpha \delta q^{[a]}_\alpha A_\alpha}  \left| Q_a \right>, \\
\left| Q_b + \delta Q_b \right>  &=& e^{\sum_\alpha \delta q^{[b]}_\alpha A_\alpha}  \left| Q_b \right>,
\label{eq:statevaria}
\end{eqnarray}
where now $\delta q^{[a]}_\alpha$ and $\delta q^{[b]}_\alpha$ may also contain a fluctuating part.

The aim of the present section is to show that, given a class of trial states, a hierarchy of 
Monte-Carlo formulations can be systematically obtained, written as 
\begin{eqnarray}
\left\{
\begin{array} {c}
\delta q^{[a]}_\alpha = \delta q^a_\alpha + \delta \xi^{[2]}_\alpha + \delta \xi^{[3]}_\alpha+ \cdots  \\  
\delta {q^{[b]}_\alpha}^* = \delta {q^b_\alpha}^* + \delta \eta^{[2]}_\alpha + \delta \eta^{[3]}_\alpha+ \cdots   
\end{array}
\right.
\label{eq:dqdq}
\end{eqnarray}
where the second, third... terms represent stochastic variables added on top of the self-consistent
evolution. These random terms are optimized to not only insure that the average evolution of $\left< A_\alpha \right>$ 
matches the exact evolution
at each time step but also that the average evolutions of higher moments 
$\left< A_\alpha A_\beta \right>$, $\left< A_\alpha A_\beta A_\gamma \right>$,... follow the exact Ehrenfest 
dynamics.  
       
\subsubsection{Link between stochastic process and observables evolution}

\noindent{{\bf  Step 1: deterministic evolution}}
\label{sec:step1}

Assuming first that stochastic contributions $\xi^{[i]}_\alpha$ and $\eta^{[i]}_\alpha$ are
 neglected in eq. (\ref{eq:dqdq}), 
we show how variational principles described previously can be used for mixed densities given by eq. (\ref{eq:dab}).
It is worth noticing that variational principles have also been proposed to 
estimate transition amplitudes \cite{Bla86} (see also discussion in \cite{Bal85}). 
In that case, different states are used in the left and right hand side of the action. 
This situation is similar to the case we are considering. 
We are interested here
in the short time evolution of the system, therefore we 
disregard the time integral in equation (\ref{eq:varia}) and consider directly the action    
\begin{eqnarray}
S = {\rm Tr} \left( \left\{ i \hbar \partial^\triangleright_t - i \hbar \partial^\triangleleft_t - H  \right\} D \right). 
\end{eqnarray}
Starting from the above action,  different aspects discussed in section \ref{sec:varia} can be generalized to 
the case of densities formed of trial states couples. For instance, the minimization with respect to the variations
$\left< \delta Q_b \right|$ and $\left| \delta Q_a \right>$ leads to the two conditions 
\begin{eqnarray}
\left\{
\begin{array} {c}
i \hbar \left< Q_b \left| A_\alpha \right| d Q_a \right> = \left< Q_b \left| A_\alpha  H \right| Q_a \right>, \\
\\
i \hbar \left< d Q_b \left| A_\alpha \right| Q_a \right> = \left< Q_b \left|  H A_\alpha \right| Q_a \right>,
\end{array}
\right.
\end{eqnarray}
from which we deduce that
\begin{eqnarray}
i \hbar \frac{d}{dt} \left< A_\alpha  \right> = \left< \left[ A_{\alpha}, H \right] \right>,
\end{eqnarray}
where $\left< A_\alpha  \right> = \left< Q_b \left| A_\alpha \right| Q_a \right>$. Therefore, the minimization of the action again insures that the exact Ehrenfest evolution is followed by the $A_\alpha$ observable
over one time step. Similarly, the evolution of both $\left| Q_a \right>$ and $\left| Q_b \right>$ are given by \footnote{For simplicity, we 
consider here non-necessarily normalized states.}  
\begin{eqnarray}
\left\{
\begin{array} {ccc}
\left| dQ_a \right> &=& \sum_\alpha d q^{a}_\alpha A_\alpha \left| Q_a \right>=
\frac{dt}{i \hbar} {\cal P}_1 H\left| Q_a \right> \nonumber \\
\\
\left< dQ_b \right| &=& \left< Q_b \right|\sum_\alpha d {q^{b}_\alpha}^*A_\alpha = -\frac{dt}{i \hbar} 
\left< Q_b \right|H {\cal P}_1 \nonumber
\end{array}
\right.
\end{eqnarray}  
where ${\cal P}_1$ now reads
\begin{eqnarray}
{\cal P}_1 = \sum_{\alpha \beta} A_\alpha \left| Q_a \right>C^{-1}_{  \alpha \beta} 
\left< Q_b \right|A_\beta.
\end{eqnarray}
In opposite to previous section, ${\cal P}_1$ cannot be interpreted as a projector 
onto the space of observable. Indeed, $C_{\alpha \beta } = \left< Q_b\left| A_\alpha A_\beta \right| 
Q_a\right>$ is not anymore a metric for that space. 
However, the 
total Hamiltonian can still be split into two parts
\begin{eqnarray}
H &=& {\cal P}_1 H + (1-{\cal P}_1) H =  H {\cal P}_1+  H(1-{\cal P}_1)
\end{eqnarray}
the first part being responsible for the mean-field deterministic evolution.

\noindent{{\bf Step 2 : Introduction of Gaussian stochastic processes:}}

In this section, it is shown 
that the description of the dynamics can 
be further improved by introducing diffusion in the Hilbert space of trial states.
We consider that the evolutions of $ q^{[a]}_\alpha$ and $ q^{[b]}_\alpha$ now read 
\begin{eqnarray}
d q^{[a]}_\alpha &=& d q^a_\alpha + d \xi^{[2]}_\alpha , \nonumber \\
d {q^{[b]}_\alpha}^* &=& d {q^b_\alpha}^* + d \eta^{[2]}_\alpha , \nonumber 
\end{eqnarray}
where $d \xi^{[2]}_\alpha$ and $d \eta^{[2]}_\alpha$ correspond to two sets 
of stochastic gaussian variables  
with mean values equal to zero and variances verifying 
\begin{eqnarray}
d\xi^{[2]}_\alpha  d\xi^{[2]}_\beta &=& d \omega_{\alpha \beta} , ~~~ 
d\eta^{[2]}_\alpha d\eta^{[2]}_\beta = d \sigma_{\alpha \beta} , ~~~
d\xi^{[2]}_\alpha  d\eta^{[2]}_\beta = 0 \nonumber
\end{eqnarray}
We assume that $d \omega_{\alpha \beta}$ and $d  \sigma_{\alpha \beta}$ are proportional to 
$dt$. The advantage of introducing the Monte-Carlo method 
can be seen in the average 
evolutions of the states. Keeping only linear terms  in $dt$ in eq. (\ref{eq:statevaria}) gives for instance 
\begin{eqnarray}
\overline {\left| dQ_a \right>} &=& \Big\{\sum_\alpha d q^{a}_\alpha A_\alpha \nonumber \\
&&\hspace*{0.5cm}+ \sum_{\alpha < \beta} 
d \omega_{\alpha \beta } \left( A_\alpha A_\beta + A_\beta A_\alpha \right) \Big\}\left| Q_a \right>. 
\label{eq:dq1dq2}
\end{eqnarray}
Mean field approximation leads to an approximate treatment 
of the dynamics associated to effective Hamiltonian which can only be written as a linear superposition of the $A_\alpha$ (see Eq.
(\ref{eq:meanlin})).
Last expression underlines that, while the states remain 
in a simple class of trial states, the average evolution can now simulate the evolution with an effective Hamiltonian containing not only
linear but also quadratic terms in $A_\alpha$. 

The goal is now to take advantage of this generalization and reduce further the distance between the 
average evolution and the exact one. The most natural generalization of mean-field is to
minimize the average action 
\begin{eqnarray}
S = \overline{{\rm Tr}\left( \left\{ i \hbar \partial^\triangleright_t - i \hbar \partial^\triangleleft_t 
- H  \right\} D \right)} ,
\label{eq:actionaver}
\end{eqnarray} 
with respect to the variations of different parameters, i.e. 
$\delta q^{a}_\alpha$, $\delta {q^{b}_\alpha}^*$,  $\delta \omega_{\alpha \beta}$
and $\delta \sigma_{\alpha \beta}$. In the following, 
a formal solution of the minimization procedure is obtained. The variational principle applied 
to stochastic process generalizes the deterministic case by imposing that not only that 
expectation values $\left< A_\alpha \right>$ but also the second moments 
$\left< A_\alpha A_\beta \right>$, follow the Ehrenfest theorem prescription.   

\noindent{{\bf Effective Hamiltonian dynamics deduced from the minimization: }}

The variations with respect to $\delta {q^{b}_\alpha}^*$ and $\delta \sigma_{\alpha \beta}$ 
give two sets of coupled equations between $dq^a_\alpha$ and $d\omega_{\alpha \beta}$. 
The formal solution of the minimization can however be obtained by making an appropriate change on the variational 
parameters prior to the minimization. In the following, the notation 
$B_{\nu} = A_\alpha A_\beta + A_\beta A_\alpha$ is introduced where $\nu$ denotes $(\alpha,\beta)$ with $\alpha < \beta$. Starting from the general form of the effective evolution
(\ref{eq:dq1dq2}), we  dissociate the part which contributes to the evolution of the $\left<A_\alpha \right>$ 
from the rest. This could be done by introducing the projection operator 
${\cal P}_1$. Equation (\ref{eq:dq1dq2}) then reads
\begin{eqnarray}
\overline {\left| d Q_a \right>} &=& \left\{\sum_\alpha  d {z^a_\alpha} A_\alpha + \sum_{\nu} d \omega_{\nu} 
(1-{\cal P}_1)B_\nu
\right\}\left| Q_a \right>,
\label{eq:dq1dq2new}
\end{eqnarray}  
where the new set of parameters $d {z^a_\alpha}$ are given by 
\begin{eqnarray}
d {z^a_\alpha} = d {q^a_\alpha} + \sum_{\beta \nu}  
d \omega_{\nu} C^{-1}_{\alpha \beta} \left< Q_b \left| A_\beta B_\nu  \right| Q_a \right>.
\end{eqnarray}  
Similarly, the average evolution $\left< dQ_b \right|$ transforms into
\begin{eqnarray}
\overline {\left< d Q_b \right|} &=& \left< Q_b \right| \left\{\sum_\alpha  d {z^b_\alpha}^* A_\alpha 
+ \sum_{\nu} d \sigma_{\nu} B_\nu(1-{\cal P}_1)
\right\},
\label{eq:dq1dq2new2}
\end{eqnarray}  
where $d {z^b_\alpha}$ is given by 
\begin{eqnarray}
d {z^b_\alpha}^* = d {{q^b_\alpha}^*} + \sum_{\beta \nu}  d \sigma_{\nu} 
 \left< Q_b \left|  B_\nu A_\beta \right| Q_a \right>C^{-1}_{\beta \alpha}.
\end{eqnarray}
In the following, we write $B'_\nu = (1-{\cal P}_1)B_\nu$ and $B''_\nu = B_\nu(1-{\cal P}_1)$. 
The great interest of this transformation is to have $\left< A_\alpha B'_\nu \right> = 0$ and 
$\left< B''_\nu A_\alpha \right>=0$ for all $\alpha$
and $\nu$. Accordingly, the variations with 
respect to $\delta {{z^b_\alpha}^*}$ and $\delta {{z^a_\alpha}}$ lead to 
\begin{eqnarray}
\left\{
\begin{array} {c}
i \hbar \overline{\left< Q_b \left| A_\alpha \right| d Q_a \right>} = \left< Q_b \left| A_\alpha  H \right| Q_a \right> \\
\\
i \hbar \overline{\left< d Q_b \left| A_\alpha \right| Q_a \right>} = \left< Q_b \left|  H A_\alpha \right| Q_a \right>,
\end{array}
\right.
\end{eqnarray}
leading to closed equations for the variations $dz^a_\alpha$ and ${dz^b_\alpha}^*$
that are decoupled from the evolution of $d \omega_{\nu} $ and $d \sigma_{\nu}$. These equations
are identical to the ones derived in step 1 and can be again inverted as 
\begin{eqnarray}
\sum_\alpha  d {z^a_\alpha} A_\alpha \left| Q_a \right> = \frac{dt}{i\hbar} {\cal P}_1 H \left| Q_a \right>, \\
\left< Q_b \right| \sum_\alpha  d {z^b_\alpha}^* A_\alpha  = -\frac{dt}{i\hbar} \left< Q_b \right|
H {\cal P}_1. 
\end{eqnarray}  
On the other hand, the variations with respect to $\delta \sigma_\nu$ and $\delta \omega_\nu$ lead to
\begin{eqnarray}
\left\{
\begin{array} {c}
i \hbar \overline{\left< Q_b \left| B''_\nu \right| d Q_a \right>} = \left< Q_b \left| B''_\nu  H \right| Q_a \right>, \\
\\
i \hbar \overline{\left< d Q_b \left| B'_\nu \right| Q_a \right> } = \left< Q_b \left|  H B'_\nu \right| Q_a \right>,
\end{array}
\right.
\end{eqnarray} 
which again gives closed equations for $d \omega_\nu$ and $d \sigma_\nu$.
These equations can be formally integrated by introducing the two projectors 
${\cal P}_2$ and ${\cal P}'_2$ associated respectively to the subspaces of operators $B_\nu(1-{\cal P}_1)$
and $(1-{\cal P}_1)B_\nu$. ${\cal P}_2$ differs from ${\cal P}'_2$ due to the fact that $B_\nu$ operators 
and $A_\alpha$ operators do not a priori commute.
Then, the effective evolution given by eq. (\ref{eq:dq1dq2}) becomes
\begin{eqnarray}
\overline{\left| dQ_a \right>} &=& \frac{dt}{i\hbar} 
\left(\sum_\alpha  d {z^a_\alpha} A_\alpha + (1-{\cal P}_1) \sum_{\nu} d \omega_{\nu} B_\nu \right) \left| Q_a \right> \nonumber \\
&=&  \frac{dt}{i\hbar} \left( {\cal P}_1 + {\cal P}_2 \right)H \left| Q_a \right>, 
\end{eqnarray}   
while 
\begin{eqnarray}
\overline{\left< dQ_b \right|} &=& -\frac{dt}{i\hbar} \left< Q_b \right| H\left( {\cal P}_1 + {\cal P}'_2 \right). 
\end{eqnarray}
In both cases, the first part corresponds to the projection of the exact dynamics on the space of 
observable $\left< A_\alpha\right>$.
The second term corresponds to the projection on the subspace of the observable $\left< A_\alpha A_\beta \right>$ 
"orthogonal" to the space of the $\left< A_\alpha \right>$.

\noindent{{\bf Interpretation in terms of observable evolution:}}

The variation with respect to an enlarged set of parameters does a priori completely determine the deterministic 
and stochastic evolution of the two trial state vectors. The associated average Schroedinger evolution 
corresponds to a projected dynamics. 
The interpretation of the solution obtained by variational principle 
is rather clear in terms of observable evolution. Indeed, from the two 
variational conditions, we can easily deduce that 
\begin{eqnarray}
\overline{ d\left< A_\alpha  \right>} &=& \left< \left[ A_{\alpha}, H \right] \right>, \nonumber \\
\overline{d\left< B_\nu  \right>} &=&  \frac{dt}{i\hbar} \left< \left[ B_{\nu}, H \right] \right>. \nonumber 
\end{eqnarray}
In summary, using the additional parameters associated with the stochastic contribution as variational parameters
for the average action given by eq. (\ref{eq:actionaver}), one can further reduce the distance between the 
simulated evolution and the exact solution. When gaussian noises are used, this 
is equivalent to impose that the evolution of the correlations between operators $A_\alpha$ obtained by averaging over 
different stochastic trajectories also matches the exact evolution.

\noindent{{\bf Step 3: Generalization}}

If the Hamiltonian $H$ applied to the trial state can be written as a quadratic Hamiltonian in terms of $A_\alpha$ and 
if the trial states form an over-complete basis of the total Hilbert space, then the above procedure provides 
an exact stochastic reformulation of the problem.
If it is not the case, the above methods can be generalized  by introducing higher 
order stochastic variables. Considering now the more general form
\begin{eqnarray}
\left\{
\begin{array} {c}
\delta q^{[a]}_\alpha = \delta q^a_\alpha + \delta \xi^{[2]}_\alpha + \delta \xi^{[3]}_\alpha+ \cdots \nonumber \\  
\delta {q^{[b]}_\alpha}^* = \delta {q^b_\alpha}^* + \delta \eta^{[2]}_\alpha + \delta \eta^{[3]}_\alpha+ \cdots \nonumber   
\end{array}
\right.
\end{eqnarray}  
we suppose now that the only non vanishing moments for $d\xi^{[k]}_\alpha$ and $d\eta^{[k]}_\alpha$ are the moments 
of order $k$ (which are then assumed to be proportional to $dt$). 
For instance, we assume that $d\xi^{[3]}_\alpha$ verifies 
\begin{eqnarray}
\overline{d\xi^{[3]}_\alpha} = \overline{d\xi^{[3]}_\alpha d\xi^{[3]}_\beta}&=& 0 , \\
\overline{d\xi^{[3]}_\alpha d\xi^{[3]}_\beta d\xi^{[3]}_\gamma} &\neq& 0.
\end{eqnarray}
Then without going into details, the method presented in step 2 can be generalized.
The average evolutions of the trial states will be given by
\begin{eqnarray}
\overline{\left| d Q_a \right>}  
&=& \frac{dt}{i\hbar}\left\{ {\cal P}_1  + {\cal P}_2 + {\cal P}_3 + \cdots \right\} H \left| Q_a \right> \nonumber \\
\overline{\left< d Q_b \right|}  &=& -\frac{dt}{i\hbar}\left< Q_b \right|  H \left\{ {\cal P}_1  + {\cal P}'_2 + {\cal P}'_3 + \cdots \right\} \nonumber
\label{eq:stocgen}
\end{eqnarray}
where the first terms contain all the information on the evolution of the $\left< A_\alpha \right>$, the 
second terms contain all the information on the evolution of the $\left< A_\alpha A_\beta \right>$ which is not
accounted for 
by the first term, the third terms contain all the information on the evolution of the $\left< A_\alpha A_\beta A_\gamma 
\right>$ which is not contained in the first two terms, ... 
The procedure described here gives an exact Monte-Carlo formulation of a given problem if the Hamiltonian 
$H$ applied on $\left| Q_a \right>$ or $\left< Q_b \right|$ 
can be written as a polynomial of $A_\alpha$. If the polynomial is of order $k$, then the sum stops 
at ${\cal P}_k$.

\subsubsection{Summary and discussion on applications}

Considering a restricted class of trial state vectors associated to a set of observable $A_\alpha$, a hierarchy of stochastic 
approximations can be obtained. The method discussed here insures that at
the level $k$ of the hierarchy, all moments of order $k$ or below of the observable $A_\alpha$ evolve 
according to the exact Ehrenfest equation over short time. 
The Monte-Carlo formulation might  becomes exact if the Hamiltonian applied to the trial state writes as a 
polynomial of the $A_\alpha$ operators. 
  
Aside of the use of variational techniques, we end up with the following important conclusion:
{\it Given an initial density $D=\left| Q_a \right>\left< Q_b \right|$ where both states belongs to a
given class of trial states associated to a set of operators $A_\alpha$, we can always 
find a Monte-Carlo process which preserves the specific form of $D$ and insures that expectations values of 
all moments of the $A_\alpha$ up to a certain order $k$ evolve in average according to the Ehrenfest theorem 
associated to the exact Hamiltonian at each time step and along each trajectory.}  

This statement is referred to as the "existence theorem" in ref. \cite{Lac07}. Such a general statement 
is very useful
in practice to obtain stochastic processes.
Indeed, the use of variational techniques might become rather complicated 
due to the large number of degrees of freedom involved. 
An alternative method is to take advantage of the natural link made
between the average effective evolution deduced from the stochastic evolution 
and the phase-space dynamics. Indeed, according 
to the existence theorem, we know that at a given  
level $k$ of approximation, the dynamics of each trial state can be simulated by an average effective 
Hamiltonian insuring that all moments of order $k$ or below matches the exact evolution.
In practice, it is easier to express the exact evolution of the moments and then "guess" the associated stochastic 
process. Many examples taken from general quantum mechanics, atomic physics, interacting bosons or fermions 
have been given in ref. \cite{Lac07}. 

As an illustration, let us come back to the problem of interacting fermions with a two-body Hamiltonian. Assuming that at a given time step, the exact density can be recovered 
by averaging over an ensemble 
of densities  
\begin{eqnarray}
D = \left| \Phi_a \right>\left< \Phi_b \right|,
\label{eq:phiaphib}
\end{eqnarray}
where both states correspond to SD states. If we denote by $\left\{ \left| \beta_i \right> \right\}_{i=1,N}$ 
and $\left\{ \left| \alpha_i \right> \right\}_{i=1,N}$ the set of $N$ single-particle states, we assume in 
addition that for each couples of SD, associated singles-particle wave-functions verify  
$\left< \beta_j \left.  \right| \alpha_i \right> = \delta_{ij}$. Accordingly, the one-body density matrix
associated to a given $D$ reads \cite{Low55,Lac05,Lac06a} 
\begin{eqnarray}
\rho_1 = \sum_i \left| \alpha_i \right> \left< \beta_i \right|.
\label{eq:rhofermion}
\end{eqnarray}     
It can be easily verified that $\rho^2_1 = \rho_1$ and $Tr(D)=1$.
For each $D$ given by eq. (\ref{eq:phiaphib}), the two-body density 
writes as $\rho_{12} = (1-P_{12}) \rho_1 \rho_2$. 
The evolution of $\rho_1$ and $\rho_{12}$ over one time step are given by the 
two first equations of the BBGKY hierarchy which reads in that case
\begin{eqnarray}
i \hbar \frac{d}{dt}\rho_1 &=& \left[h_{MF},\rho_1 \right] ,
\label{eq:ro1fermion}\\  
i\hbar \frac{ d }{dt } \rho_{12} &=& \left[ h_{MF}(1)+h_{MF}(2),\rho_{12} \right] \nonumber \\
&+& ( 1 - \rho_1)(1- \rho_2)v_{12} \rho_{1}\rho_{2} \nonumber \\
&-&\rho_{1}\rho_{2}v_{12} ( 1 - \rho_1)(1 - \rho_2).
\label{eq:ro12fermion}
\end{eqnarray}
Again, decomposing the interaction as a sum over separable terms built from a complete set 
of hermitian operators $O_n$ (Eq. (\ref{eq:dvoo})), the previous expression can be 
simulated by a stochastic dynamics in phase-space given by 
\begin{eqnarray}
d \rho_1 &=& \frac{dt}{i\hbar} \left[h_{MF}, \rho_1  \right] + \sum_{n} d\xi^{[2]}_n (1-\rho_1) 
O_n \rho_1 \nonumber \\
&+& \sum_{n} d\eta^{[2]}_n  
\rho_1 O_n (1-\rho_1) ,
\label{eq:dphifer}
\end{eqnarray}
where $d\xi^{[2]}_\lambda$ and $d\eta^{[2]}_\lambda$ are two sets of independent stochastic variables with mean zero and 
verifying $d\xi^{[2]}_n d\xi^{[2]}_{n'} = \delta_{nn'} \frac{dt}{i\hbar} \lambda_n$  and 
$d\eta^{[2]}_n d\eta^{[2]}_{n'} = -\delta_{nn'} \frac{dt}{i\hbar} \lambda_n$. This stochastic master equation is exact 
and can equivalently be replaced by a Stochastic Schr\"odinger equation for single-particle 
wave-functions given by
\begin{eqnarray}
d \left| \alpha_i \right> &=& \Big( \frac{dt}{i\hbar} h_{MF} + \sum_n d\xi^{[2]}_n (1-\rho_1) 
O_n  \Big) \left| \alpha_i \right>, \nonumber \\
d \left< \beta_i \right| &=& \left< \beta_i \right| 
\Big( -\frac{dt}{i\hbar} h_{MF} + \sum_n d\eta^{[2]}_n  O_n (1-\rho_1) \Big). \nonumber 
\end{eqnarray}  
This stochastic equation preserves the property 
$\left< \beta_j \left.  \right| \alpha_i \right> = \delta_{ij}$.
Therefore, it corresponds in many-body space to a Monte-Carlo procedure which 
transforms the initial set of densities into another set of densities with identical properties. 

Using the present method, quantum Monte-Carlo approach to a closed system can be rather easily guessed. 
Application of QMC remains very challenging. First, in most physical cases, statistical fluctuations 
around the mean trajectory become very large for long time evolutions. As a consequence, the 
number of trajectories necessary to properly describe the problem increases very fast and prevent from 
using such a technique. Specific methods, that explicitly use the QMC flexibility, can however be proposed 
to reduce statistical fluctuations \cite{Lac05}. Second, implementation of QMC requires to solve non-linear stochastic 
equations. It turns out that trajectories can make large excursion in unphysical regions of the phase-space 
leading to unstable trajectories (also called spikes). This is a problem which seems to be recurrent
in the context of quantum stochastic mechanics both with Stochastic Schroedinger Equation \cite{Car01} or
stochastic evolution in phase-space \cite{Gar00}. Therefore, to take full advantage of these techniques 
one should develop specific numerical methods. 
This has been done for instance in refs. \cite{Car01,Pli01,Deu02} using the fact 
that stochastic equations are generally not unique.          

\section{Summary}

In this review, we have summarized some of the possible ways to extend TDHF, some of them are able 
to incorporate pairing correlations (like TDHFB or TDDM) whereas others concentrates on direct nucleon-nucleon
collisions (ETDHF, STDHF) or initial correlation effects (SMF). Table \ref{tab:approches} gives an overview 
of the theory introduced here while figure \ref{fig:stochastic} illustrates the differences between 
the three stochastic methods, namely Stochastic TDHF, SMF and QMC.  
While very promising applications of these theories to the nuclear many-body problem 
remain very challenging and some of the above theories have never been used. A first difficulty is
the computational effort required to treat time dependent methods beyond mean-field. However, besides numerical 
difficulties, more fundamental problems persist.   
Indeed, a second critical aspects which has not been discussed here is that all applications 
of dynamical quantum transport theories to nuclear reactions are nowadays 
possible thanks to the introduction of effective interactions (essentially Skyrme like). 
These interactions have led to the more 
general concept of Energy Density Functional (EDF) and are expected, in a similar way as Density Functional 
Theory (DFT) in condensed matter, to incorporate most of the correlations already at the mean-field level. 
Then, the very notion of "mean-field " and/or "beyond mean-field" framework becomes ill defined. 
All theories presented in this chapter
(extended, stochastic...) start from a Many-Body Hamiltonian. In the EDF context, such an 
Hamiltonian, although it exists, is not simply connected to the EDF itself. As a consequence, the Hamiltonian 
derivation could only serve as a guideline and a proper formulation in the EDF framework is mandatory. Large debates
exist nowadays on the validity and foundation of the nuclear EDF applied to static properties of nuclei.


\section*{Acknowledgment}  
We would like to thank K. Washiyama and B. Yilmaz that contributed significantly 
to the application of SMF to heavy-Ion collisions. We also thank D. Gambacurta and G. Scamps 
for discussion and 
collaboration on superfluid systems.
This work is supported in part by the US DOE Grant No. DE-FG05-89ER40530. 

\appendix


\section{Density matrices}
\label{app:dens}
In this appendix, some relationship and definition related to densities in many-body systems and
that are useful in this article, are summarized.
Given a many-body state $\left| \Psi  \right>$. We define the k-body density as 
\begin{eqnarray}
\left< k'\cdots 1'\left| \rho_{1,\cdots,k} \right| 1 \cdots k \right> = \left< \Psi \left| a^+_1 \cdots a^+_k a_{1'} 
\cdots a_{k'}   \right| \Psi \right>
\end{eqnarray} 
This is equivalent to define the $1$, $2$, $3$-body density as (note the $1/k!$ factor compared to \cite{Low55}) 
\begin{eqnarray}
\begin{array} {l}
\displaystyle \rho_1 (x_1 | x'_1) = A \int \Psi(x_1,\cdots,x_A) \Psi^*(x_1,\cdots,x_A) d{(2 \cdots A)}, \nonumber \\
\nonumber \\
\displaystyle \rho_{12} (x_1,x_2 | x'_1,x'_2) =  A(A - 1) \nonumber \\
\displaystyle \hspace*{1cm} \times \int \Psi(x_1,x_2\cdots,x_A) \Psi^*(x_1,x'_1,\cdots,x_A) d{(3 \cdots A)} ,
\nonumber \\
\cdots  
\end{array}
\label{eq:rr2} 
\end{eqnarray} 
Here $A$ is the number of particles and the notation 
\begin{eqnarray}
d{(k \cdots A)} = dx_k \cdots dx_A dx'_k \cdots dx'_A ,
\end{eqnarray} 
is used. 
With this relations, densities are normalized as 
\begin{eqnarray}
Tr(\rho_{1,\cdots, k}) = \frac{A!}{(A-k)!},
\end{eqnarray}
and verifies the recurrence relation
\begin{eqnarray}
\rho_{1\cdots k} = \frac{1}{A-k} Tr_{k+1} \rho_{1,\cdots,k+1}.
\end{eqnarray}

\subsection{Two and three-body Correlations}

$C_{12}$ denotes the two-body correlation matrix and is defined from 
\begin{eqnarray}
C_{12} = \rho_{12} - \rho_1 \rho_2 (1-P_{12})
\end{eqnarray} 
where $P_{12}$ is the permutation operator ($P_{12}| i~j \rangle = | j~i \rangle$)
Properties of $C_{12}$ are essentially those of $\rho_{12}$ : 
\begin{itemize}
\item {\bf Hermiticity}: 
\begin{eqnarray}
\left\langle {\mathbf x}_1 , {\mathbf x}_2 | C_{12} | {\mathbf x'}_1 , {\mathbf x'}_2  \right\rangle &=&
\left\langle {\mathbf x'}_1 , {\mathbf x'}_2 | C_{12} | {\mathbf x}_1,{\mathbf x}_2  \right\rangle^* 
\end{eqnarray}  
\item {\bf Anti-symmetry}: 
\begin{eqnarray}
\left\langle {\mathbf x}_1 , {\mathbf x}_2 | C_{12} | {\mathbf x'}_1 , {\mathbf x'}_2  \right\rangle &=& - 
\left\langle {\mathbf x}_1 , {\mathbf x}_2 | C_{12} | {\mathbf x'}_2 , {\mathbf x'}_1  \right\rangle 
\end{eqnarray}
\item {\bf Consistency between the two and one-body density}: due to the fact that the one-body density can be obtained
from the two-body density, $C_{12}$ should fulfill some relations. We have (using ${\rm Tr} (\rho_1) = N$)
\begin{eqnarray}
{\rm Tr}_2 \rho_{12} = {\rm Tr}_2 C_{12} + N \rho_1 - \rho^2_1,
\end{eqnarray}  
since we should also have 
\begin{eqnarray}
{\rm Tr}_2 \rho_{12} = (N-1) \rho_1,
\end{eqnarray}
we finally deduce the consistency relation 
\begin{eqnarray}
Tr_2 C_{12} &=& -\rho_1(1 - \rho_1).
\end{eqnarray}
\end{itemize}

The three-body correlation is defined as 
\begin{equation}
\begin{array}{ll}
\displaystyle C_{123}= & \rho _{123}-\rho _{1}\rho _{2}\rho
_{3}\left( 1-P_{12}\right) \left( 1-P_{13}-P_{23}\right) \\
&-\rho _{1}C_{23}\left( 1-P_{12}-P_{13}\right) \nonumber \\
\displaystyle &-\rho_{2}C_{13}\left( 1-P_{21}-P_{23}\right) \\ 
\displaystyle & -\rho _{3}C_{12}\left( 1-P_{31}-P_{32}\right) \end{array}
\label{eq:c123}
\end{equation}
Similarly to the two-body case, the three-body correlation matrix verifies a large 
number of properties associated to anti-symmetry, particle number conservation...



\section{Correlations between observables and Projection techniques}
\label{sec:acp}

To properly introduce the projection onto a subspace of observables, the notion 
of independence and correlation between observables should be first discussed. The 
strategy followed here is essentially the same as in the Principal 
Component Analysis (PCA) used in statistical analysis.   
Let us consider a set of operators $\{ A_{\alpha} \}$ and a 
density $D$ describing the properties of a system at a given time which is interpreted as a probability. The $\{ A_{\alpha} \}$
form a subset of the total space of observables. In the following, It is shown
how any other observables can be projected out on this subset. Part of the 
method presented here has been used to introduce stochastic mean-field approaches 
in closed system in ref. \cite{Lac07}.
Readers that are not interest in technical details may skip this part 
and directly jump toappendix \ref{sec:projecteddyn}. \\

\noindent {\bf Creation of an independent set of operators in the $\{ A_{\alpha} \}$ subspace:} 
Observables $\{ A_{\alpha} \}$ are not necessarily statistically
independent from each others with respect to the state $D$. In the following, we will just say that they are $D$-correlated 
or $D$-independent in the opposite case. To measure correlation between observables, we introduce the  
variance-covariance 
matrix defined as \footnote{It is worth mentioning that the strict equivalent of statistical mechanics
would be the symmetric quantity:
\begin{eqnarray}
C'_{\alpha \beta} = \frac{1}{2} \langle A_\alpha A_\beta+A_\beta A_\alpha   \rangle -  
\langle A_\alpha \rangle \langle A_\beta  \rangle.
\end{eqnarray}
Strictly speaking, only the above quantity can be regarded as a 
scalar product. However, as it will become clear in the following, it is 
more convenient to define the non-symmetric $C_{\alpha\beta}$.   
 }:
\begin{eqnarray}
C_{\alpha \beta} =  \langle A_\alpha A_\beta  \rangle -  \langle A_\alpha \rangle \langle A_\beta  \rangle ,
\label{eq:corstat}
\end{eqnarray} 
has non zero off-diagonal matrix elements. We assume here that the $\{ A_\alpha \}$ are hermitian operators 
implying that 
$C$ is also hermitian. 
Below, the different notations: 
\begin{eqnarray}
C_{\alpha \beta} &=& C(A_\alpha,A_\beta) = \langle\langle  A_\alpha |  A_\beta \rangle\rangle
\label{eq:scalar}
\end{eqnarray} 
will be used. In the following, it is 
assumed that $C$ is not singular. Note that, if it is the case, it does only mean that
their is redundant information and that the subset of observables can be further reduced. 
$C$ could be diagonalized by a unitary transformation $U$ and has only positive 
eigenvalues denoted by $\lambda_\alpha$\footnote{Note that the $\lambda_i$ measure the information content with respect 
to $D$ of the new 
operators $e_\alpha$.}. 
It is then convenient to
introduce a new set of operators $\{e^\dagger_\alpha\}$, defined from the relationship
\begin{eqnarray}
e_\alpha =  \frac{1}{\sqrt{\lambda_\alpha}} \sum_\beta U^{-1}_{\alpha \beta} (A_\beta - \left< A_\beta \right>) 
\label{eq:newop}
\end{eqnarray}   
It is worth to mention that these operators are explicitly dependent on the density $D$.
Using this definition, we have $\langle e_\alpha \rangle = \langle e^\dagger_\beta \rangle= 0$ while 
\begin{eqnarray}
\langle \langle e^\dagger_\alpha | e_\beta \rangle \rangle = 
\frac{1}{\sqrt{\lambda_\alpha \lambda_\beta}}\left( U^{-1} C U \right)_{\alpha \beta} = \delta_{\alpha \beta} 
\end{eqnarray}
Therefore, couples of operators $(e^\dagger_\alpha ,e_\beta)$ are $D$-independent.
We also have the inverse relation
\begin{eqnarray}
A_\alpha - \left< A_\alpha \right>&=&   \sum_\beta \sqrt{\lambda_\beta}U_{\alpha \beta}e_\beta =  
\sum_\beta \sqrt{\lambda_\beta} e^\dagger_\beta U^{-1}_{\beta \alpha};
\end{eqnarray}
provided that the $\{ A_{\alpha} \}$ are hermitian operators. \\

\noindent{\bf Projection of observables: } With the aid of eq. (\ref{eq:scalar}) and new operators 
(eq. \ref{eq:newop}), any 
observable, denoted by $B$ could be projected onto the subspace of the $\{A_\alpha \}$. 
Let us now consider a new operator $B$ and assume that it is eventually partially correlated to $A_\alpha$, the new operator
\begin{eqnarray}
B^{\perp} = B- \sum_\alpha e_\alpha
\langle \langle e^\dagger_\alpha|  B  \rangle \rangle
\end{eqnarray}
is statistically independent of the $\{ A_\alpha \}$ with respect to $D$ ($D$-independent). 
First, $B^{\perp}$  verifies $ \langle B^{\perp} \rangle = \langle B \rangle$, while 
for any operator $e^\dagger_\beta$, we have 
\begin{eqnarray}
\langle \langle B^\perp | e^\dagger_\beta   \rangle \rangle &=& 
\langle \langle B | e^\dagger_\beta   \rangle \rangle  - \sum_\alpha 
\langle \langle B | e^\dagger_\alpha   \rangle \rangle 
\langle \langle e^\dagger_\beta   | e_\alpha \rangle \rangle = 0
\end{eqnarray}  
due to $\langle \langle e^\dagger_\beta   | e_\alpha \rangle \rangle = \delta_{\alpha \beta}$.
 Since, the $A_\alpha$ are linear combination of the $e^\dagger_\beta$, 
$\langle \langle B^\perp | A_\alpha  \rangle \rangle = 0$ for any $\alpha$. 
The new operator can also directly be expressed in terms of the operator $A_\alpha$, we finally obtain
\begin{eqnarray}
B^{\perp} = B- \sum_{\alpha \beta} (A_\alpha - \langle A_\alpha \rangle)C^{-1}_{ \alpha \beta} 
\langle \langle A_\beta | B   \rangle \rangle, 
\label{eq:statdec}
\end{eqnarray}
or written differently
\begin{eqnarray}
B &=& B^{\parallel} + B^{\perp} \hspace*{1.cm} 
\end{eqnarray}
with
\begin{eqnarray}
B^{\parallel}  = \sum_{\alpha \beta} (A_\alpha - \langle A_\alpha \rangle)C^{-1}_{ \alpha \beta} 
\langle \langle A_\beta | B   \rangle \rangle .
\end{eqnarray}
Therefore, using the same technique as the Principal Component Analysis, any operators can be decomposed into 
two operators, the second one is statistically independent from the observables $\{ A_\alpha \}$, while the first 
one could be written as a linear combination of the $\{ A_\alpha \}$ and contains all the information on the correlation 
between $B$ and the latter observables. Properties of the two operators $B^{\parallel}$ and $B^{\perp}$ are:
\begin{eqnarray}
\langle B^{\parallel}  \rangle &=& 0, \hspace*{0.9cm} {\rm and} \hspace*{0.5cm} 
\langle \langle B^{\parallel} | A_\alpha  \rangle\rangle  =  
\langle \langle B | A_\alpha  \rangle \rangle, \\
\langle B^{\perp} \rangle &=& 
\langle B \rangle  ,  \hspace*{0.5cm} {\rm and} \hspace*{0.5cm} 
\langle \langle B^{\perp} | A_\alpha  \rangle \rangle =  0 ,
\hspace*{1.cm} 
\end{eqnarray}  
valid for any $\alpha$. In the limit where $B$ is fully described in the subspace of the $\{ A_\alpha \}$, then $B^{\perp}$ 
simply identifies with a number $\langle B \rangle$. If on opposite case, $B$ is statistically independent from these observables, 
$B^\parallel = 0$.   \\

\begin{figure}
\resizebox{0.45\textwidth}{!}{
  \includegraphics{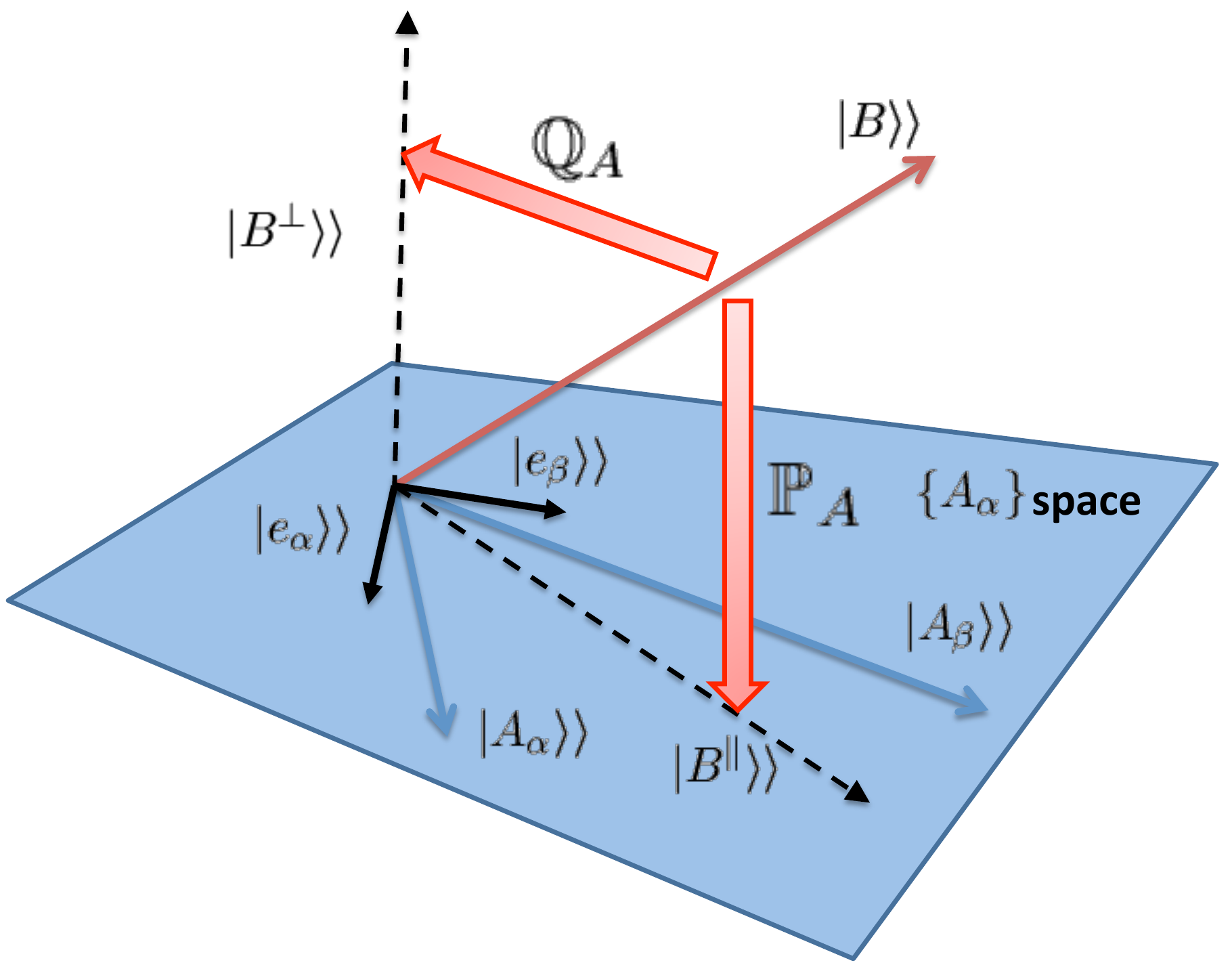}
}
\caption{Schematic representation of the projection technique assuming only two observables 
$(A_\alpha , A_\beta )$. Different entities introduced in the text are shown.  }
\label{fig:proja}      
\end{figure}

\noindent{\bf Projection operators:}
Using the notation $| B \rangle\rangle$, two projectors denoted respectively by $\mathbb{P}_A$ and $\mathbb{Q}_A$, 
can be introduced with 
\begin{eqnarray}
| B^{\parallel} \rangle\rangle &=& \mathbb{P}_A | B \rangle\rangle, \nonumber \\
| B^{\perp} \rangle\rangle &=& \mathbb{Q}_A | B \rangle\rangle = (1-\mathbb{P}_A) | B \rangle\rangle,
\end{eqnarray} 
with the convention 
\begin{eqnarray}
\mathbb{P}_A \equiv  \sum_\alpha | e^\dagger_\alpha   \rangle \rangle 
\langle \langle e^\dagger_\alpha   | = \sum_{\alpha \beta} | A_\alpha   \rangle \rangle C^{-1}_{\alpha \beta}  
\langle \langle A_\beta  | .
\end{eqnarray}
It could be easily checked that $\mathbb{P}^2_A = \mathbb{P}_A$ and $\mathbb{Q}^2_A = \mathbb{Q}_A$ and therefore verify 
standard properties of projectors. The projection onto the subspace of $\{ | A \rangle\rangle \}$ 
is illustrated  
in figure \ref{fig:proja}.

\section{Mean-field as a projected dynamics}
\label{sec:projecteddyn} 

As we have seen in section \ref{sec:meanfield}, the mean-field approach is a powerful 
method to select relevant degrees of freedoms (DOF) and provide an optimal evolution of them 
in the absence of knowledge of others irrelevant DOF at least for short time. 
It is often said that mean-field theory  "corresponds to the optimal projected evolution onto a sub-space 
of the total observable manifold". The present appendix goal is to illustrate what is hidden behind this sentence. 
In section \ref{sec:ehrenfest}, we already have shown
that variational principle used in combination with variational states given by (\ref{eq:variastate}) leads 
to the exact evolution of the $\{ A_\alpha \}$ over short time. It is shown in the present appendix that mean-field 
evolution corresponds to a projected dynamic onto the subspace of variables $\{ A_\alpha \}$ where the projection 
is nothing but the statistical projection in appendix \ref{sec:acp}.

\subsection{Projector associated to mean-field}

Trial states given by eq. (\ref{eq:variastate}) are not a priori normalized along the path. To enforce normalization, an additional 
parameter is generally added \cite{Fel00}. Equivalently, one could slightly modify equation (\ref{eq:variastate}) as
\begin{eqnarray}
| {\mathbf Q} + \delta {\mathbf Q}  \rangle = 
e^{\sum_\alpha \delta q_\alpha (A_\alpha - \langle  A_\alpha \rangle)} | {\mathbf Q}  \rangle.
\label{eq:variastateq}
\end{eqnarray}  
This automatically insures a constant normalization of the state along the path. 
Accordingly equation (\ref{eq:evolq}) now becomes:
\begin{eqnarray}
i\hbar \langle  {\mathbf Q}  | (A_\alpha - \langle  A_\alpha \rangle )  | \dot {\mathbf Q}  \rangle 
&=& \langle  {\mathbf Q}  | (A_\alpha - \langle  A_\alpha \rangle ) H | {\mathbf Q}  \rangle, \label{eq:meanlin}
\end{eqnarray} 
where $ | \dot {\mathbf Q}  \rangle $ could be written in terms of the $\{ q_\alpha \}$ evolutions as 
\begin{eqnarray}
| \dot {\mathbf Q}  \rangle  &=& \sum_\alpha \dot q_\alpha (A_\alpha - \langle  A_\alpha \rangle ) | {\mathbf Q}  \rangle. 
\end{eqnarray} 
The evolution of ${\mathbf Q}$ given above is nothing but an approximate evolution with an effective Hamiltonian written in terms of 
a linear combination of the $\{ A_\alpha \}$ operators. Combining these equations leads to
\begin{eqnarray}
i\hbar \sum_{\beta} \dot q_\beta C_{\alpha \beta} &=&  \langle  {\mathbf Q}  | 
(A_\alpha - \langle  A_\alpha \rangle ) H | {\mathbf Q}  \rangle, 
\end{eqnarray}  
where $C$ denotes the correlation matrix whose components are defined in eq. (\ref{eq:corstat}). Inverting the equation  
to obtain $\dot q_\alpha$ explicitly finally gives:  
\begin{eqnarray}
i\hbar | \dot {\mathbf Q}  \rangle  &=&  \Big\{
\sum_{\alpha \beta} (A_\alpha - \langle  A_\alpha \rangle ) | {\mathbf Q}  \rangle ~C^{-1}_{\alpha \beta}~
\langle {\mathbf Q} | (A_\beta - \langle  A_\beta \rangle ) \Big\} H | {\mathbf Q}  \rangle \nonumber \\
&=&  \Big\{ \sum_{\alpha \beta} (A_\alpha - \langle A_\alpha \rangle)C^{-1}_{ \alpha \beta} 
\langle \langle A_\beta | H   \rangle \rangle   \Big\}  | {\mathbf Q}  \rangle \nonumber \\
&=& H^{\parallel}(t) | {\mathbf Q}  \rangle  .
\label{eq:mfdyncomplete}
\end{eqnarray} 
Therefore, the mean-field evolution is indeed 
equivalent to a projected dynamics onto a sub-space containing the relevant information on selected 
observables. More generally, for any observable that are eventually out of the relevant subspace, mean-field 
will provide the best approximation retaining only the optimal path for the $\{ \langle A_\alpha \rangle \}$ 
observables. The projected dynamic corresponds to an effective mean-field Hamiltonian, denoted hereafter simply by 
$H_{\rm MF}(t)$. This Hamiltonian writes as a linear combination of the $\{ A_\alpha \}$ operators and 
identifies with the projected part of $H$ onto the relevant space. Note that same conclusion can be drawn using a slightly different approach based on Liouville 
representation \cite{Bal99}.
 
In a pure density case, i.e. $D = | {\mathbf Q} \rangle\langle {\mathbf Q} |$, 
one can further precise the approximation made by introducing a projector 
${\cal P}_A$ directly acting in Hilbert space 
\begin{eqnarray}
{\cal P}_A (t) &=& 
\sum_{\alpha \beta} (A_\alpha - \langle  A_\alpha \rangle ) | {\mathbf Q}  \rangle ~C^{-1}_{\alpha \beta}~
\langle {\mathbf Q} | (A_\beta - \langle  A_\beta \rangle ).  
\end{eqnarray} 
According to eq. (\ref{eq:mfdyncomplete}), we simply have $H_{\rm MF}(t) = {\cal P}_A(t) H$. 

\section{Mean-field from Thouless theorem}
\label{app:mfthouless}

In this appendix, the mean-field equation are deduced 
by applying the Hamiltonian (\ref{eq:hphi}) without the residual interaction term 
directly to an initial state written as $ \prod_\alpha  a^{\dagger}_\alpha \left| - \right\rangle$
where $\left| - \right\rangle$ is a single-particle vacuum. 
Using the fact that $e^{-\frac{dt}{i\hbar} 
 H_{MF}} e^{\frac{dt}{i\hbar} 
 H_{MF}} = 1$  and $e^{\frac{dt}{i\hbar} 
 H_{MF}}  \left| - \right\rangle = \left| - \right\rangle$
\begin{eqnarray}
e^{\frac{dt}{i\hbar} 
 H_{MF}} \left| \Psi \right\rangle &=& 
e^{\frac{dt}{i\hbar} 
 H_{MF}} \prod_\alpha  a^{\dagger}_\alpha \left| - \right\rangle \nonumber \\
&=& e^{\frac{dt}{i\hbar} 
 H_{MF}} a^{\dagger}_{\alpha_1} 
e^{-\frac{dt}{i\hbar} 
 H_{MF}} e^{\frac{dt}{i\hbar} 
 H_{MF}} a^{\dagger}_{\alpha_2} e^{\frac{dt}{i\hbar} 
 H_{MF}} \nonumber \\
 && \cdots e^{-\frac{dt}{i\hbar} 
 H_{MF}} a^{\dagger}_{\alpha_N} e^{-\frac{dt}{i\hbar} 
 H_{MF}} e^{+\frac{dt}{i\hbar} 
 H_{MF}}\left| - \right\rangle . \nonumber  
\end{eqnarray}
Considering the transformation of each creation operator separately, we have 
\begin{eqnarray}
 e^{\frac{dt}{i\hbar} 
 H_{MF}}  a^{\dagger}_{\alpha}  e^{-\frac{dt}{i\hbar} 
 H_{MF}} &=& a^{\dagger}_{\alpha} + \frac{dt}{i\hbar} [ H_{MF}, a^{\dagger}_{\alpha}] + O(dt) \nonumber \\
&=& a^{\dagger}_{\alpha} + \frac{dt}{i\hbar} \sum_i  a^\dagger_i \left\langle i \left| h[\rho] \right| \alpha 
\right\rangle + O(dt) \nonumber \\
&\equiv& a^{\dagger}_{\alpha + d \alpha } + O(dt) ,
\end{eqnarray}
where the expression of the mean-field operator defined in eq. (\ref{eq:hphi}) has been used and where 
"i" refers to the complete original basis. From the above identity, we see that, the propagated many-body 
state writes:
\begin{eqnarray}
\left| \Psi(t+dt) \right\rangle \propto \prod_{\alpha} a^{\dagger}_{\alpha + d \alpha } 
\left| - \right\rangle,
\end{eqnarray}  
where, using $\sum_i \left| i \right\rangle\left\langle i \right|=1$, the single-particle 
states evolves according to 
\begin{eqnarray}
i\hbar \frac{d\left| \alpha \right\rangle}{dt} = h[\rho] \left| \alpha \right\rangle,
\end{eqnarray}
which is nothing but the standard mean-field evolution.



\section{Ito calculus}
\label{sec:ito}

\subsection{Basic rules for Ito stochastic calculation}
We are considering a stochastic evolution 
\begin{equation}
dx=a(x)dt+b(x)\xi \left( t\right) dt
\end{equation}
which could be integrated as 
\begin{equation}
x\left( t\right) =x_{0}+\int a(x)dt+\int b(x)\xi \left( t\right) dt.
\end{equation}
The problem is to define the second integral. We define the stochastic
function 
\begin{equation}
W\left( t\right) =\int_{0}^{t}\xi \left( t\right) dt
\end{equation}
The fondamental formula is 
\begin{equation}
\begin{array}{ll}
\left\langle W\left( t\right) W\left( t^{\prime }\right) \right\rangle & 
=\int_{0}^{t}\int_{0}^{t^{\prime }}\left\langle \xi \left( s\right) \xi
\left( s^{\prime }\right) \right\rangle dsds^{\prime } \\ 
& =\int_{0}^{\min \left( t,t^{\prime }\right) }\delta \left( s-s^{\prime
}\right) ds=\min \left( t,t^{\prime }\right).
\end{array}
\end{equation}
Thus, we have 
\begin{equation}
dx=a(x)dt+b(x)dW.
\end{equation}
In the Ito rule for the evaluation of integrals, we define 
\begin{equation}
\int G\left( t\right) dW=\sum_{i=1}^{n}G\left( t_{i-1}\right) \left( W\left(
t_{i}\right) -W\left( t_{i-1}\right) \right).
\end{equation}

{\bf Rules} :

\begin{equation}
\left\{ 
\begin{array}{l}
dWdW=dt \\ 
dW^{2+N}=0
\end{array}
\right. ,
\end{equation}
If $f$ is a function of $W$ such that $f=f\left( W,t\right) $, we have the
rule $df=\left[ \frac{\partial f}{\partial t}+\frac{1}{2}\frac{\partial ^{2}f%
}{\partial W^{2}}\right] dt+\frac{\partial f}{\partial W}.$

{\bf Example }: 
\begin{equation}
\begin{array}{ll}
d\left( \exp \left( W\right) \right) & =\exp \left( W+dW\right) -\exp \left(
W\right) \\ 
& =\exp \left( W\right) \left( dW+\frac{1}{2}dt\right).
\end{array}
\end{equation}

{\bf Correlation formula} :

Let $G\left( t\right) $ (and $F\left( t\right) $) be a non-anticipating
function, i.e. a function independent on the future, i.e. independent of $%
W\left( s\right) -W\left( t\right) $ if $s>t$. Then, we have the property 
\begin{equation}
\left\langle \int_{0}^{t}G\left( s\right) dW\left( s\right)
\int_{0}^{t}F\left( s^{\prime }\right) dW\left( s^{\prime }\right)
\right\rangle =\int_{0}^{t}\left\langle G\left( s\right) F\left( s\right)
\right\rangle ds \nonumber
\end{equation}

{\bf Changing variables} : If $f$ depends on a variable $x$ that evolves
according to the above stochastic equation, then 
\begin{equation}
df\left[ x\right] =\left\{ a\left( t\right) f^{\prime }\left( t\right) +%
\frac{1}{2}b^{2}\left( t\right) f^{\prime \prime }\left( t\right) \right\}
dt+b\left( t\right) f^{\prime }\left( t\right) dW. \nonumber
\end{equation}

\subsection{Differentiation of specific operators}

\subsubsection{Entropy}

We consider the entropy associated to the density $D$ defined as 
\begin{eqnarray}
S(D) = -Tr (D \ln D).
\end{eqnarray}
In a mean-field approach, since we are considering a specific form for the density along the path, $S(D)$ 
identifies with the one-particle entropy
\begin{eqnarray}
S(\rho) = -Tr \left( \rho \ln(\rho) + (1-\rho) \ln (1-\rho) \right)
\end{eqnarray} 
Let us derive here the variation of entropy associated with the stochastic one-body density. 
We start from 
\begin{eqnarray}
dS(\rho) &=& S(\rho+d\rho) -S(\rho).
\end{eqnarray}
For any operator such that $u^2 = u$, we have 
\begin{eqnarray}
d\left[u \ln u  \right] = du \ln(u) + du - \frac{du ^2}{2} + \frac{du ^2}{u}.
\end{eqnarray}
We finally obtain
\begin{eqnarray}
dS(\rho) &=& - Tr \left\{ d\rho \ln\left( \frac { 1- \rho} {\rho} \right) + d\rho^2 \left[ 1-\frac{ 1 }{\rho } -\frac{ 1 }{1- \rho } \right]\right\} \nonumber
\end{eqnarray} 
Using the fact that $ d\rho^2 = 0 + O(dt)$, we finally obtain  
\begin{eqnarray}
dS(\rho) &=& - Tr \left\{ d\rho \ln\left( \frac { 1- \rho} {\rho} \right)\right\}
\end{eqnarray}   
Therefore, the average evolution of $S\left( \rho \right)$ is equal to zero.

\subsection{evolution of $df^{-1}$}

 {\bf Expression of }$df$ $^{-1}$:{\bf \ }using $f^{-1}f=1$, we have $%
df^{-1.}f+f^{-1}.df+df^{-1}.df=0$, we thus have 
\begin{eqnarray}
df^{-1}&=&-f^{-1}.df.\left( f+df\right) ^{-1}=-f^{-1}.df.\left(
1+f^{-1}df\right) ^{-1}f^{-1} \nonumber \\
&=&-f^{-1}.df.f^{-1}+f^{-1}.df.f^{-1}df.f^{-1} .
\end{eqnarray}

\subsection{Evolution of a determinant}

{\bf Evolution of }$\det \left( f\right) ${\bf \ : }using 
\begin{eqnarray}
\det\left( f+df\right) =\det \left( f\right) \det \left( 1+f^{-1}df\right),
\end{eqnarray} 
and 
\begin{eqnarray}
\det \left( 1+M\right) =1+Tr\left( M\right) -\frac{1}{2}\left( Tr\left(
M^{2}\right) -Tr\left( M\right) ^{2}\right) + \ldots \nonumber 
\end{eqnarray}
we obtain 
\begin{eqnarray}
\det \left( 1+f^{-1}df\right) &=& 1+{\rm Tr}\left( f^{-1}df\right) -\frac{1}{2}%
\big\{ {\rm Tr}\left( f^{-1}df.f^{-1}df\right) \nonumber \\
&-& {\rm Tr} \left( f^{-1}df\right)^{2}\big\} . 
\end{eqnarray}



\end{document}